\documentclass[aps,prmat,reprint]{revtex4-2}
\usepackage{amsmath}
\usepackage{amssymb}
\usepackage{cases}
\usepackage{upgreek}
\usepackage{graphicx}
\usepackage{xcolor}
\begin{document}

\title{Chemical pressure and vacancies in crystals under nonhydrostatic stress}

\author{Michiel Sprik}
\email{ms284@cam.ac.uk}
\affiliation{Yusuf Hamied Department of Chemistry, University of Cambridge, Lensfield Road, Cambridge CB2 1EW, United Kingdom}

\begin{abstract} Inhomogeneous stress is a driving force for diffusion of vacancies in crystals. The other way around a non-uniform distribution of vacancies induces stress. The accepted theory of composition-stress coupling in crystals is Larch\'{e}-Cahn (LC) theory. Composition is defined in terms of the population of lattice sites with a limit of one-particle per site.  Coupling is modelled by adding a compositional strain term to the elastic strain in the constitutive relation for stress.  An alternative  mechanism, proposed here,  is letting the site binding energy vary with spatial (deformed) density. This generates a chemical pressure adding to the elastic stress.  The governing equations for this model are derived using non-equilibrium continuum thermodynamic methods.  The gradient of the chemical pressure has a dual function  acting both as the drift force for migration and as an effective internal one body force in the Cauchy equation for the elastic stress.  The practical evaluation presented in the paper is restricted to equilibrium properties. We examine the effect of  externally applied non-hydrostatic stress, either in the form of surface tractions or a one-body force density (gravitation).  The model system is a one component crystal with a fixed number of lattice sites.  The number of particles can be variable but is always smaller than the number of lattice sites.  The linear elastic response is modelled by the standard Lam\'{e} stress tensor.  The results for systems deformed by surface tractions are in qualitative agreement with LC theory allowing for differences in the expression for the elastic moduli. Deviations are more serious for the crystal deformed by gravitation.    

\end{abstract}

\maketitle

\section{Introduction} \label{sec:intro}

Diffusion in solids is one of the pillars of modern materials science\cite{balluffi05,divinski14,nix16}. It may seem at first a paradox. How can mass transport be reconciled with the rigidity of solids? For crystals this question was resolved by the concept of a stable lattice. Diffusion is pictured as the hopping of particles to neighbouring empty sites leaving a vacancy behind.  Both the number of particles and the number of lattice are conserved. By implication also the number of vacancies is conserved and the process can be equally represented as diffusion of vacancies. This picture was formalized by Larch\'{e} and Cahn in terms of a network constraint\cite{cahn73}. There is however a limit. Eventually the vacancies reach a defect in the lattice where sites can be added or removed. Creation or destruction of lattice sites  in closed systems necessarily generates or eliminates vacancies. The difference is that vacancies are mobile while lattice sites are not.  Vacancies play therefore a crucial role in the thermodynamics and kinetics of crystals despite their low concentration\cite{balluffi05}.  

The literature on diffusion of vacancies and their interaction with defects is huge. Examples are creep, the migration of grain boundaries and climbing of dislocations\cite{nix16}. The question remains however what drives transport of vacancies between defects. It was agreed  that this cannot only be concentration gradients (Fick's law). Diffusion can be directed by stress as well. Non-uniform distributions of vacancies can also create stress. The coupling of vacancies and stress was modelled similar to composition-stress coupling in binary alloys. One option is Eshelby's continuum mechanics model\cite{eshelby61,balluffi05,nix14}. The minority component is treated as a solute and the other as a solvent matrix.  Their interaction is represented as an elastic misfit of a spherical inclusion in an elastic continuum.  The alternative is inspired by the observation that the effective lattice parameter in binary alloys is linear in the composition (Vegard's law). This was captured in a model by superimposing elastic strain and composition strain.   The elastic strain is,  as usual, the gradient of displacement. Composition strain is proportional to the variation in site occupation coupling composition to elastic deformation. This is the essence of the Larch\'{e}-Cahn (LC)  composition-strain model \cite{cahn73, cahn73,cahn78,cahn82,cahn85} which has become a standard approach in material science  validated by numerous equilibrium \cite{mullins84,mullins85,sekerka89,mishin10,mishin12,cahn13,mishin16}  and non-equilibrium studies\cite{mishin13,mishin15, voorhees24} (for a review see Ref.~\citenum{voorhees04}).

The present investigation explores the effect of modifying the LC composition strain model as proposed in a previous publication\cite{sprik25}.   Chemo-mechanical interaction is represented as a coupling to spatial (deformed) density rather than occupation only. Occupation is the number of particles per lattice cell and is preserved under deformation of the cell. Spatial density, while proportional to occupation, in addition scales with the volumetric strain.  The result  is an internal density dependent chemical pressure which alters the LC thermomechanical response. In Ref.~\citenum{sprik25} we only considered  uniform crystals under hydrostatic stress ignoring shear deformation. These restriction will be lifted in the present work.

A further deviation from standard LC theory is the use of non-equilibrium continuum thermodynamics. LC derived the equilibrium thermodynamics by applying the Gibb's variational principle. Diffusion is modelled by a phenomenological generalization of Fick's law.  The advantage  of non-equilibrium thermodynamics is consistent treatment of transport opening a new  perspective on the chemical potential of a crystal which has proven a troublesome concept. The diffusive flux is proportional to the gradient of the chemical potential as obtained from the inhomogeneous free energy density. We closely follow the textbook by Gurtin, Fried and Anand (GFA)\cite{gurtin10}, which is the main source for the theoretical development presented here. In the application to LC theory we were led by a recent paper by Sinisa Mesarovic (SM)\cite{mesarovic16}.  Building on earlier work by Berdichevsky \cite{berdichevsky97} and Garikipati et al.~\cite{garikipati01} Mesarovic showed how to formally separate the unbound diffusive motion of particles from displacement of the lattice sites.

Theory and modelling of the coupling of stress to composition and diffusion have a long history.  The process is not only of paramount importance in metallurgy but also in the softmatter physics of  the swelling of gels\cite{zhao08,doi09}.  Particularly relevant is the theory developed by Baek and Srinivasa\cite{srinivasa2004} for diffusion and absorption of liquids in elastomers.  Rather than a mixture of liquid and solid the gel is viewed as a one-component single phase system supporting elasticity and migration of mass \cite{srinivasa2004,pence2011,steigmann2013,anand10,anand11,anand15,morro16}. This is similar to a LC crystal. The complication is that the  continuum thermodynamics of gels is constrained by the requirement that total volume is the sum of the absorbed liquid  and dry solid. This socalled swelling constraint is not relevant for transport in LC solids. This is a major simplification\cite{srinivasa2004,steigmann2013, morro16} enabling a direct connection of the Baek-Srinivasa approach to the diffusion of particles in a crystal as conceived by Mesarovic\cite{mesarovic16,berdichevsky97,garikipati01}.

The present work extends the author's 2025 JCP paper on the thermodynamics of vacancies\cite{sprik25} in uniform systems under hydrostatic pressure.  The 2025 publication was preceded by two studies focusing on the liquid-solid interface at phase equilibrium\cite{sprik21c,sprik24}.  As explained in the introduction of Ref.~\citenum{sprik25} these two earlier papers failed to appreciate the fundamental nature of the LC lattice constraint.  Even more serious Ref.~\citenum{sprik21c} overlooked the significance of accretion at solid-liquid interfaces as already discussed by Gibbs(see for example  Ref.~\citenum{sekerka15book} ch.~14). This was rectified in Ref.~\citenum{sprik24}.  Solid-liquid phase equilibrium is not an issue for the present work on single phase crystals.   However, certain elements of the kinematics of diffusion as treated in Ref.~\citenum{sprik21c} are not entirely mistaken and will be given a proper non-equilibrium continuum thermodynamics basis in section \ref{sec:mass}. 

The structure of the paper is as follows. We begin in section \ref{sec:preview} with a brief summary of the governing equations entirely in the spatial (Eulerian) representation.  This is followed by a formal proof  in the framework of Lagrangian rational continuum thermodynamics spread out over sections \ref{sec:kindyn}-\ref{sec:nonequitherm}.  Returning to current space the theory is linearized in section \ref{sec:linap} using the well-known Lam\'{e} constitutive model for the stress tensor. We also address the problem of a suitable definition of a reference state for a deformable system with vacancies.  The theory is illustrated in section \ref{sec:examples} by working through three elementary examples of crystals  under non-hydrostatic stress. The model systems are all taken form the book of Lubarda and Lubarda\cite{lubarda20}. We conclude with some general comments and an outlook in section \ref{sec:conclook}.

  \section{Preview of theory}\label{sec:preview}
  
  \subsection{Factorization of density} \label{sec:rhofact}
  
  The fundamental premise of Larch\'{e}-Cahn (LC) theory of crystals is the preservation of lattice structure.  The lattice sites are the markers quantifying deformation and are conserved whether occupied or empty.  This elevates the density of lattice sites to an independent thermodynamic degree of freedom next to the density of  particles.  The combination of these two conserved variables defines a  third conserved variable,  the  occupation of a site.   Indicating the density of sites by $s$, the density of particles by $\rho$ and the occupation by $n$ these three variables are related by a product rule
 \begin{equation}
\rho = s n =  \frac{n}{v_c}
\label{rhosn}
 \end{equation}
 where $v_c = 1/s$ is the volume of a lattice cell in the deformed crystal.  In a non-uniform system these variables are spatial fields obtained by averaging the corresponding microscopic quantities over some suitable mesoscopic volume\cite{cahn85,mullins85}.  LC theory is a continuum thermodynamics theory and the implementation of such a coarse graining procedure is not of direct concern.
  
The thermodynamic state of the crystal is specified by selecting two of the three fields as independent variables.  The common practice in materials science is to choose site density $s$ and occupation $n$ treating particle density as a dependent variable. The total number of sites M and particles $N$ are obtained by integration of $s$ and $\rho$ over volume $V$ using Eq.~\ref{rhosn}
\begin{equation}
M = \int_V s dv, \quad N  = \int_V s n dv  
\label{intsn}
\end{equation}
 Only crystals with single site occupation will be considered. Interstitial atoms are excluded. Hence,  $N \le M$ and the difference $N_v = M-N$ is the number of vacancies.  $N_v$ is small in realistic systems, typically in the order of $10^{-4}$ in metals.  The $N \rightarrow M$  ideal crystal limit is singular requiring matching changes in $M$\cite{sprik25}.  This is the process of Gibbs accretion which is beyond the scope of the present work.  $M$ is held fixed and $N_v > 0$. 
 
Ratio's of extensive  quantities are intensive thermodynamic degrees of freedom. A complementary extensive quantity is needed to specify the total system size. This is taken care of in Eq.~\ref{intsn} by the integration  volume $V$.  Volume is an essential state variable in deformable crystals. Similarly, exchange of particles with the environment changes $N$ in open systems.  In contrast, the number of sites is strictly fixed.  It remains the same under deformation and removal of particles.  $M$ is therefore the natural choice for quantifying system size of LC crystals.   Note  that $M$ is in fact an observable.  Diffraction experiments carried out on an (approximately) uniform crystals give an estimate of cell volume $v_c = V/M$.   Assuming  $V$ has been measured by some other experimental method,  the number of lattice sites is the ratio $V/v_c$.   This means that  $N_v$ is an observable as well because particle number $N$ can in principle be obtained by determining the mass of the crystal.

 \subsection{Lattice free energy} \label{sec:flat}
 The total Helmholtz free energy $F$ is the integral over a free energy density $\psi$
\begin{equation}
F = \int_V \psi  dv 
\end{equation}
 Following Ref.~\citenum{sprik25} the free energy density is resolved in three contributions
\begin{equation}
\psi = s f_{\mathrm{s}} + s f_{\mathrm{e}} - \rho I_{\mathrm{b}}
\label{psilat}
\end{equation}
The first term is the entropy generated by partial occupation $0 \le n \le 1$ of the sites.  
\begin{equation}
f_{\mathrm{s}}  
 = k_{\mathrm{B}} T \left( n \ln n + \left(1-n\right) \ln \left(1-n\right) \right)
 \label{psilats}
\end{equation}
Indicating the population of vacancies by $c = 1- n$, the second term can be written as $c \ln c$  mirroring the $n \ln n$ term. This is a well-known feature of lattice gas models  reflecting the Fermi-Dirac type statistics of occupation\cite{nix16}. 

The second contribution in Eq.~\ref{psilat}  is the elastic energy. In Ref.~\citenum{sprik25} only hydrostatic pressure was considered and the elastic response was restricted 
to volume expansion
\begin{equation}
f_{\mathrm{e}}  = \frac{\alpha_j}{2} \left( J -1 \right)^2
\label{psilate}
\end{equation}
where $J = V/V_{\mathrm{R}}$ is the ratio of the volume $V$ of the deformed system relative to some reference volume $V_{\mathrm{R}}$.  $J -1$ is the volume strain.  In the present work we also account for shear deformation.  However the central point made in Ref~\citenum{sprik25} remains valid. In the presence of vacancies, deformation is described in terms of the displacement of lattice sites. The elastic energy is  independent of particle number as illustrated by Eq.~\ref{psilate}.  The consequence is that elastic energy is skipped in thermodynamic derivative expressions for the chemical potential. This is a major obstacle for the formulation of the thermodynamics of crystals and will be again the dominant theme of the present work. 

The third term in Eq.~\ref{psilat} is not standard in LC theory.  $I_{\mathrm{b}}$ is the average binding energy represented as the energy required for removal of a particle from a lattice site.  $I_{\mathrm{b}} > 0$,  hence the negative sign in Eq.~\ref{psilat}.  The crucial difference with  the other two terms is that $I_{\mathrm{b}}$ is an energy per particle  while $f_{\mathrm{s}}$ and $f_{\mathrm{e}}$ are energies per site. This is why $I_{\mathrm{b}}$ is multiplied by $\rho$ instead of $s$.   A further distinction is that $I_{\mathrm{b}}$ is fully determined by the local density $\rho$.  Interpreting Eq~\ref{psilat} as the free energy density of a lattice gas, $I_{\mathrm{b}}(\rho)$ can be regarded as a mean field approximation of the interaction between particles occupying neighbouring sites. This short range interaction is to be distinguished from the long range elastic interactions described by $f_{\mathrm{e}}$. The physical picture of Eq.~\ref{psilat} is that of a mean field compressible lattice lattice gas\cite{domb56,essam70,oitmaa75,salinas87}, or elastic  cell gas (Lennard-Jones Devonshire) model\cite{fisherm16,cerdeirina17}. Coupling to a  flexible lattice introduces correlations between spins (occupation) transforming short range to long range (non-additive) interactions\cite{ruffo09,ruffo14,dellago20,dellago21}. 
   
   \subsection{Balance equations and transport} \label{sec:balintro}
The pressure $p_{\mathrm{b}}$ generated by $I_{\mathrm{b}}$  plays a crucial role in the formulation of LC theory presented in this work.  For the uniform systems of Ref.~\citenum{sprik25}  this pressure was defined as a derivative with respect to volume at fixed particle number of the site binding energy $- N I_{\mathrm{b}}$
\begin{equation}
 p_{\mathrm{b}} =  \left(\frac{\partial \left( N  I_{\mathrm{b}} \right) }{\partial V}\right) _N
\end{equation}
Temperature is held constant.  For the non-uniform system this is refined to a local pressure of the form 
\begin{equation}
p_{\mathrm{b}} = - \rho^2 \frac{d I_{\mathrm{b}}(\rho)}{d \rho}
\label{pblat}
\end{equation}
 $p_{\mathrm{b}}$ was referred to in Ref.~\citenum{sprik25} as molecular pressure but is renamed here to chemical pressure.   $p_{\mathrm{b}}$ provides the coupling between strain and occupation.  It can do so because its argument, particle density $\rho$,  is the product of occupation and  inverse volume of the elastically deformed lattice cells (Eq.~\ref{rhosn}). This  endows the gradient of $p_{\mathrm{b}}$ with a double function,  appearing both in the mechanical and chemical equilibrium equation.   Cauchy's equation is resolved in 
\begin{equation}
\mathrm{div} \mathbf{T}_{\mathrm{e}} = \mathrm{grad} \, p_{\mathrm{b}}
\label{cauchintro}
\end{equation}
where $\mathbf{T}_{\mathrm{e}}$ is the Cauchy stress tensor derived from the elastic energy $f_{\mathrm{e}}$ according to rules of continuum mechanics.  There is no external one-body force in Eq.~\ref{cauchintro}. Seemingly taking its place is an internal force density $\mathrm{grad} \, p_{\mathrm{b}}$ generated by the occupation dependent molecular interactions. 

The gradient of the chemical pressure appears a second time in the expression for the gradient of the chemical potential. It acts as a mechanical drift force
\begin{equation}
\mathrm{grad} \,\mu = \frac{\mathrm{grad}\, p_{\mathrm{b}}}{\rho}  +  \frac{ k_{\mathrm{B}}T}{n\left(1-n\right)} \mathrm{grad} \, n
\label{chemintro}
\end{equation}
The last term is the gradient of the entropic chemical potential derived from Eq.~\ref{psilats}.  The chemical potential is homogeneous  in equilibrium systems undisturbed by external forces. Setting the left hand side of Eq.~\ref{chemintro} to zero gives a chemical balance equation coupled to the force balance Eq.~\ref{cauchintro}. Solving this set of equations, given proper boundary conditions,  should determine the occupation and deformation. 
The liquid like (Eulerian) setting of the chemical pressure suggests that our LC model could be interpreted as a faithful realization of the Gibbs liquid in solid model of solid thermodynamics. 

 In the framework of the kinematic and constitutive assumptions outlined above Eqs.~\ref{cauchintro} and \ref{chemintro}  are exact.  This will be verified by a formal derivation using the methods of non-linear rational continuum thermodynamics\cite{gurtin10,mesarovic16}. LC theory in its original formulation is based on Gibbs's variational principle\cite{cahn73,cahn85}. The chemical potential is treated as a Lagrange multiplier. The non-equilibrium framework gives a different perspective. The focus is on diffusive flux and Eq.~\ref{chemintro} is actually the zero flux residue of a deformation coupled Fick's law.  The full expression for the (spatial) diffusive flux $\mathbf{h}$ including the effect of an externally applied potential $\phi_0$ is a key result of the present work 
 \begin{equation}
\mathbf{h} = - M_{m} J^{-1} \mathbf{B} \, \mathrm{grad}  \left( \mu +  \phi_0 \right)
\label{fickintro}
\end{equation}
 $M_{m} $ is the (scalar) isotropic material mobility. As pointed by LC\cite{cahn82,cahn85}, the mobility matrix in current space is in general anisotropic distorted by  strain. In Eq.~\ref{fickintro} this subtle effect is quantified by the left Cauchy-Green tensor $\mathbf{B}$ (the volumetric stretch $J = \sqrt{\det \mathbf{B}}$). 
  
The continuum thermodynamics derivation, presented in the theory part of the paper,  is virtually copied from the textbook of GFA.  We decided to include this lengthy and admittedly technical verification to give the special dual role of chemical pressure a rigorous foundation. Composition-strain coupling in this specific form was not considered by GFA.  A further issue is the treatment of external forces.  $\phi_0$ in Eq.~\ref{fickintro} is directly added to the chemical potential $\mu$  driving the diffusive flux $\mathbf{h}$. The chemical potential of a zero flux state  cannot be homogeneous but must match (minus) the applied potential.  This observation will be immediately recognized by those familiar with classical density functional theory\cite{evans79,hansen13}.  In non-equilibrium continuum thermodynamics the same chemical balance  requires careful separation between diffusive and convective current\cite{mesarovic16} which is not treated  by GFA in sufficient detail for our purpose. 

\subsection{Eigenstrain and reference pressure} \label{sec:eigenintro}

The chemical pressure defined in Eq.~\ref{pblat} is a mean field approximation of short range interaction between occupied lattice sites.  Repulsive forces put the network of elastic bonds under tension. The volumetric strain in Eq.~\ref{psilate} in that case is tensile ($J>1$).  This is an intrinsic effect called eigen strain persisting even under zero applied pressure. As a result a nominally stress free state is not strain free and is not suitable as a reference for deformation. A way to resolve this is postulating the existence of a state without strain at finite external pressure. This pressure, indicated by $p_0$, must vanish in systems without composition strain interaction. $p_0$ can therefore be used as the parameter determining the coupling strength. Clearly eigen strain varies with composition requiring in practice a choice of a reference value $n^0$ for occupation $n$.   The ideal crystal ($n^0 =1$) would seem the natural choice but this value is excluded because the entropy Eq.~\ref{psilats} diverges for $n \rightarrow 1$.  The vacancy population $c^0$ in the reference state must be larger than zero.  The selection  of $n^0$ is therefore constrained by a realistic value of $c^0$ but other than that arbitrary.  

This is worked out in detail in the linear approximation in the second half of the paper.  Defining the relative vacancy population
\begin{equation}
\xi = \frac{c - c^0}{n^0} 
\label{xidef}
\end{equation}
it will be shown in section \ref{sec:linap} that the linearized pressure can be written as 
\begin{equation}
p = p_0 - \left( \alpha_j  + p_0 \right) \Delta J - p_0 \xi 
\label{pintro}
\end{equation}
For simplicity elastic forces are still described by the volume stress Eq.~\ref{psilate}.   $\Delta J = J -1$ is the volumetric strain.  The pressure is hydrostatic. This will be later amended to the isotropic Lam\'{e} model. The chemical potential at the same level of approximation is found to be
\begin{equation}
\mu = \mu_0 - \left( \frac{k_{\mathrm{B}} T}{c^0} + \frac{p_0}{\rho^0} \right) \xi - \left(\frac{p_0}{\rho^0} \right) \Delta J
\label{muintro}
\end{equation}
$\rho^0 = n^0 / v_{\mathrm{c}}^0$ is the particle density in the reference state with $v_{\mathrm{c}}^0$ the corresponding cell volume.  $\mu^0$ is the reference for chemical potential.  Setting $p_0 = 0$ decouples pressure and chemical potential. To obtain Eqs.~\ref{pintro} and \ref{muintro} $p_0$ was treated as a small parameter. This is the level of approximation used in the examples of section \ref{sec:examples}. The theory of section \ref{sec:balintro} is however fully nonlinear.

\section{Kinematics and dynamics}\label{sec:kindyn}

\subsection{Lattice site based transport theorem} \label{sec:reynolds}
Migration of atoms in a moving lattice seemingly creates a conflict in one-component systems.  Diffusive motion is unbound.  Particles can go anywhere given enough time. How can this freedom be reconciled with the strictly bound displacement of elastic deformation?  There is a way around this in lattice based theories. The material points defining deformation no longer follow mass but are identified with the lattice sites. Diffusive flux can then be unambiguously  defined in terms of a transport theorem convecting with the lattice motion\cite{gurtin10}.   A complete lattice theory must also  include creation and elimination of lattice sites at peripheries and defects.  This was worked out in detail by Sinisa Mesarovic (SM)\cite{mesarovic16} in a theory of diffusional creep continuing on earlier work by Berdichevsky\cite{berdichevsky97} and Garikipati\cite{garikipati01}. 

In the present study the integrity of the  lattice is rigorously conserved.  Lattice bonds don't break even when connecting to an unoccupied site. The displacement of empty sites  still contributes to the elastic energy.  This is the basis of the lattice constraint of LC theory\cite{cahn73}.   The material reviewed in this section can be regarded as the kinematic underpinning of LC theory.  Following the presentation in GFA,  the flow $\mathbf{v}$ in the definition of the material time derivative of field $\varphi$ 
\begin{equation}
\frac{D \varphi}{Dt} = \frac{\partial \varphi}{\partial t}+ \mathbf{v} \cdot \mathrm{grad} \varphi \equiv  \overset{\centerdot}{\varphi} 
  \label{overdot}
 \end{equation}
 is now associated with the motion of lattice sites.  The second equality introduces the overdot notation commonly used in the continuum mechanics literature as a compact notation for the material time derivative. The  density of lattice sites  is conserved which in this convention is expressed as 
 \begin{equation}
\overset{\centerdot}{s} = - s\, \mathrm{div} \, \mathbf{v} 
 \label{sitebal}
 \end{equation}
 Eq.~\ref{sitebal} must be relaxed for a description of evolving lattices at interfaces, which is a key application of LC theory\cite{cahn85}, but outside the scope of the paper. 
  
 Rather than defining specific variables per unit of mass as in conventional continuum mechanics,  the thermodynamics of crystals is based on relations between quantities per site. A  field $\varphi$ is represented by site variable  $f$ defined as
\begin{equation}
\varphi = s f
\label{phisf}
\end{equation}
where $s$ is the site density. This scheme was already applied to the number density in Eq.~\ref{rhosn}. It leads to a modification of the transport identity. Adopting the formal notation of GAF the integral of $\varphi$ over a convecting subdomain $\mathcal{P}_t$ (also called "body part") is written as
\begin{equation}
\Phi\left(\mathcal{P}_t\right) = \int_{\mathcal{P}_t} \varphi \, dv =   \int_{\mathcal{P}_t} s f \, dv
\label{PhiPt}
\end{equation} 
 $dv$ is a spatial volume element. The rate of change of $\Phi$ is given by the transport identity
 \begin{equation}
\frac{d }{dt}  \Phi\left(\mathcal{P}_t\right) = \int_{\mathcal{P}_t} s \overset{\, \centerdot}{f}  \, dv  
\label{transps}
\end{equation} 
This lattice variant of the Reynolds theorem plays a central role in the irreversible continuum thermodynamics of crystals as pointed by Mesarovic \cite{mesarovic16}. The proof proceeds as for the Reynolds theorem with $s$ replacing $\rho$ making use of the site conservation relation Eq.~\ref{sitebal}.

\subsection{Mass balance and diffusive flux} \label{sec:mass}

As a first application of the transport theorem Eq.~\ref{transps} we will look at mass balance. In integral form this involves the time evolution of the number of particles $N\left(\mathcal{P}_t\right) $ in the convecting lattice region $\mathcal{P}_t$.  Using  Eq.~\ref{rhosn} 
\begin{equation}
N\left(\mathcal{P}_t\right) = \int_{\mathcal{P}_t} \rho dv =   \int_{\mathcal{P}_t} s n \, dv
\label{NPt}
\end{equation}
The transport identity gives for the time derivative
  \begin{equation}
\frac{d}{dt} N\left(\mathcal{P}_t\right) = \int_{\mathcal{P}_t} s \overset{\centerdot}{n}  \, dv  
\label{dNPdt}
\end{equation}
 A lattice subdomain with vacancies is an open system. Particles can enter or leave while the number of lattice sites remains the same
  \begin{equation}
 \int_{\mathcal{P}_t} s \overset{\centerdot}{n}  \, dv  = - \int_{\partial \mathcal{P}_t} \mathbf{h} \cdot \mathbf{n} \, da +  \int_{\mathcal{P}_t} h \, dv
\label{surflux}
\end{equation} 
Exchange with the environment is accounted for by a net flux through the boundary $\partial \mathcal{P}_t$ of $\mathcal{P}_t$. The flux is written as a surface integral of the normal component of a vector $\mathbf{h}$ ($da$ is an infinitesimal surface area).  The minus sign is because the normal $\mathbf{n}$ points outward.  Following GFA (section 61) we have also added a supply term in the form of a volume integral over a source density $h$.  While unphysical for a unreactive system one could imagine this process to represent the formal insertion of particles in a grand-canonical statistical mechanics scheme.

The moving subdomain $\mathcal{P}_t$ can be chosen arbitrarily, which allows us to localize Eq.~\ref{surflux} using Gauss theorem.
\begin{equation}
s \overset{\centerdot}{n} = - \mathrm{div} \mathbf{h} +h 
\label{occubal}
\end{equation}
Eq.~\ref{occubal} can be converted right away  to a continuity equation for the population of vacancies.  Substituting  \mbox{$c = 1-n$} we obtain
\begin{equation}
s \overset{\centerdot}{c} = \mathrm{div} \mathbf{h} -h
\label{vacbal}
\end{equation}
Particles can only move to empty sites near to it. The flux of vacancies is strictly opposite to the flux of particles. which is  reflected in the change of sign of the divergence in Eq.~\ref{vacbal} compared to Eq.~\ref{occubal}.

 Diffusive flux $\mathbf{h}$ is defined relative to the motion of lattice sites  $s \mathbf{v}$ of Eq.~\ref{sitebal}.  The  net continuity equation for  particle density $\rho$ is found by evaluating the material time derivative of $\rho$ applying the chain rule to Eq.~\ref{rhosn} and substituting Eqs.~\ref{sitebal} and \ref{massbal}. This  gives
\begin{equation}
\overset{\, \centerdot}{\rho} = -\rho \mathrm{div} \, \mathbf{v} - \mathrm{div}\, \mathbf{h}
\end{equation}
 Using the definition of the material derivative Eq.~\ref{overdot} this can be rearranged to
\begin{equation}
\frac{\partial \rho}{\partial t}  + \mathrm{div} \left(  \rho \mathbf{w}  \right) = 0
\label{massbal}
\end{equation}
with  the velocity field $\mathbf{w}$  defined as
\begin{equation}
\mathbf{w} = \mathbf{v}+ \rho^{-1} \mathbf{h}
\label{defw}
\end{equation}
Particle motion is a superposition of deformation and diffusion.  Eq.~\ref{defw} can be compared to the  separation of the motion of liquid and the polymer skeleton in the continuum theory of the swelling of elastomers\cite{srinivasa2004,steigmann2013}.  We adopted this approach in our 2021 study of diffusion in deformable one-component solids treating the lattice as a fictitious elastomer\cite{sprik21c}.   We now make the claim that this heuristic scheme can be justified by the lattice based transport theorem Eq.~\ref{surflux}.

\subsection{Force balance and work rate} \label{sec:work}
GFA as well as SM  treat stress coupled vacancy diffusion in the limit of instantaneous response of strain.  The kinetic energy and momentum density are set to zero. While this is outlined in detail in section 60 of GFA, we summarize the force and power balance in this approximation. These equations are needed later to formulate the free energy imbalance.  Without inertial contributions the Cauchy's equation is reduced to
\begin{equation}
\mathrm{div} \mathbf{T} + \mathbf{b} = 0
\label{bcauchy}
\end{equation} 
where $\mathbf{T}$ is the Cauchy stress tensor and $\mathbf{b}$  the one-body force density.  In absence of inertial effects $\mathbf{b}$ is equal to the applied force density. One-body forces are normally derived from an external potential $ \phi_0(\mathbf{x})$.   The question is what is the relevant density.  We will take the view that this is the particle density $\rho$, not the site density $s$
\begin{equation}
\mathbf{b} =  \rho \,  \,\mathbf{b}_0  = - \rho \, \mathrm{grad} \, \phi_0 
\label{defb0}
\end{equation}
External surface forces (tractions) are accounted for in regular Cauchy free boundary conditions 
\begin{equation}
\mathbf{T}  \mathbf{n} = \mathbf{t}_0
\label{scauchy}
\end{equation}
As will be shown in section \ref{sec:Tcon}, $\mathbf{T}$ is a function of both the lattice strain and the particle density.

Having set up the force balance we continue the GFA derivation of the mechanical laws for the LC  crystal and write out the corresponding mechanical work rate 
\begin{equation}
\mathcal{W}_0\left( \mathcal{P}_t \right) =  \int_{\partial \mathcal{P}_t} \mathbf{t}_0 \cdot \mathbf{v} da + 
 \int_{\mathcal{P}_t} \mathbf{b} \cdot \mathbf{w}  dv
 \label{extpower}
\end{equation}
The first term is the work per unit of time exerted by the traction forces  on the crystal surface. The volume term accounts for the work by the one body force. Eq.~\ref{extpower} again reflects the distinction between the dynamics of migrating particles and lattice sites.  The velocity of lattice sites is determined by the elastic deformation rate $\mathbf{v}$ (see Eq.~\ref{sitebal}).  The velocity $\mathbf{w}$ of particles of Eq.~\ref{defw} contains an additional contribution due to diffusion. Combining with Eq.~\ref{defb0} we have
\begin{equation}
\int_{\mathcal{P}_t} \mathbf{b} \cdot \mathbf{w}  dv = \int_{\mathcal{P}_t} \mathbf{b}_0 \cdot \left( \rho \mathbf{v} + \mathbf{h} \right) dv
\label{bwdv}
\end{equation}
This generalized expression for one-body work also appears in the paper by  Epstein and Goriely \cite{epstein12} but is left out by SM who sets $\mathbf{b}_0 = 0$. 

The surface term in Eq.~\ref{extpower} is converted to a volume integral following standard procedure in continuum mechanics (see GFA). Substituting Eq.~\ref{scauchy}  gives
\begin{equation}
 \int_{\partial \mathcal{P}_t} \mathbf{t}_0 \cdot \mathbf{v} da 
 =  \int_{\partial \mathcal{P}_t} \mathbf{Tn} \cdot \mathbf{v} da
 \label{t0vda}
\end{equation}
Applying the Gauss theorem
\begin{equation}
  \int_{\partial \mathcal{P}_t} \mathbf{Tn} \cdot \mathbf{v} da = \int_{\mathcal{P}_t} \left( \mathbf{v} \cdot \mathrm{div} \mathbf{T}
 + \mathbf{T} : \mathrm{grad} \mathbf{v} \right) dv
\end{equation}
 together with the Cauchy equation (Eq.~\ref{bcauchy}) gives
 \begin{equation}
  \int_{\partial \mathcal{P}_t} \mathbf{Tn} \cdot \mathbf{v} da = \int_{\mathcal{P}_t} \left( \mathbf{T}: \mathrm{grad} \mathbf{v}
  - \mathbf{b} \cdot  \mathbf{v} \right) dv
  \label{Tnvda}
\end{equation}
The "double dot" product of tensors $\mathbf{A}$ and $\mathbf{B}$ is the contraction $\mathbf{A}:\mathbf{B} = \mathrm{tr} \mathbf{A}^{\mathrm{T}}\mathbf{B} = A_{ij}B_{ij}$ (repeated indices are summed over). 
Finally substituting Eqs.~\ref{bwdv},\ref{t0vda} and \ref{Tnvda} in Eq.~\ref{extpower} for the mechanical work rate we obtain
\begin{equation}
\mathcal{W}_0\left( \mathcal{P}_t \right) =   \int_{\mathcal{P}_t} \left( \mathbf{T}: \mathrm{grad} \mathbf{v}
  + \mathbf{b}_0 \cdot  \mathbf{h} \right) dv
\label{powerbal}
\end{equation}
The integrand proportional to the strain rate $\mathrm{grad} \mathbf{v}$ is the well-known expression for deformation related work (see GFA). The work done by the external one-body potential on the diffusing particles adds a separate term proportional to the diffusion flux $\mathbf{h}$. This effect seems to be missing from the presentation by GFA but will be needed later to make a connection to classical density functional theory in section \ref{sec:mobil}. 

\subsection{Chemical  free energy imbalance } \label{sec:muflux}
Following GFA the chemical potential is defined in analogy with heat conduction in  non-uniform systems. The integral Clausius-Duhem inequality
is written as
\begin{equation}
\frac{d}{dt}  \int_{\mathcal{P}_t} s \psi \, dv \le \mathcal{W}_0\left( \mathcal{P}_t \right) + \mathcal{T}\left( \mathcal{P}_t \right)
\label{globalcdh}
\end{equation}
where $\psi$ is the free energy per site and $\mathcal{W}_0\left( \mathcal{P}_t \right)$ is the mechanical work rate of Eq.~\ref{powerbal}. 
The term $\mathcal{T}\left(\mathcal{P}_t\right)$  describes the input of  the energy carried by the particles entering $\mathcal{P}_t$ by diffusion.  
\begin{equation}
 \mathcal{T}\left( \mathcal{P}_t \right) = - \int_{\partial \mathcal{P}_t} \mu \mathbf{h} \cdot \mathbf{n} \, da +  \int_{\mathcal{P}_t}  \mu h\, dv
 \label{Tdifa}
\end{equation}
To understand this expression compare to the particle number balance Eq.~\ref{surflux}. The particle supply term $h$ multiplied by the chemical potential $\mu$ accounts for the increase in energy by particle insertion. The same amount of energy per particle can equally added by the flux particle flux $\mathbf{h}$ keeping in mind that $\mathcal{P}_t$ is a moving control volume.  All of this proceeds isothermally. Heat flux terms are ignored. Expressions similar to Eq.~\ref{Tdifa} interpreting the chemical potential in terms of chemical energy  transport can already be found in de textbook by de Groot and Mazur \cite{degrootmazur11} who credit the idea to Eckart. 

As in section \ref{sec:work} the next step is changing over from surface to volume integration applying the Gauss divergence theorem to the flux term
\begin{equation}
- \int_{\partial \mathcal{P}_t} \mu \mathbf{h} \cdot \mathbf{n} \, da = -  \int_{ \mathcal{P}_t} \left( \mu \mathrm{div} \mathbf{h}
+ \mathbf{h} \cdot \mathrm{grad} \mu \right) dv
\end{equation}
The $\mathrm{div} \mathbf{h}$ term is replaced by the (material) occupation time derivative using Eq.~\ref{occubal}. The $\mu h$ contribution cancels out against the supply term in Eq.~\ref{Tdifa} and we are left with  
\begin{equation}
  \mathcal{T}\left( \mathcal{P}_t \right) =  -  \int_{\mathcal{P}_t}  \left( \mathbf{h} \cdot \mathrm{grad} \mu - s \mu  \overset{\centerdot}{n} \right) dv
  \label{tdifv}
  \end{equation}
  Note that this result is valid in open as well as  closed system ($h = 0$). 
  
Substituting in Eq.~\ref{globalcdh} and using Eq.~\ref{powerbal} for $\mathcal{W}_0$ yields after applying once more the transport identity Eq.~\ref{transps}
\begin{multline}
 \int_{\mathcal{P}_t} s \overset{\centerdot}{\psi} \, dv \le \int_{\mathcal{P}_t} \left( \mathbf{T}: \mathrm{grad} \mathbf{v} 
 +  s \mu  \overset{\centerdot}{n}  \right. \\
 \left. -\mathbf{h} \cdot \left(\mathrm{grad} \mu - \mathbf{b}_0 \right) \right)  dv
 \label{dtglobalcdh}
\end{multline}
Localizing we obtain for the (isothermal) free energy imbalance  
\begin{equation}
s \left(  \overset{\centerdot}{\psi} -  \mu \overset{\centerdot}{n} \right) - \mathbf{T} : \mathrm{grad} \mathbf{v}
 + \mathbf{h} \cdot \left(\mathrm{grad} \mu  - \mathbf{b}_0 \right)  \le 0
\label{freimb}
\end{equation}
which is the expression given in GFA extended with a coupling of $\mathbf{h}$ to $\mathbf{b}_0$.  The origin of the $- \mathbf{h} \cdot \mathbf{b}_0$  term is the distinction between particle migration and convective motion of the material points attached to the lattice sites as explained in section \ref{sec:mass}. 

\section{Referential representation}\label{sec:lagrange}

\subsection{Material coordinates and deformation gradient}\label{sec:defgrad}
The balance and imbalance laws presented in section \ref{sec:kindyn} must be supplemented with constitutive relations.  This is implemented by GFA using hyperelastic theory. Constitutive relations are obtained as thermodynamic derivatives of a stored free energy density.  The functional form of the stored energy function is the single unifying constitutive assumption of the theory.  The complication is that  the energy density is defined in material space.  This means that, in order to continue our derivation, all relations postulated or derived in section \ref{sec:kindyn} must be transformed ("pulled back") to reference space. We could have started in reference space from the beginning, but chose not to because this would have obscured the essence of the non-equilibrium thermodynamic reasoning. 

The position of a material particle in reference space is given by the Cartesian vector $\mathbf{X}=\left(X_1,X_2,X_3\right)$.  Deformation maps  $\mathbf{X}$  to the position   $\mathbf{x}$  of the particle at time $t$ in current space.    
\begin{equation}
\mathbf{x} = \boldsymbol{\chi}_t \left(\mathbf{X} \right)
\label{chimap}
\end{equation}
where $\boldsymbol{\chi}_t$ is the common notation for the time dependent deformation map. $\boldsymbol{\chi}_t$ defines the deformation gradient
 \begin{equation}
 \mathbf{F} \left( \mathbf{X}, t \right) = \nabla \boldsymbol{\chi}_t \left( \mathbf{X} \right)
 \label{Fdef}
 \end{equation}
 In explicit coordinate notation
\begin{equation}
F_{ij} = \frac{\partial x_i}{\partial X_j}
\end{equation}
The matrix $\mathbf{F}$ transforms infinitesimal displacements $d \mathbf{X}$ in reference space into infinitesimal displacement $d\mathbf{x} = \mathbf{F}d \mathbf{X}$ in current space. This relation holds even if the displacement of $\mathbf{x}$ relative to $\mathbf{X}$ is finite.  $\mathbf{F}$  must however be invertible and have a positive definite determinant 
\begin{equation}
J = \mathrm{det} \mathbf{F} > 0
\label{Jdef}
\end{equation}
Eq.~\ref{Fdef} also introduces a symbol for reference space gradients. As chosen by GFA, this  is the familiar  $\nabla$ operator.   Gradients
considered in section  \ref{sec:kindyn} all involved differentiating wrt spatial coordinates $x_i$.  The two gradients are interconverted according to
\begin{equation}
\nabla \varphi = \mathbf{F}^{\mathrm{T}} \mathrm{grad} \varphi, \quad 
  \mathrm{grad} \varphi =  \mathbf{F}^{-\mathrm{T}} \nabla \varphi
\label{grade2l}
\end{equation}
where $\mathbf{F}^{\mathrm{T}}$ is the transpose of $\mathbf{F}$ (in component notation $F^{\mathrm{T}}_{ij} = F_{ji}$).  In general 
$\mathbf{F}^{\mathrm{T}} \neq \mathbf{F}^{-1}$.   However 
\begin{equation}
 \left( \mathbf{F}^{\mathrm{T}}\right)^{-1} = \left( \mathbf{F}^{-1}\right)^{\mathrm{T}} \equiv \mathbf{F}^{-\mathrm{T}}
 \end{equation}
which is how the inverse transformation in Eq.~\ref{grade2l} is obtained. 

Material variables will be indicated by appending a subscript R to the corresponding spatial variables. The transformation can be carried put by equating volume and surface integrals using  the geometric relations
\begin{equation}
dv = J dv_{\mathrm{R}}, \quad \mathbf{n} da = J \mathbf{F}^{-\mathrm{T}} \mathbf{n}_{\mathrm{R}} da_{\mathrm{R}}
\label{dvdae2l}
\end{equation}
Applying this procedure to Eq.~\ref{PhiPt} for the per site expression of the volume integral of a spatial field $\varphi$
\begin{equation}
\Phi\left(\mathcal{P}_t\right) = \int_{\mathcal{P}_t} s f dv = \int_{\mathcal{P}} s_{\mathrm{R}} f dv_{\mathrm{R}}
\label{Phie2lx}
\end{equation}
with $\mathcal{P}= \boldsymbol{\chi}_t ^{-1} \mathcal{P}_t$ the material domain evolving into $\mathcal{P}_t$. The reference  site density 
\begin{equation}
s_{\mathrm{R}} = J s 
\label{se2l}
\end{equation}
 is the inverse of the volume of a lattice cell  in reference space and is a material constant.  Following GFA factors $s_{\mathrm{R}}$ are absorbed in the definition of the "pull back"  $\varphi_{\mathrm{R}}$ of  $\varphi$ turning $\varphi_{\mathrm{R}}$ into a reference space density
\begin{equation}
\varphi_{\mathrm{R}} =  s_{\mathrm{R}} f
\label{phie2l}
\end{equation}
Substituting  in Eq.~\ref{Phie2lx} leads to 
\begin{equation}
\Phi\left(\mathcal{P}_t\right) = \int_{\mathcal{P}_t} \varphi dv = \int_{\mathcal{P}} \varphi_{\mathrm{R}} dv_{\mathrm{R}}
\label{Phite2l}
\end{equation}
 $\varphi_{\mathrm{R}}$ is a reference space density.  Returning to Eq.~\ref{transps} and substituting Eq.~\ref{Phite2l} we have
 \begin{equation}
\int_{\mathcal{P}_t} s \overset{\, \centerdot}{f}  \, dv = \frac{d }{dt}  \Phi\left(\mathcal{P}_t\right) = 
\int_{\mathcal{P}} \frac{\partial \varphi_{\mathrm{R}}}{\partial t}  dv_{\mathrm{R}}
\end{equation}
The counter part of the material time derivative $\dot{f}$ is a simple partial time derivative in reference space. As material quantities are identified by the subscript R we can continue to use the convenient overdot notation for partial time derivatives in  reference space 
 \begin{equation}
\int_{\mathcal{P}_t} s \overset{\, \centerdot}{f}  \, dv = \int_{\mathcal{P}} \overset{\, \centerdot}{\varphi}_{\mathrm{R}} dv_{\mathrm{R}}
\label{fdote2l}
\end{equation}
This is the rule for changing a lattice sum to an integral over material space volume.

 \subsection{Reference mass and force balance}\label{sec:refbal}
  With Eq.~\ref{Phite2l} we seem to be back to the conventional Lagrangian formulation of solid continuum mechanics. The crucial difference  is that the lattice sites have taken over the role of material particles determining the deformation velocity $\mathbf{v}$. As a result particle density in reference space is left free to vary with time.  Using Eq.~\ref{fdote2l} this can be expressed as 
 \begin{equation}
 \int_{\mathcal{P}_t} s \overset{\, \centerdot}{n}  \, dv = \int_{\mathcal{P}}  \overset{\, \centerdot}{\rho}_{\mathrm{R}} \, dv_{\mathrm{R}}  
\label{dNPdR}
\end{equation}
The evolving material number density is written as
\begin{equation}
 \rho_{\mathrm{R}} = s_{\mathrm{R}} n
  \label{ne2l}
  \end{equation}
with the understanding that the material site density $s_{\mathrm{R}}$ is a constant (constitutive) parameter. 

 The corresponding particle  flux is found by equating the surface integrals over  subdomains in current and reference space using Eq.~\ref{dvdae2l}
\begin{equation}
\int_{\partial \mathcal{P}_t} \mathbf{h} \cdot \mathbf{n} \, da = 
\int_{\partial \mathcal{P}} \mathbf{h}_{\mathrm{R}} \cdot \mathbf{n}_{\mathrm{R}} \,da_{\mathrm{R}}
\end{equation}
This gives
\begin{equation}
\mathbf{h}_{\mathrm{R}} = J \mathbf{F}^{-1}\mathbf{h}
\label{he2l}
\end{equation}
The supply term in eq.~\ref{occubal} is a regular volume density and therefore with  Eq.~\ref{dvdae2l}
\begin{equation}
h_{\mathrm{R}} = J h
\end{equation}
Assembling terms and applying the Gauss theorem in reference space we find for the mass balance expressed in material variables
\begin{equation}
  \overset{\centerdot}{\rho}_{\mathrm{R}} = - \mathrm{Div} \mathbf{h}_{\mathrm{R}} + h_{\mathrm{R}}
\label{massbalR}
\end{equation}
The operator $\mathrm{Div}$ denotes a divergence in reference space. 
Eq.~\ref{massbalR} is the proper balance law for vacancy diffusion in a rigid lattice. The particles migrate by changing lattice site. A source term for removing/inserting particles is also included.  

Next is the material representation of the Cauchy Eq.~\ref{bcauchy}
\begin{equation}
\mathrm{Div} \mathbf{T}_{\mathrm{R}} + \mathbf{b}_{\mathrm{R}} = 0 
\label{rcauchy}
\end{equation}
$\mathbf{T}_{\mathrm{R}}$ is the (first) Piola stress tensor\cite{gurtin10}
\begin{equation}
\mathbf{T}_{\mathrm{R}} = J \mathbf{T} \mathbf{F}^{-\mathrm{T}}
\label{Tpiola}
\end{equation}
The one-body force  $\mathbf{b}$, as defined in  Eq.~\ref{defb0}, is the spatial density of a force  derived from a potential $\phi_0$.  The potential is the same in deformed and reference space, the forces are different. Defining  $\mathbf{b}_{0\mathrm{R}} = - \nabla \phi_0$ the relation is found by applying the transformation rule Eq.~\ref{grade2l}  \begin{equation}
\mathbf{b}_{0\mathrm{R}}= \mathbf{F}^{\mathrm{T}} \mathbf{b}_0
\label{b0e2l}
\end{equation}
 leading to the material force density
\begin{equation}
 \mathbf{b}_{\mathrm{R}} = J  \mathbf{F}^{\mathrm{T}} \mathbf{b} =  \rho_{\mathrm{R}}  \mathbf{b}_{\mathrm{0R}}
 \label{bRrhoR}
\end{equation}
Recalling Eq.~\ref{ne2l} we see that the external  one-body force couples the Piola stress tensor to occupation.  As will be shown below,  composition strain interaction gives rise to an  effective internal one-body force establishing a similar  coupling mechanism in crystals deformed by surface tractions without external one-body forces. 
 
 \subsection{Equations of state}\label{sec:cnrt}

The free energy imbalance relation Eq.~\ref{freimb} is a  central equation in the rational thermodynamics scheme applied here.  Given a set of thermodynamic state variables the free energy imbalance relation is used to derive the corresponding state equations.  These relations can be regarded as  a non-equilibrium generalization of the derivative relations imposed by the Gibbs  fundamental equation. This is the procedure proposed by Coleman and Noll (CN) and carries their names\cite{Coleman1963}. It is discussed at length in GFA.  For finite deformation the CN approach is usually implemented in reference space as is done here. Formulated in terms of material variables Eq.~\ref{freimb} becomes
  \begin{equation}
    \overset{\centerdot}{\psi}_{\mathrm{R}} - \mu \overset{\centerdot}{\rho}_{\mathrm{R}} -\mathbf{T}_{\mathrm{R}} :  \overset{\centerdot}{\mathbf{F}}
    + \mathbf{h}_{\mathrm{R}} \cdot \left(\nabla \mu - \mathbf{b}_{0\mathrm{R}} \right) \le 0
    \label{freimbR}
  \end{equation}
The proof is rather technical. An outline can be found in Appendix \ref{sec:cnr}. This derivation is taken from GFA to which we refer for further details. 

The next step in the CN scheme is specifying the independent thermodynamic state variables. The  choice suggested by Eq.~\ref{freimbR} is the combination of material space particle density $\rho_{\mathrm{R}}$ and deformation gradient matrix $\mathbf{F}$. However as explained in every treatise on finite deformation $\mathbf{F}$ is not directly suitable as the state variable for strain because of conflict with the principle of  reference frame indifference.  The common approach is to use the right Cauchy-Green (CG) tensor
\begin{equation}
\mathbf{C} = \mathbf{F}^{\mathrm{T}}\mathbf{F}
\label{CGdef}
\end{equation}
 The tensor  $\mathbf{C}$ is invariant under rotation of the spatial reference frame.  Moreover,  contrary to $\mathbf{F}$,  CG  is symmetric ($\mathbf{C}^{\mathrm{T}} = \mathbf{C}$).  Its determinant is directly related to $J$ as defined in Eq.~\ref{Jdef}
\begin{equation}
\det \mathbf{C} = J^2
\label{detCG}
\end{equation}
Having modified the strain rate in Eq.~\ref{freimbR} the conjugate stress tensor must be adjusted accordingly.  The corresponding force is the second Piola stress tensor 
\begin{equation}
\mathbf{T}_{\mathrm{RR}} = \mathbf{F}^{-1} \mathbf{T}_{\mathrm{R}}  
\label{TRRdef}
\end{equation}
changing the internal mechanical work term in Eq.~\ref{freimbR} to
\begin{equation}
\mathbf{T}_{\mathrm{R}} : \overset{\centerdot}{\mathbf{F}} = \tfrac{1}{2} \mathbf{T}_{\mathrm{RR}} : \overset{\centerdot}{\mathbf{C}} 
\end{equation}
For a detailed explanation we again refer to GFA. 

The second state variable in Eq.~\ref{freimbR} is the reference density $\rho_{\mathrm{R}}$ which we will represent in terms of the occupation $ n = \rho_{\mathrm{R}}/s_{\mathrm{R}}$.  The time derivative of material free energy function $\psi_{\mathrm{R}}$ with $n$ and $\mathbf{C}$  as arguments
\begin{equation}
\psi_{\mathrm{R}} = \hat{\psi}_{\mathrm{R}} \left( n, \mathbf{C} \right)
\label{psihatdef}
\end{equation}
can now be expanded as
\begin{equation}
  \overset{\centerdot}{\psi}_{\mathrm{R}}
   =\left( \frac{\partial \hat{\psi}_{\mathrm{R}}}{\partial n}\right)_{\mathbf{C}} \overset{\centerdot}{n} + 
   \left( \frac{\partial \hat{\psi}_{\mathrm{R}}}{\partial \mathbf{C}}\right)_n \overset{\centerdot}{\mathbf{C}}  
\end{equation}
Substituting in   Eq.~\ref{freimbR}  gives
\begin{multline}
\left( \frac{\partial \hat{\psi}_{\mathrm{R}}}{\partial n} -s_{\mathrm{R}} \mu \right) \overset{\centerdot}{n} +
\left( \frac{\partial \psi_{\mathrm{R}}}{\partial \mathbf{C}} -\tfrac{1}{2} \mathbf{T}_{\mathrm{RR}} \right) : \overset{\centerdot}{\mathbf{C}}
\\  + \mathbf{h}_{\mathrm{R}} \cdot \left( \nabla \mu- \mathbf{b}_{0\mathrm{R}} \right) \le 0
\end{multline}
The CN argument requires this inequality to apply to an arbitrary admissible constitutive process $\dot {n}, \dot{\mathbf{C}}$.  This sets the prefactors to zero. The result is a derivative state equation for the chemical potential
\begin{equation}
\mu  = \frac{1}{s_{\mathrm{R}}} \left(\frac{\partial \hat{\psi}_{\mathrm{R}}}{\partial n} \right)_{\mathbf{C}} 
\label{mucn}
\end{equation}
and for the stress tensor
\begin{equation}
\mathbf{T}_{\mathrm{RR}}   =  2 \left( \frac{\partial \hat{\psi}_{\mathrm{R}}}{\partial \mathbf{C}} \right)_{n}
\label{TRRcn}
\end{equation}
The residual transport inequality
\begin{equation}
\mathbf{h}_{\mathrm{R}} \cdot \left(\nabla \mu - \mathbf{b}_{0\mathrm{R}} \right) \le 0
\end{equation}
is satisfied assuming the linear flux equation
\begin{equation}
\mathbf{h}_{\mathrm{R}}  = - \mathbf{M}\, \left( \nabla \mu - \mathbf{b}_{0\mathrm{R}} \right)
\label{fickcn}
\end{equation}
where $\mathbf{M}$ is the in principle anisotropic material mobility tensor.  The linear flux relation Eq.~\ref{fickcn} can be interpreted as an externally biased variant of Fick's law. To repeat,  all of this can be considered regular rational continuum thermodynamics procedure once we have accepted the GFA  approach to diffusion Eq.~\ref{Tdifa}.  An important point in this respect,  as stressed by GFA,  is that chemical potentials are not transformed by a pull back to reference space (the gradient however is).

\section{Model thermodynamics}\label{sec:thermlast}
\subsection{Free energy density}\label{sec:constimod}

The constitutive model for  the stored energy function $\psi_{\mathrm{R}}$ introduced in section \ref{sec:cnrt} is based on Eq.~\ref{psilat} for the uniform system. There are again three contributions
\begin{equation}
\hat{\psi}_{\mathrm{R}} \left( n, \mathbf{C} \right) = \hat{\psi}_{\mathrm{sR}} \left( n \right) + \hat{\psi}_{\mathrm{eR}} \left(\mathbf{C} \right)
+ \hat{\psi}_{\mathrm{bR}} \left( n, \mathbf{C} \right) 
\label{psiR3}
\end{equation}
$\hat{\psi}_{\mathrm{sR}} \left( n \right)$ is the inhomogeneous generalization  of the entropy term of Eq.~\ref{psilats}. This is a free energy per site and all we have to do to convert it to a reference space density is multiplying by the reference space site density $s_{\mathrm{R}}$.
\begin{equation}
\hat{\psi}_{\mathrm{sR}} (n) =  s_{\mathrm{R}}  k_{\mathrm{B}} T  \left( n \ln n + \left(1-n\right) \ln \left(1-n\right) \right)
 \label{psisR}
  \end{equation}
 The second term $\hat{\psi}_{\mathrm{eR}} \left(\mathbf{C} \right)$ is the material elastic energy density, which is left for moment unspecified but will eventually be approximated by the Lam\'{e} model of linear elastic theory (see section \ref{sec:lamehook}).  The important point is that  $\hat{\psi}_{\mathrm{eR}} \left(\mathbf{C} \right)$ is not depending on the occupation \mbox{field $n$}.  Similarly, the entropy term of Eq.~\ref{psisR} is not 
 affected by deformation as quantified by the the CG tensor $\mathbf{C}$.  
 
 The occupation-strain coupling is due to the third term, the density dependent site binding energy  $I_{\mathrm{b}}\left(\rho\right)$ introduced in section \ref{sec:flat}. Turning this energy per site into a free free energy density we multiply again by the $s_{\mathrm{R}}$
\begin{equation}
\hat{\psi}_{\mathrm{bR}}  \left( n, \mathbf{C} \right) = - s_{\mathrm{R}} n I_{\mathrm{b}} \left( \rho \right)
 \end{equation} 
The spatial number density is a dependent variable.  The functional dependence on the independent variables $n$ and $\mathbf{C}$ is found combining
 Eqs.~\ref{rhosn}  and \ref{se2l} 
 \begin{equation}
 \rho = \frac{s_{\mathrm{R}}n }{J} = s_{\mathrm{R}}n \left( \det \mathbf{C} \right)^{- 1/2}
 \label{rhob}
 \end{equation}
 where in the last step we have used Eq.~\ref{detCG}. The material coupling term can now be written as 
 \begin{equation}
\hat{\psi}_{\mathrm{bR}}  \left( n, \mathbf{C} \right) = 
- s_{\mathrm{R}} n I_{\mathrm{b}} \left(  \frac{s_{\mathrm{R}} n}{J\left(\mathbf{C}\right)}\right)
 \label{psibR}
\end{equation}
with $J\left(\mathbf{C}\right)$ given by Eq.~\ref{detCG}. 

Comparing  to LC theory in the original formulation, the composition strain  is captured by the variation of the site binding energy with the spatial (deformed) density $\rho$. To quantify this dependence we introduce the density derivative function
\begin{equation}
\gamma_{\mathrm{b}}\left(\rho\right)  = - \rho  \left(\frac{I_{\mathrm{b}}\left(\rho\right)}{d \rho} \right)
\label{gambdef}
\end{equation}
While $I_{\mathrm{b}} > 0$,  the sign of $\gamma_{\mathrm{b}}$ can be positive (repulsive) or negative (attractive) as discussed in Ref.~\citenum{sprik25}. 

\subsection{Chemical potential}\label{sec:mucon}
The evaluation of the chemical potential consists of determining the thermodynamic derivate Eq.~\ref{mucn}. This is little different from the procedure in Ref.~\citenum{sprik25}. $\mu$ is separated in an entropy and site binding term
\begin{equation}
\mu = \mu_{\mathrm{s}}+ \mu_{\mathrm{b}}
\label{musb}
\end{equation}
$\mu_{\mathrm{s}}$ is obtained by differentiating Eq.~\ref{psisR}. We recognize the familiar Langmuir-type chemical potential of lattice gas models
\begin{equation}
\mu_{\mathrm{s}}  =  \frac{1}{s_{\mathrm{R}}} \left(\frac{\partial \hat{\psi}_{s \mathrm{R}}}{\partial n} \right)
 = k_{\mathrm{B}} T \ln \left( \frac{n}{1-n} \right)
 \label{mus}
 \end{equation}
The divergence $\mu_{\mathrm{s}} \rightarrow \infty$ in the limit of the ideal crystal ($n=1$) effectively imposes the LC lattice constraint.  

The second term in Eq.~\ref{musb} is the partial derivative of the site binding energy  Eq.~\ref{psibR}
\begin{equation}
\mu_{\mathrm{b}} = \frac{1}{s_{\mathrm{R}}} \left(\frac{\partial \hat{\psi}_{b \mathrm{R}}}{\partial n} \right)
= - I_{\mathrm{b}} +\gamma_{\mathrm{b}}
\label{mub}
\end{equation}
with $\gamma_{\mathrm{b}}$ defined in Eq.~\ref{gambdef}. $\mu_{\mathrm{b}}$ will be given the name "binding chemical potential".  As anticipated the strain energy $\psi_{e\mathrm{R}}\left(\mathbf{C}\right)$  makes no explicit contribution to the chemical potential.  Still  $\mu$ indirectly depends on deformation through Eq.~\ref{rhob}.  

\subsection{Cauchy stress tensor}\label{sec:Tcon}
 The Cauchy stress tensor is recovered from the material thermodynamic derivative relation Eq ~\ref{TRRcn} by transforming the second Piola stress tensor back to the current frame.  This is product of the inverse of Eq.~\ref{Tpiola} and  \ref{TRRdef}
 \begin{equation}
\mathbf{T} = J^{-1} \mathbf{FT}_{\mathrm{RR}}\mathbf{F}^{\mathrm{T}}
\label{crr2e}
\end{equation}
 Again there are two contributions
\begin{equation}
\mathbf{T} = \mathbf{T}_{\mathrm{e}} + \mathbf{T}_{\mathrm{b}} 
\label{Teb}
\end{equation}
The first term is purely elastic stress.  The second term is the chemical stress generated by the site binding energy $I_{\mathrm{b}}$.  There is no site entropy related stress.  Occupation is invariant under deformation and therefore also the Langmuir entropy Eq.~\ref{psisR}.

The elastic stress tensor is obtained from the model of the material stored energy $\hat{\psi}_{\mathrm{eR}}$ 
using Eqs.~\ref{TRRcn} and \ref{crr2e} 
\begin{equation}
\mathbf{T}_{\mathrm{e}}   = 2 J^{-1} \mathbf{F}  \left( \frac{\partial \hat{\psi}_{\mathrm{eR}}}{\partial \mathbf{C}} \right) \mathbf{F}^{\mathrm{T}}
\label{cerr2e}
\end{equation}
We will not use this equation  but directly model Cauchy stress in the small deformation approximation.   This is the subject of section \ref{sec:lamehook}.  For the  binding energy related stress we will follow the formal route
\begin{equation}
\mathbf{T}_{\mathrm{b}}   = 2 J^{-1} \mathbf{F}  \left( \frac{\partial \hat{\psi}_{\mathrm{bR}}}{\partial \mathbf{C}} \right) \mathbf{F}^{\mathrm{T}}
\label{cbrr2e}
\end{equation}
As derived in appendix \ref{sec:molstress} 
\begin{equation}
 \mathbf{T}_{\mathrm{b}} = - \rho \gamma_{\mathrm{b}} \mathbf{1}
 \label{cbgb}
\end{equation} 
with $\gamma_{\mathrm{b}}$  given in Eq.~\ref{gambdef}.  This leads us to define a chemical pressure
\begin{equation}
p_{\mathrm{b}}\left( \rho \right) =  \rho \gamma_{\mathrm{b}}\left( \rho \right)
\label{pbgamb}
\end{equation}
and identify $\mathbf{T}_{\mathrm{b}}$ with the tensor isotropic tensor
\begin{equation}
 \mathbf{T}_{\mathrm{b}} = - p_{\mathrm{b}} \mathbf{1}
 \label{cbpb}
\end{equation}
Returning to the preview we have verified  that pressure  introduced in  Eq.~\ref{pblat} without further argument is the same as the properly defined chemical pressure of Eq.~\ref{pbgamb}.  Also it should be mentiond that the decomposition of the stress tensor Eq.~\ref{Teb} in an elastic and isotropic density dependent component is anticipated again in the Baek-Srinivasa paper\cite{srinivasa2004} on the absorption of fluid in elastomers (see also Ref.~\citenum{morro16} section 4). 

Stress in elastic solids is in general not hydrostatic.  The stress tensor consists of an isotropic and deviatoric component indicated by a tilde\cite{gurtin10}
\begin{equation}
\mathbf{T} = - p_{\mathrm{m}} \mathbf{1} + \widetilde{\mathbf{T}}
\label{devdef}
\end{equation}
$p_{\mathrm{m}}$  is minus the mean normal stress
\begin{equation}
p_{\mathrm{m}} = - \frac{1}{3} \mathrm{tr} \mathbf{T},   \quad
 \label{pmdef}
\end{equation}
and will be referred to in this work as the Cauchy pressure.   Similarly we will define an elastic pressure
\begin{equation} 
p_{\mathrm{em}} = - \frac{1}{3} \mathrm{tr} \mathbf{T}_{\mathrm{e}}  = p_{\mathrm{m}} - p_{\mathrm{b}} 
\label{pem}
\end{equation}
Mean normal stress can no longer be equated with  mean normal elastic stress.   This will be recurring theme in the examples presented in section \ref{sec:examples}. 

\subsection{Chemical pressure} \label{sec:pbchem}
The chemical pressure introduced in Eq.~\ref{pbgamb} and chemical potential term $\mu_{\mathrm{b}}$ of Eq.~\ref{mub} are 
local thermodynamic quantities  derived from the site binding energy $I_{\mathrm{b}}$.   Substituting Eq.~\ref{pbgamb} in Eq.~\ref{mub} and rearranging we find 
\begin{equation}
p_{\mathrm{b}} = \rho \mu_{\mathrm{b}} + \rho I_{\mathrm{b}} 
\label{pbmub}
\end{equation}
Eq.~\ref{pbmub} resembles the hydrostatic pressure equation for liquids with the $\rho I_{\mathrm{b}}$ term interpreted as (minus) the free energy density.  A number of useful and familiar properties follow.  The first the response relation
\begin{equation}
\frac{d p_{\mathrm{b}}}{d \rho} = \rho \frac{d \mu_{\mathrm{b}}}{d \rho}
\label{bcompress}
\end{equation}
which is of the same form as the compressibility theorem for liquids\cite{hansen13}. The origin of this Gibbs-Duhem type  connection is the functional form of the site binding energy $I_{\mathrm{b}}$ depending only on $\rho$, the actual density\cite{sprik25}. 

The spatial gradients show a similar local thermodynamic relationship.   Using the chain rule the gradient of the binding chemical potential Eq.~\ref{mub} can be expressed in terms $\gamma_{\mathrm{b}}$ and its spatial gradient. 
\begin{equation}
\mathrm{grad} \mu_{\mathrm{b}} = \left(\frac{\gamma_{\mathrm{b}}}{\rho} \right) \mathrm{grad} \rho + \mathrm{grad} \gamma_{\mathrm{b}}
\end{equation}
Next, substituting Eq.~\ref{pbgamb} for the chemical pressure 
\begin{equation}
\mathrm{grad} \mu_{\mathrm{b}}  = \left(\frac{p_{\mathrm{b}}}{\rho^2}\right) \mathrm{grad} \rho + \mathrm{grad}\left( \frac{p_{\mathrm{b}}}{\rho} \right)
\end{equation}
we arrive at a Gibbs-Duhem equation for chemical pressure and binding chemical potential
 \begin{equation}
\mathrm{grad} \, p_{\mathrm{b}} = \rho \,\mathrm{grad} \, \mu_{\mathrm{b}}
\label{GdHb}
 \end{equation}
 The thermodynamic  relations Eqs.~\ref{bcompress} and \ref{GdHb} are characteristic for the number density of liquids and are consistent with the  diffusion equations  of section \ref{sec:reynolds}.

\section{Transport}\label{sec:nonequitherm}

  \subsection{Deformation modulated mobility} \label{sec:mobil}
 In view of  the  simplicity of the expressions for chemical potential and Cauchy stress in sections \ref{sec:mucon} and \ref{sec:Tcon} one could query whether the burdensome Lagrangian derivation of section \ref{sec:lagrange}  is really necessary.  Indeed these results are almost intuitive and can be regarded as a faithful realization of Gibbsian concept of a solid with a liquid component. However this is less clear for the transport properties implied by Eq.~\ref{fickcn}.  Transforming the material gradient and flux to spatial variables using Eqs.~\ref{grade2l}, \ref{he2l} and  \ref{b0e2l} we obtain
\begin{equation}
\mathbf{h} = - J^{-1} \mathbf{FMF}^{\mathrm{T}} \left( \mathrm{grad} \, \mu - \mathbf{b}_0 \right)
\end{equation}
Assuming a default  isotropic reference mobility tensor \mbox{$\mathbf{M} = M \mathbf{1}$} with $M$ a fixed constitutive constant the diffusion equation  in the deformed system is  simplified to
\begin{equation}
\mathbf{h} = - M J^{-1} \mathbf{B} \left( \mathrm{grad} \, \mu - \mathbf{b}_0 \right)
\label{fickCb}
\end{equation}
with the tensor $\mathbf{B}$ defined as
\begin{equation}
\mathbf{B} =\mathbf{F} \mathbf{F}^{\mathrm{T}}
\label{BGdef}
\end{equation}
$\mathbf{B}^{\mathrm{T}} = \mathbf{B}$ and therefore $\mathbf{B}$ is symmetric.  

 The tensor Eq.~\ref{BGdef}  is known as the left Cauchy-Green tensor, the spatial counterpart of $\mathbf{C}$ of Eq.~\ref{CGdef}\cite{gurtin10}. The determinants are the same $\det \mathbf{B} = \det \mathbf{C} = J^2$
and so are the eigenvalues. These eigenvalues have an important geometric interpretation. They are the squares of the principle stretches $\lambda_1,  \lambda_2$ and  $\lambda_3$.  Observed in actual space,  deformation  modulates mobility.  The effect is demonstrated already by simple isotropic dilatation without interference of one-body forces ($\mathbf{b}_0=0$).   Setting $\lambda_1 = \lambda_2 = \lambda_3 = \lambda$ and substituting in Eq.~\ref{fickCb} produces something with closer resemblance to Fick's law 
\begin{equation}
\mathbf{h} = - \lambda M  \mathrm{grad} \, \mu 
\label{difstretch}
\end{equation}
The mobility is seemingly scaled up by the stretch.  Increase in distance in the deformed system $\lambda > 1$  accelerates  the diffusion rate.   Sensitivity of diffusion to strain is investigated by LC in their 1982 paper\cite{cahn82} from more phenomenological perspective.   A comparison is outside the scope of the present paper. 

What is relevant here are the implications of Eq.~\ref{fickCb} for equilibrium.  In theories of irreversible thermodynamics equilibrium is defined as a state with vanishing flux\cite{degrootmazur11}. Setting $\mathbf{h}= 0$ we obtain a balance relation for the gradient of the chemical potential and the external force.  For potential forces Eq.~\ref{defb0} this leads to
\begin{equation}
\mathrm{grad} \mu = -  \mathrm{grad} \phi_0  \; \rightarrow  \; \mu = \mu_0 - \phi_0 
\label{chembal}
\end{equation} 
 where $\mu_0$ is an integration constant. For physical chemists with a background in classical density functional theory (cDFT) this looks obvious. It is the cDFT equilibrium condition in the local density approximation (LDA)\cite{hansen13}.  The chemical potential in Eq.~\ref{chembal} is an intrinsic chemical potential.   What is gained with the non-equilibrium thermodynamics derivation is a formal proof that the chemical potential in Fick's law Eq.~\ref{fickCb} and the equilibrium condition Eq.~\ref{chembal} are the same thermodynamic function.  This is often presented as an assumption in variational thermodynamics\cite{cahn85}.

 \subsection{Cauchy-Nernst-Plank equation} \label{sec:cnp}
 Having worked out the expressions for the chemical potential and stress tensor in section \ref{sec:thermlast} we are ready to investigate how strain is coupled to occupation.  As pointed out in section \ref{sec:refbal} the external one-body force density, being proportional to the reference number density (Eq.~\ref{bRrhoR}),  is doing just that, as is also clear from Eq.~\ref{chembal}.  However, there is also an intrinsic coupling mechanism.  Hence,  we set $\mathbf{b} = 0$ in the Cauchy equation Eq.~\ref{bcauchy} and resolve the stress tensor in an elastic and  chemical component according to Eq~\ref{Teb} using  \ref{cbpb} for the hydrostatic chemical stress tensor. This gives
  \begin{equation}
   \mathrm{div} \mathbf{T}_{\mathrm{e}} =  \mathrm{grad} \, p_{\mathrm{b}} 
   \label{dce2dpb}
   \end{equation}
  The left hand side is the elastic force density.   Deformed by surface tractions only,  the Cauchy stress tensor of an ideal elastic medium is fully transverse.   The elastic force density vanishes. In Eq.~\ref{dce2dpb},  even without external one body forces,  the elastic force density is finite.  The atomic forces generated by the chemical pressure act as an internal one-body force.    Eq.~\ref{dce2dpb} is equivalent to the set of field equations\cite{dirisio24}.
 \begin{gather}
\mathrm{div} \,  \mathrm{div} \mathbf{T}_{\mathrm{e}} = \Delta p_{\mathrm{b}}
\nonumber
\\[2pt] 
 \mathrm{curl}\,  \mathrm{div} \mathbf{T}_{\mathrm{e}} = 0 
 \label{fieldeq}
 \end{gather}
 where $\Delta$ is the spatial Laplace operator.   Fortunately,  these formidable partial differential equations are not needed for the problems in  the example section \ref{sec:examples}.   The stress fields are linear functions of the position coordinate  and therefore automatically satisfy Eqs.~\ref{fieldeq} (see also the discussion in  section \ref{sec:conclook}).  
   
 Complementing the force balance equation the second equilibrium condition under zero applied body force is uniformity of  the chemical potential (Eq.~\ref{chembal} with $\phi_0 = 0$). To see how this couples occupation and deformation we must evaluate the two terms in Eq.~\ref{musb} making up the chemical potential.
Differentiating the entropic term  Eq.~\ref{mus} delivers  the expected driving force for diffusion due to concentration gradients
\begin{equation}
\mathrm{grad} \, \mu_{\mathrm{s}} =\frac{ k_{\mathrm{B}}T}{n\left(1-n\right)} \,\mathrm{grad} n
\label{dmus}
\end{equation}
As observed before, the ideal crystal limit $n \rightarrow1 $ is ill behaved and should be avoided (see section \ref{sec:matref}). Under mechanical equilibrium conditions  the gradient of chemical pressure and elastic forces match allowing us  to insert  Eq.~\ref{dce2dpb} in Eq.~\ref{GdHb} 
 \begin{equation}
    \mathrm{div} \mathbf{T}_{\mathrm{e}} =   \rho \,\mathrm{grad} \, \mu_{\mathrm{b}}
  \end{equation}
  Combining with Eq.~\ref{dmus} we find
  \begin{equation}
  \mathrm{grad} \, \mu =  \frac{1}{\rho} \,  \mathrm{div} \mathbf{T}_{\mathrm{e}} +
   \frac{ k_{\mathrm{B}}T}{n\left(1-n\right)} \mathrm{grad} \, n
  \end{equation}
  The divergence of the elastic stress tensor is the driving  force in a Cauchy-Nernst-Planck equation for particle diffusion. Even with $\mathrm{div} \mathbf{T}$ vanishing,  $\mathrm{div} \mathbf{T}_{\mathrm{e}}$
 is not zero because of the effect of the chemical pressure.  
  
   In absence of external one-body forces $ \mathrm{grad} \, \mu = 0$. The applied surface tractions generate a population profile with a spatial gradient
  \begin{equation}
  \mathrm{grad} \, n = - \frac{n \left(1-n \right)}{ \rho k_{\mathrm{B}}T } \mathrm{div} \mathbf{T}_{\mathrm{e}}
  \label{nprofpe}
  \end{equation}
 Intuitively one might have expected  the gradient of the mean normal pressure $p_{\mathrm{m}}$ (Cauchy pressure) of Eq.~\ref{pmdef} to play that role.  This impression is not  mistaken.  $\mathrm{grad} \, p_{\mathrm{b}}$  is first order in $\gamma_{\mathrm{b}}$ of Eq.~\ref{gambdef} and therefore also $\mathrm{div}  \mathbf{T}_{\mathrm{e}}$.  The coefficient $\gamma_{\mathrm{b}}$  quantifies the strain-composition coupling strength.  In the small coupling limit the population profile can be approximated by  a linear function of $p_{\mathrm{m}}$ multiplied  by  $\gamma_{\mathrm{b}}$.   This will be shown in section \ref{sec:linap}.  Eq.~\ref{nprofpe} is the general formulation exposing more openly the structure of the theory.

\section{Linearization}\label{sec:linap}
\subsection{Reference state and eigenstrain}\label{sec:matref}
The specification of an appropriate reference state for deformation requires a decision how to deal with eigenstrain.  This was mentioned in section \ref{sec:eigenintro} and we will now fill in the details. The root of the problem the reference for the occupation. Why did we have to adopt a reference  $n^0< 1$ and not the ideal state $£n^0 =1$? The reason is the singularity in the entropic term in the chemical potential.  Eq.~\ref{mus} diverges in the limit of zero vacancy population $c=1-n $ driving the chemical potential to infinity.  Oddly, the ideal crystal is not a suitable thermodynamic reference.  In a closed system the average population $n^0 = N /M$ is fixed.  We can therefore use the small but finite average vacancy population $c^0 = 1- n^0 $ as the reference for $c$.  The awkward consequence is that a state under  zero applied pressure is not strain free and cannot be used as the reference for deformation. The origin of the internal stress is the chemical pressure. This problem already arose in our study of the uniform system\cite{sprik25} and we will proceed in the same way. 

 The chemical pressure remains finite in the limit of vanishing volumetric strain ($J=1$). External pressure is matching the internal pressure and is therefore also finite.  There is in principe no objection to declaring a  non-zero total pressure state to be the reference for deformation.   We already assumed a uniform reference $n^0$ for population. With now also the reference density  $\rho^0$ fixed we expand Eq.~\ref{pbgamb} for the chemical pressure to first order in the particle density difference $\Delta \rho = \rho -\rho^0$.  
\begin{equation}
p_{\mathrm{b}}  = p_0 + \left(\gamma_{\mathrm{b}}^0 + \chi_{\mathrm{b}}^0\right) \Delta \rho
 \label{pbgambref}
\end{equation}
where $\gamma_{\mathrm{b}}^0 = \gamma_{\mathrm{b}}\left(\rho^0\right)$. The reference value  $p_0$  of the chemical pressure is proportional to $\gamma_{\mathrm{b}}^0$
\begin{equation}
p_0 = p_{\mathrm{b}}(\rho^0) =\rho^0 \gamma_{\mathrm{b}}^0
\label{pb0def} 
\end{equation}
$ \chi_{\mathrm{b}} $ is an effective susceptibility 
\begin{equation}
 \chi_{\mathrm{b}} = \rho  \left(\frac{d \gamma_{\mathrm{b}}}{d \rho} \right)
\end{equation}
containing a second order derivative of $I_{\mathrm{b}}$. 
$\chi_{\mathrm{b}}^0 = \chi_{\mathrm{b}}(\rho^0)$ is the value at the reference density.  

We now return to density factorization of Eq.~\ref{rhosn}.  Occupation is conserved in closed systems.  The particle density $\rho$ varies inversely proportional to $J$, the volumetric stretch.  $\rho = n^0 s_{\mathrm{R}} = \rho^0$ in the strain free reference state ($\Delta J =  J - 1 = 0$).  The strain at zero pressure is the eigenstrain $\Delta J_0$.  Volume has changed and with it the particle density.  At the small risk of confusion this density will be indicated by $\rho_0$ to be distinguished from the reference density $\rho^0$.  
\begin{equation} 
\rho_0 =\frac{\rho^0}{1+ \Delta J_0} = \rho^0 \left(1 -\Delta J_0 \right)
\label{rholin0}
\end{equation}
The second equality is not an identity but a linear approximation.  Substituting in Eq.~\ref{pbgambref} gives an expression for the residual  chemical pressure after releasing the external pressure. Using Eq.~\ref{pb0def} we obtain
\begin{equation}
p_{\mathrm{b}}\left(\rho_0\right)  = p_0 \left( 1 - \Delta J_0 \right)
 \label{pbgamb0}
\end{equation}
$\chi^0_{\mathrm{b}}$ has been neglected  which is reasonable because it is a higher order effect.
 
The system is in mechanical equilibrium.  Without externally applied pressure,  the hydrostatic chemical pressure Eq.~\ref{pbgamb0}  is fully compensated by the elastic stress which is equally determined by $\Delta J_0$.   Equating these two pressures should give an estimate of the eigenstrain given a model for the elastic forces. 
This calulation  was carried out in  our previous study\cite{sprik25} for the elementary volumetric harmonic potential Eq.~\ref{psilate}.  It may be instructive to briefly review this calculation to get a better feeling of the physics of eigenstrain.  The elastic stress $\sigma_j = \alpha_j \Delta J$  is hydrostatic and linear in the volumetric strain $\Delta J = J -1 $.
At zero external pressure the volumetric stress is matched by the chemical pressure of Eq.~\ref{pbgamb0}.   Setting $p_{\mathrm{b}} - \sigma_{\mathrm{j}} = 0$ gives an equation for $\Delta J_0$.  Working this out we obtain
\begin{equation}
\Delta J_0 = \frac{  p_0} {\alpha_j + p_0}
\label{deltaj0}
\end{equation}
which reproduces our previous  result\cite{sprik25} in a somewhat different notation.   The key parameter controlling strain-composition coupling is $p_0 \propto \gamma_{\mathrm{b}}^0$.   Eigenstrain disappears in the limit  $p_0  \rightarrow 0$ (note such a system still contains vacancies).  

\subsection{Small population changes} \label{sec:chemref}
 From this section on the discussion is restricted to linear theory.  To proceed we start by dropping the "0" superscript where it has become redundant.  In particular $\gamma_{\mathrm{b}}^0 \rightarrow \gamma_{\mathrm{b}}$. 
The vacancy population represented by $\xi$  of Eq.~\ref{xidef} and volume strain $\Delta J$ are the independent field variables.  The first order variation of the spatial density $\rho$ in terms of these variables is found by linearization of Eq.~\ref{rhosn}. Using Eq.~\ref{se2l} for the site density we obtain
\begin{equation}
\Delta \rho =-  \rho^0 \left(\xi + \Delta J \right)
\label{rholin}
\end{equation}
Substituting  in Eq.~\ref{pbgambref} gives the two variable linear approximation to the fluctuation of the  chemical pressure  
\begin{equation}
 \Delta p_{\mathrm{b}}  = p_{\mathrm{b}} -  p_0 = - p_0 \left( \xi + \Delta J \right) 
\label{pblin}
\end{equation}
We have assumed that  the effect of the secondary response $\chi_{\mathrm{b}}$ can be neglected. This will be from now on  standard procedure. 

Given  the linear form of the chemical pressure the  corresponding approximation for the binding chemical potential  Eq.~\ref{mub} follows directly from the relation Eq,~\ref{bcompress} between the density derivatives.  
\begin{equation}
\mu_{\mathrm{b}} = \mu_{\mathrm{b}}^0 + \gamma_{\mathrm{b}}  \frac{\Delta \rho}{\rho^0}
\label{mubgambref}
\end{equation}
with $\mu_{\mathrm{b}}^0 =  \mu_{\mathrm{b}}\left(\rho^0\right)$.  Using Eq.~\ref{rholin}  we find for the molecular chemical potential relative to its reference value
\begin{equation}
\Delta \mu_{\mathrm{b}}   = \mu_{\mathrm{b}}  - \mu_{\mathrm{b}}^0 = -  \gamma_{\mathrm{b}} \left( \xi + \Delta J \right)
\label{mublin}
\end{equation}
Comparing Eqs.~\ref{pblin} and \ref{mublin} we see that $ \Delta p_{\mathrm{b}} = \rho^0 \Delta \mu_{\mathrm{b}}$ consistent with  Eq.~\ref{bcompress}. 

  Linearization of the  entropic component  of the chemical potential of Eq.~\ref{mus} is easy
\begin{equation}
\mu_{\mathrm{s}} = \mu_{\mathrm{s}}^0 - \chi_{\mathrm{s}}  \xi
\label{muslin}
\end{equation}
 $\mu_{\mathrm{s}}^0$ is the  reference entropy per site.  
\begin{equation}
\mu_{\mathrm{s}}^0 = k_{\mathrm{B}}T  \ln \left(\frac{n^0}{c^0} \right)
\label{muslin0}
\end{equation}
We have also introduced the entropic susceptibility
\begin{equation}
\chi_{\mathrm{s}} = \frac{k_{\mathrm{B}}T}{c^0}  
\label{chis0}
\end{equation}
Occupation is preserved under deformation. $\chi_{\mathrm{s}}$ is not a function of volumetric strain $\Delta J$.  On the other hand there is the $c^0$ in the denominator causing $\mu$ to increase without limit approaching the vacancy free ideal crystal. Moreover, because $c>0$, $\Delta c$  has a lower bound of $-c^0$, which can be potentially be violated in a linear  theory if $c^0$ is too small.  

Adding the binding  contribution Eq.~\ref{mublin} gives
\begin{equation}
\Delta \mu = - \left(\chi_{\mathrm{s}}+  \gamma_{\mathrm{b}}\right) \xi - \gamma_{\mathrm{b}} \Delta J
\label{mulin}
\end{equation}
with $\Delta \mu = \mu - \mu_{\mathrm{s}}^0 - \mu_{\mathrm{b}}^0$ denoting the deviation of the full chemical potential relative to the reference value. $\Delta \mu$  is linear in the volume strain. Elastic energy which, as  a quadratic function of strain, makes no distinction between tensile and compressive strain. This difference is crucial for transport.  It provides a driving force for migration of vacancies from regions with small cell volume ($\Delta J < 0$) to regions with cells of larger volume with the same occupation. 

 \subsection{Small deformation approximation} \label{sec:lamehook}

The Cauchy stress tensor was separated  in section  \ref{sec:Tcon} in an elastic component $\mathbf{T}_{\mathrm{e}}$ and a hydrostatic chemical pressure tensor $\mathbf{T}_{\mathrm{b}} = - p_{\mathrm{b}} \mathbf{1}$ (Eqs.~\ref{Teb} and \ref{cbpb}). The elastic stress tensor was left without further specification.  This is what we will do now.  We will use  the standard Lam\'{e} model in its elementary  cubic isotropic formulation.  
\begin{equation}
\mathbf{T}_{\mathrm{e}} = 2 \mu_{\mathrm{e}} \boldsymbol{\epsilon} +\lambda_{\mathrm{e}} \left(\mathrm{tr} \boldsymbol{\epsilon} \right) \mathbf{1}
\label{Tlame}
\end{equation}
where $\boldsymbol{\epsilon}$ is the linear strain tensor
\begin{equation}
\epsilon_{ij} =    \frac{1}{2} \left( \frac{\partial u_i}{\partial x_j} + \frac{\partial u_j}{\partial x_i} \right)
\label{straineps}
\end{equation}
The trace is the volumetric strain 
\begin{equation}
\Delta J =  \epsilon_{kk} = \mathrm{tr} \boldsymbol{\epsilon} 
\label{traceps}
\end{equation}
We already have a small deformation approximation for the chemical pressure.   This is Eq.~\ref{pblin} which is linear in both the population variable $\xi$ and strain $\Delta J$.  Changing over to the component representation we have for the full Cauchy stress tensor
\begin{equation}
 T_{ij} =  2 \mu_{\mathrm{e}} \epsilon_{ij} + \left( \left( \lambda_{\mathrm{e}} +  p_0  \right) \epsilon_{kk}  
 - p_0\left( 1 - \xi \right) \right) \delta_{ij}
\label{LLLdx}
 \end{equation}
 with $p_0 = \rho^0 \gamma_{\mathrm{b}}$ as defined in  Eq.~\ref{pb0def}.   The chemo-mechanical generalization of the Lam\'{e} stress tensor Eq.~\ref{LLLdx} is the basis of the strain-composition coupling theory developed here.  The Cauchy stress  is explicitly dependent on the chemical degree of freedom $\xi$.  A further modification is a fixed correction term $p_0$ added to the Lam\'{e} $ \lambda$ modulus.   

Under chemical equilibrium conditions $\xi$ is a function of the deformation state and particle number (or in open systems the imposed chemical potential, see section \ref{sec:closop}).  Following the LC protocol we could express  $\xi$ in terms of $\Delta J$ and substitute in Eq.~\ref{LLLdx}. The result is a regular stress-strain relation as used in linear elasticity with modified effective  moduli\cite{cahn73,cahn85}. The alternative is a semi-inverse scheme. Mechanic equilibrium is  solved starting from an anstaz for the full stress field (elastic plus chemical).  The strain is then obtained from the inverse of Eq.~\ref{LLLdx}.
 \begin{equation}
\epsilon_{ij} = \frac{1}{E} \left( \left(1 + \nu \right) T_{ij} - \nu T_{kk} \delta_{ij} \right) +\frac{p_0}{3K} \left(1 - \xi \right) \delta_{ij}
\label{Hookdx}
\end{equation}
Eq.~\ref{Hookdx} is a form of Hook's law with  elastic moduli corrected for the presence of vacancies
\begin{gather}
E = \frac{\mu_{\mathrm{e}} \left( 3\left(\lambda_{\mathrm{e}} + p_0\right) + 2 \mu_{\mathrm{e}}\right)}
{\lambda_{\mathrm{e}} + p_0 + \mu_{\mathrm{e}}}
\nonumber \\[4pt] 
\nu = \frac{\lambda_{\mathrm{e}} + p_0}
{2 \left(\lambda_{\mathrm{e}} + p_0+ \mu_{\mathrm{e}}\right)}
\nonumber \\[4pt] 
K = \lambda_{\mathrm{e}} + p_0 + \frac{2}{3} \mu_{\mathrm{e}}  = K_{\mathrm{e}} + p_0
\label{Kdx}
\end{gather}
 As compellingly argued by Lubarda and Lubarda (LL) the semi-inverse route is often more convenient for mechanical engineering applications which have to satisfy free boundary conditions\cite{lubarda20}.  This is also the method used in the examples of section \ref{sec:examples} which all have been adapted from the LL textbook.  We will make repeatedly use of various relationships between  linear elastic moduli without explicitly stating the equations.  A complete list can be found in LL section 3.5.2 (pg.~65).
 
 As a first critical test we consider the uniform crystal studied in Ref.~\citenum{sprik25} as briefly reviewed in section \ref{sec:preview}.  The elastic model now includes  a non-zero shear stress modulus.  The system is closed.  $\xi$ is  directly determined by the number of particles $N_0$ specifying the size of the system.  $N_0$ in general is different from the number of particles $N^0$ defining the  reference system.  Hence according to definition of $\xi$ of Eq.~\ref{xidef} the vacancy population variable
\begin{equation}
\xi_0 = 1 - \frac{N_0}{N^0} = 1 - \frac{n_0}{n^0}
\label{xi0}
\end{equation}
can be either positive or negative.   The crystal is deformed by a finite hydrostatic applied pressure $p$.  The Cauchy stress  is uniform  $T_{xx} = T_{yy} =T_{zz} = -p$.  The normal strains are obtained by substituting in Eq.~\ref{Hookdx} setting $\xi = \xi_0$. This gives for the volumetric strain
\begin{equation}
\Delta J = \frac{1}{K}   \left(p_0\left(1 - \xi_0  \right)  - p\right)
\label{dJxp}
\end{equation}
The strain Eq.~\ref{dJxp} at $p=0$   should correspond to the eigenstrain. 
\begin{equation}
\Delta J_0 = \frac{p_0}{K} \left(1 - \xi_0 \right)
\label{DJ0hook}
\end{equation}
$n_0 = n^0$ in reference state.  $\xi_0$ vanishes.  Eq.~\ref{DJ0hook} gives  $\Delta J_0 = p_0/K$. Indeed,  substituting  Eq.~\ref{Kdx} for $K$ with $\mu_{\mathrm{e}} = 0$  reproduces Eq.~\ref{deltaj0}. 

Eigenstrain is an intrinsic deformation.  Volumetric strain  induced by external pressure is therefore more conveniently specified relative to the eigenstrain.   Accordingly Eq.~\ref{dJxp} is reformulated as 
\begin{equation}
\Delta J - \Delta J_0 = - p/K 
\label{dJpun} 
\end{equation}
We now have values for $\xi$ and $\Delta J$.  This is what is needed to evaluate the linearized chemical pressure 
as given in Eq.~\ref{pblin}.   Substituting and rearranging  we obtain 
 \begin{equation}
p_{\mathrm{b}} - K \Delta J_0 = \left(\frac{ p_0}{K} \right) p
\label{selfstress0}
\end{equation}
Formally Eq.~\ref{selfstress0}  should have included a  $p_0 \Delta J_0$ term which has been discarded as a higher order correction  in $p_0$.  The offset $K \Delta J_0$ is the residual stress at $p=0$ and is interpreted as selfstress.  

In section \ref{sec:Tcon} we made a distinction between  elastic pressure $p_{\mathrm{em}}$ and Cauchy pressure  $p_{\mathrm{m}}$(Eq.~\ref{pmdef}).   According to Eq.~\ref{pem}  chemical and elastic pressure are both partial pressures adding up to the Cauchy pressure.    In the linear approximation we can be more specific about this decomposition.   Under the hydrostatic conditions  the Cauchy pressure  $p_{\mathrm{m}}$ is the external pressure $p$.   Substituting Eq.~\ref{selfstress0} in Eq.~\ref{pem}  gives to lowest in $p_0$
 \begin{equation}
 p_{\mathrm{em}} + K \Delta J_0 = \left(1 - \frac{p_0}{K} \right)p 
 \label{pelast0}
 \end{equation}
 In the non-interacting limit $p_0 \rightarrow 0$ the chemical pressure tends to zero.   There is no difference between elastic and Cauchy pressure as is the defining property of an ideal crystal.  The key message is that for $p_0 \neq 0$  mechanical equilibrium equalizes  external pressure and Cauchy pressure,  not external and  elastic pressure.

 \subsection{Open crystals and absorption} \label{sec:closop}
   Exchange of particles with a reservoir under strict conservation of lattice sites (network constraint) leads to fundamental differences between the thermodynamics of solids and liquids even for systems under hydrostatic pressure\cite{sprik25}.  The  unique contribution of LC is a simple theory to understand and compute these effects\cite{cahn73}. Switching to chemical potential control is relatively easy in the linear approximation. Eq.~\ref{mulin} can be inverted to an expression for vacancy population as a function of the chemical potential and volumetric deformation
\begin{equation}
\xi  = -\beta c^0 \tau_{\mathrm{s}} \Delta \mu -   \tau_{\mathrm{b}} \Delta J
\label{dcdmu}
\end{equation}
with the response  coefficients $\tau_{\mathrm{s}}$ and $\tau_{\mathrm{b}}$ defined as
\begin{equation}
\tau_{\mathrm{s}} = \frac{\chi_{\mathrm{s}}}{\chi_{\mathrm{s}} +  \gamma_{\mathrm{b}} }, \quad 
\tau_{\mathrm{b}} = \frac{\gamma_{\mathrm{b}}}{\chi_{\mathrm{s}} +  \gamma_{\mathrm{b}} } 
\label{taudef}
\end{equation}
and $\beta= 1/k_{\mathrm{B}}T$. 
Substituting,  Eq.~\ref{LLLdx} is transformed to
  \begin{equation}
 T_{ij} = 2 \mu_{\mathrm{e}} \epsilon_{ij} + \left( \left( \lambda_{\mathrm{e}} +  \lambda_{\mathrm{b}}  \right) \epsilon_{kk}  - \left( p_0 + \alpha \Delta \mu \right) \right)\delta_{ij}
\label{LLLmu}
\end{equation}
with modified $\lambda$ modulus and chemical coupling $\alpha$
 \begin{equation}
 \lambda_{\mathrm{b}} =  p_0 \tau_{\mathrm{s}}, \quad   \alpha = \rho^0 \tau_{\mathrm{b}}
\label{lambdab}
 \end{equation}
For many applications the small coupling approximation is adequate and  is also more instructive for analysis. 
In this limit the coefficients Eq.~\ref{taudef} are reduced to
\begin{equation}
\tau_{\mathrm{s}} = 1, \quad \tau_{\mathrm{b}} = \beta c^0 \gamma_{\mathrm{b}}
\label{taulim}
\end{equation}
which implies for the parameters of Eq.~\ref{lambdab}
 \begin{equation}
  \lambda_{\mathrm{b}} =  p_0 , \quad   \alpha = \beta c^0 p_0
  \label{alphalim}
   \end{equation}
 As in section \ref{sec:lamehook} we have excluded external body forces.  Note that  $\alpha$ is measured in units of inverse volume (density) suggesting a relation to some characteristic microscopic length scale. 

 Eq.~\ref{LLLmu} can be transformed by inversion to an open system Hooke's law 
 \begin{multline}
 \epsilon_{ij}^\ast = \frac{1}{E^\ast}  \left( \left(1 + \nu^\ast \right) T_{ij} - \nu^\ast T_{kk} \delta_{ij} \right) \\
 +\frac{1}{3K^\ast} \left(p_0 + \alpha \Delta \mu \right) \delta_{ij}
 \label{Hookmu}
 \end{multline}
 with elastic parameters
\begin{gather}
E ^\ast= \frac{\mu_{\mathrm{e}} \left( 3\left(\lambda_{\mathrm{e}} + \lambda_{\mathrm{b}}\right) + 2 \mu_{\mathrm{e}}\right)}
{\lambda_{\mathrm{e}} + \lambda_{\mathrm{b}} + \mu_{\mathrm{e}}}
\nonumber \\[4pt]
\nu^\ast = \frac{\lambda_{\mathrm{e}} + \lambda_{\mathrm{b}}}
{2 \left(\lambda_{\mathrm{e}} + \lambda_{\mathrm{b}} + \mu_{\mathrm{e}}\right)}
\nonumber \label{numu} \\[4pt]
K^\ast = \lambda_{\mathrm{e}} + \lambda_{\mathrm{b}} + \frac{2}{3} \mu_{\mathrm{e}}  = K_{\mathrm{e}} +  \lambda_{\mathrm{b}}
\label{Kmu}
\end{gather}
The asterisk notation indicating open system quantities has been borrowed from LC\cite{cahn85}.  

The independent  mechanical state variable in Hook's  law is the Cauchy matrix $T_{ij}$.   In particular, $T_{ij}$  is the same stress field both in the closed form of Hook's law Eq.~\ref{Hookdx} and the open form Eq.~\ref{Hookmu}.  The strain matrix $\epsilon_{ij}$ is the dependent variable and is a function of $T_{ij}$  and  chemical potential $\Delta \mu$ in Eq.~\ref{Hookmu} and of $T_{ij}$ and vacancy population $\xi$ in Eq.~\ref{Hookdx}.   This is why $\epsilon_{ij}$ is marked by an asterisk in Eq.~\ref{Hookmu} but not the stress tensor $T_{ij}$.  We reiterate that $T_{ij}$ is the full Cauchy stress tensor,  the sum of elastic and chemical stress (Eq.~\ref{Teb}),  and must be applied in its entirety to balance external tractions at free boundary surfaces.   This statement is backed up by  the formal part of the paper. 

To find the volumetric strain as function of chemical potential and  hydrostatic pressure, we repeat the same approach which led to Eq.~\ref{dJxp}. The result has the form of a thermodynamic equation of state in the $(\mu,p)$ manifold.
\begin{equation}
\Delta J^\ast =\frac{1}{K^\ast} \left(p_0 - p + \alpha \Delta  \mu \right)
\label{Jmup}
\end{equation}
Negative $\Delta J^\ast$ (shrinking) under compression ($p>p_0$) is normal mechanical behaviour.  Expansion ($\Delta J^\ast > 0$) in response to increase in chemical potential ($\Delta \mu > 0$) is a chemical effect.   Evidently at this level of theory chemical and mechanical forces are additive.   In analogy with Eq.~\ref{DJ0hook} we can define an  open system eigenstrain
\begin{equation}
\Delta J_0^ \ast = \frac{1 }{K^\ast}\left(p_0 + \alpha \Delta \mu \right)
\label{DJ0star}
 \end{equation}
 Combining Eq.~\ref{Jmup} and \ref{DJ0star} leads to the open system equivalent of Eq.~\ref{dJpun} 
 \begin{equation}
 \Delta J^\ast - \Delta J^\ast_0 = - \frac{p}{K^\ast}
 \label{dJmup}
 \end{equation}
Note that formally Eq.~\ref{Jmup} can be regarded as a thermodynamic process in $(\mu,P,T)$ space  which is unphysical for liquids.  LC  crystals  are held together by the network constraint (see also the discussion in Ref.~\citenum{sprik25}).  

The open system moduli of Eq.~\ref{Hookmu} deviate from their closed system counterparts Eq.~\ref{Kdx}.  LC show that the difference is quadratic in the effective strain-composition coupling parameter.  This can be readily verified in our scheme for the bulkmodulus by subtracting the expression for $K$  of Eq.~\ref{Kdx} from $K^\ast$ of Eq.~\ref{Kmu}
\begin{equation}
K^\ast - K = \lambda_{\mathrm{b}} - p_0 = - p_0 \tau_{\mathrm{b}}
\label{dKast}
\end{equation}
The second equality uses Eqs~\ref{lambdab} and \ref{taudef}.    The corresponding compressibilities are related as
\begin{equation}
\frac{1}{K^\ast}  =  \frac{1}{K} \left( 1 + \frac{ p_0  \tau_{\mathrm{b}} }{K}  \right)
\label{dKinv}
\end{equation}
The coefficient $\tau_{\mathrm{b}}$ is proportional to $\gamma_{\mathrm{b}} = p^0/\rho^0$  in the small coupling approximation Eq.~\ref{taulim}. The open system is softer irrespective whether interactions are repulsive or attractive.  The  second order effect is in agreement with the LC result for  open system compressibilities\cite{cahn85}.  
However,  compared to the bare elastic bulkmodulus $K_{\mathrm{e}}$ both  $K$ and $K^\ast$ show a difference first order in $p_0$.  Such a perturbation term seems to be missing in standard LC's  theory  leaving the possibility open that this might actually represent a discrepancy.  Further comment will be deferred to the discussion in section \ref{sec:conclook}. 

The concept of chemical pressure  suggests a direct mechanistic explanation of the  enhanced compressibility of open systems.  Stretching of a lattice moves the sites further apart.  While the number of sites remains the same,  spatial density is reduced  prompting an open system to absorb particles from the reservoir.  The result is a net change in chemical pressure partially compensating for the elastic tension.   To be more precise,  compare the difference inresponse of an open and closed system to the same  infinitesimal change $d \Delta J$.  Elastic stress is entirely determined by $d\Delta J$ and is therefore  unable to differentiate between open and closed crystals.  The same holds for  the mechanical component of the chemical pressure.   For the closed uniform system the infinitesimal change $dp_{\mathrm{b}}$  directly follows by setting $d \Delta \xi = 0$ in the differential form of Eq.~\ref{selfstress}. 
\begin{equation}
dp_{\mathrm{b}} = - p_0 d \Delta J
\label{dselfstress0}
\end{equation}
Open and  closed crystal are expanded by the same fraction.  $d \Delta J^\ast = d \Delta J$ and therefore Eq.~\ref{dselfstress0} equally contributes to the change in the chemical pressure of the open system. But that is not all.

 In addition the open system absorbs (releases) a small amount of particles during the dilatation (compression) removing or creating vacancies.  The infinitesimal change  $d \Delta \xi$ in vacancy population
 is obtained from the differential form of Eq.~\ref{dcdmu}.  Setting $d \Delta \mu = 0$  gives
\begin{equation}
d \Delta \xi  = - \tau_{\mathrm{b}} d \Delta J
\label{dDxidDJ}
\end{equation}
$d \Delta \xi < 0$ for dilatation meaning less vacancies per site and therefore more particles.   Marking the  chemical pressure in the stretched open system with an asterisk  and using Eq.~\ref{selfstress}  we can write 
 $dp_{\mathrm{b}}^\ast$ as 
\begin{equation}
dp_{\mathrm{b}}^\ast  = \left( p_0 \tau_{\mathrm{b}} - p_0  \right) d \Delta J
\label{dselfstressdJ}
\end{equation}
Subtracting Eq.~\ref{dselfstress0} we obtain
\begin{equation}
dp_{\mathrm{b}}^\ast  - dp_{\mathrm{b}} = p_0 \tau_{\mathrm{b}}  d \Delta J
\label{dselfstressmu}
\end{equation}
Influx of particles due to stretching at constant chemical potential raises the chemical pressure.  Similarly outflow of particles induced by  compression reduces the chemical pressure.  The  change relative to the closed system under going the same deformation is quadratic in the composition-strain coupling strength consistent with the second order decrease in bulk modulus (Eq.~\ref{dKast}). 

The analysis above underlines an important property of chemical pressure.   The chemical response  is a secondary effect compared to the direct mechanical effect.  With  Eq.~\ref{dselfstress0} included the change in the open system chemical pressure  Eq.~\ref{dselfstressdJ} can be expressed as 
 \begin{equation}
d p_{\mathrm{b}}^\ast = - p_0 \tau_{\mathrm{s}}  d \Delta J 
\label{dpbdDJ}
\end{equation}
where Eq.~\ref{taudef}  for the $\tau$  coefficients  has been used.  In the limit $p_0 \rightarrow 0$ the coefficient $\tau_{\mathrm{s}} \rightarrow  1$ (Eq.~\ref{taulim}).  The mechanical contribution Eq.~\ref{dselfstress0} dominates.  The change  in $p_{\mathrm{b}}$  is first order in $p_0$.  

 \subsection{Non-uniform systems and vacancy profile} \label{sec:xprof}
 
A closely related observation by LC is that the linear elastic moduli in non-uniform and open systems are equivalent. This is a profound feature  that proved very convenient in applications.  Transferability is preserved in  the chemical pressure based scheme as is verified below. In principle strains and stresses in non-uniform systems  
are found by  solving the Cauchy equation generated by the stress tensor Eq.~\ref{LLLdx} 
\begin{equation}
\partial_i T_{ij} = 2 \mu_{\mathrm{e}} \partial_i \epsilon_{ij} + \left(\lambda_{\mathrm{e}} + p_0 \right) \partial_j \epsilon_{kk} + p_0 \partial_j \xi =0
\label{dLLLdx}
\end{equation}
The partial derivative with respect to $x_i$ has been abbreviated to $\partial_i$.  Eq.~\ref{dLLLdx} is a partial differential equation for the strain $\epsilon_{ij}$ and an additional independent variable  $\xi$.   A second differential equation is needed.  This is provided by  Eq.~\ref{dcdmu}.  While population $\xi$ and volumetric strain $\Delta J$ may vary in space $\mu$ remains homogeneous under chemical equilibrium conditions establishing a relation between the gradients of $\xi$ and $\Delta J = \epsilon_{kk}$.  
 \begin{equation}
\partial_i \xi = - \tau_{\mathrm{b}}  \partial_i \epsilon_{kk}
 \label{dxiddJ}
 \end{equation}
Substituting closes the force balance equation turning Eq.~\ref{dLLLdx} in a regular linear Cauchy equation.  
 \begin{equation}
2 \mu_{\mathrm{e}} \partial_i  \epsilon_{ij} + 
\left( \lambda_{\mathrm{e}}  + \lambda_{\mathrm{b}} \right) \partial_j \epsilon_{kk} = 0
\label{Cauchydx}
 \end{equation}
where we have made use of the relations between the coefficients defined in Eqs.~\ref{taudef} and \ref{lambdab}.  Indeed the modified $\lambda$  modulus in Eq.~\ref{Cauchydx} is the same as the Lam\'{e}  $\lambda$ in Eqs.~\ref{Kmu} for the open system elastic moduli. 

The discussion of inhomogeneity in this paper is restricted to crystals with fixed total number of particles.  Exchange of particles does take place, but only  between subdomains within the crystal, not with the environment.  The excess vacancy population  $\xi$ was defined in Eq.~\ref{xidef} as  the actual population $c = 1 -n$  relative to  $c^0 = 1 - n^0$ in the reference state.  For closed non-uniform systems it is more convenient to use a uniform initial state $c_0 = 1 -n_0$ as reference. This can be formalized as 
\begin{equation}
\Delta \xi = \xi - \xi_0 = \frac{1}{n^0} \left(c - c_0 \right)
\label{dxidef}
\end{equation}
where $\xi_0$ was defined in  Eq.~\ref{xi0}.  Expresion Eq.~\ref{pblin} for the chemical pressure  can be reformulated with the fluctuations of $\xi$  and the strain as argument.  Using Eqs.~\ref{DJ0hook} and \ref{dxidef}  we obtain
\begin{equation}
p_{\mathrm{b}} - K \Delta J_0 = - p_0 \left( \Delta \xi + \Delta J\right)
\label{selfstress}
\end{equation}
The offset $K \Delta J_0$ is the selfstress of Eq.~\ref{selfstress0}.   

This brings us to the question,  how to compute  $\xi$ given  a spatially varying strain field?  The strain field  could have been obtained  by solving the Cauchy differential equation  Eq.~\ref{Cauchydx}.   Alternatively  we could have used the semi-inverse method converting a stress in a strain field using  Hook's law Eq.~\ref{Hookdx}.   With the strain known Eq.~\ref{dxiddJ}  is a first order derivative relation which can be integrated to give $\xi$ provided we have a way  to fix the  integration constant.   This can be achieved by averaging Eq.~\ref{dcdmu} over body volume $V$.  The equilibrium mean value of a variable $\varphi$ is indicated by an overline
\begin{equation}
\overline{\varphi} = \frac{1}{V} \int _V \phi dv 
\label{dJav}
\end{equation}
 The total number of particles is conserved,  $\overline{\xi} = \xi_0$ before and after deformation.   Moreover $\Delta \mu$ is uniform in equilibrium (in absence of external one-body forces) and therefore \mbox{$\overline{\Delta \mu} = \Delta \mu$}.   The result is a thermodynamic sum rule for  the mean volumetric strain 
 \begin{equation}
\xi_0  = -\beta c^0 \tau_{\mathrm{s}} \Delta \mu -   \tau_{\mathrm{b}} \overline{\Delta J}
\label{dxisum}
\end{equation}
Subtracting Eq.~\ref{dcdmu} and Eq.~\ref{dxisum} cancels the dependence on the chemical potential
\begin{equation} 
\Delta \xi  = - \tau_{\mathrm{b}} \left( \Delta J - \overline{\Delta J} \right)
\label{xidJ}
\end{equation}
 The vacancy profile scales with the spatial variance of volumetric strain.   Homogenous deformation($\Delta J = \overline{\Delta J}$) has no effect on the distribution of vacancies.  Deformation by external forces of a closed system, will change the chemical potential.  Once the vacancy coupled mechanical problem has been solved, giving us a definite result for  $\overline{\Delta J}$,  the chemical potential can be evaluated by substituting in Eq.~\ref{mulin}
 \begin{equation}
\Delta \mu = - \left(\chi_{\mathrm{s}}+  \gamma_{\mathrm{b}}\right) \xi_0 - 
\gamma_{\mathrm{b}} \overline{\Delta J}
\label{mulineq}
\end{equation}
 To summarize, Eq.~\ref{mulineq} gives the equilibrium equilibrium chemical potential of a deformed closed crystal as determined by the fixed average vacancy  population $\xi_0$ and applied surface tractions. The value is relative to a finite hydrostatic reference as explained in section \ref{sec:matref}. 
 Eq.~\ref{mulineq}  can be used both for uniform and non-uniform closed systems. 
 
 \section{Examples} \label{sec:examples}
\subsection{Uniform crystal confined by parallel walls} \label{sec:cubewall}

\vspace*{2mm}
The linear theory presented in section \ref{sec:linap} is illustrated with three simple applications to crystals under non-hydrostatic stress and (partially) free boundary conditions. The model systems are all examples we found in the textbook by Lubarda and Lubarda (LL)\cite{lubarda20}.  LL solve these problems using the semi-inverse method.  The idea is to make reasonable assumptions for stress and strain,  verify  that they satisfy Hooke's or Cauchy's equation and then work backwards to obtain the deformation by integration.  We have extended this scheme to incorporate chemical equilibrium.  LL  indicate Cauchy stress  by  $\sigma_{ij}$ which is what we will do from now on.  The examples deal with normal stress only.   There are no external one-body forces in the first two examples,  the crystal is deformed by  applied surface tractions only.   In the third example we study stretching by the gravitional force. 

We begin with a uniform crystal of cubic shape.  The size is specified by length  $L_0$ of a side in the reference state.  As discussed in section \ref{sec:matref} the reference state is a strain free state under elevated hydrostatic pressure (assuming  repulsive chemical pressure).  The actual length $L$ under zero applied pressure is somewhat larger.  $L$ is obtained by multiplying  $L_0$ by a third of the volumetric eigen stretch as determined from the eigenstrain  Eq.~\ref{DJ0hook} 
 \begin{equation}
\epsilon_0   =  \frac{\Delta J_0}{3},  \quad  L =  \left(1 + \epsilon_0\right)  L_0 
\label{cubeps0}
 \end{equation} 
The initial vacancy population $\xi = \xi_0$ is set by the number of atoms according to Eq.~\ref{xi0}.  $\xi$ is a constant in a closed system which will be considered first.   The cube of initial size $L$  is placed between two smooth rigid walls at $x = \pm L/2$ and is subjected to a pressure $p$  applied in the $y$ direction. The system  is free to expand in the $z$ direction.  The model and  semi-inverse scheme  are based on LL example 3.2. 

The surface force balance determines all components of the stress tensor except in the $x$ direction
\begin{gather}
\sigma_{yy} = - p, \, \sigma_{zz} = 0, \, \sigma_{xx} \neq 0, 
\nonumber \\ \sigma_{xy}= \sigma_{yz} = \sigma_{zx} = 0
\label{cuboundstress}
\end{gather}
$\sigma_{ij}$ is the full Cauchy stress tensor,  i.e.~elastic plus chemical component (Eq.~\ref{Teb}). 
The strain in the $x$ direction is fixed by the confinement.  It is however not zero but equal to eigen strain $\epsilon_0$ of Eq.~\ref{cubeps0}.  The strain in the $y$ and $z$ direction can relax changing the shape to orthorombic without shearing.
\begin{gather}
\epsilon_{xx} = \epsilon_0, \, \epsilon_{yy} \neq 0, \, \epsilon_{zz} \neq 0, 
\nonumber \\ \epsilon_{xy}= \epsilon_{yz} = \epsilon_{zx} = 0
\label{cuboundstrain}
\end{gather}
To determine the nonvanishing stress component $\sigma_{xx}$ we apply Hooke's law Eq.~\ref{Hookdx} for the normal strain in the $x$ direction substituting  Eqs.~\ref{cuboundstress} and \ref{cuboundstrain} 
\begin{equation}
\epsilon_0 = \frac{1}{E} \left(\sigma_{xx} +\nu p \right) +  \frac{p_0}{3 K} \left( 1 - \xi_0 \right)
\label{cuboundHookdx}
\end{equation}
Subtracting Eq.~\ref{DJ0hook} for the eigenstrain  we find
\begin{equation} 
\sigma_{xx} = - \nu p 
\label{cuboundsigxx}
\end{equation}
The simple direct proportionality with applied pressure  is a consequence of the special choice of the fixed distance between the walls. 

Having all three normal stress components in hand we can compute the Cauchy pressure defined in Eq.~\ref{pmdef}. 
 \begin{equation}
 p_{\mathrm{m}} = - \frac{1}{3} \left( \sigma_{xx} + \sigma_{yy}  + \sigma_{zz} \right) = \frac{1 + \nu}{3} p
 \label{pmcube}
 \end{equation}
 Normally  $\nu < 2$  and  the mean normal pressure in the confined geometry is less than the external unidirectional pressure.  In an incompressible system ($\nu = 1/2$) it would be reduced to  half the value of $p$.   The  applied stress is non-hydrostatic  generating deviatoric stress.   Evaluation of the normal component in the $x$ direction gives 
\begin{equation}
\widetilde{\sigma}_{xx} = \sigma_{xx} + p_{\mathrm{m}} = \frac{1 - 2 \nu}{3} p
\label{devcubex}
 \end{equation}
 where we have used the tilde notation of  Eq.~\ref{devdef}.  For the  $yy$ and $zz$ components we have
 \begin{equation}
\widetilde{\sigma}_{yy} = - \frac{2 - \nu}{3} p, \qquad  \widetilde{\sigma}_{zz} =  \frac{1 +  \nu}{3} p
\label{devcubeyz}
 \end{equation}
 $\widetilde{\sigma}_{zz} = p_{\mathrm{m}}$ as required by the free boundary condition in the $z$ direction ($\sigma_{zz}= 0$).   
 
 The  strain induced by the lateral pressure is found by substituting the computed stress in Hook's law Eq.~\ref{Hookdx}. 
\begin{align}
\epsilon_{yy} - \epsilon_0  & = -\frac{1}{E}  \left(1 -\nu^2 \right)p 
\nonumber \\
\epsilon_{zz} - \epsilon_0 & = \frac{1}{E}  \left(1 +\nu\right) \nu p
\end{align}
The strain is well behaved for values of Poisson's ratio in the interval $0<\nu<1$. The volumetric strain follows by summing the normal strains
\begin{equation}
\Delta J - \Delta J_0 = -\frac{\left(1 + \nu \right)}{3 K} p = -\frac{p_{\mathrm{m}}}{K}
\label{DDJcube}
 \end{equation}
 The second equality is obtained on account of  Eq.~\ref{pmcube} confirming that the mean Cauchy pressure acts as an effective hydrostatic pressure determining the volumetric strain under non-hydrostatic conditions.  
 
The continuum mechanics of Eqs.~\ref{cuboundstress} -\ref{DDJcube} is copied from the derivation given by  LL for an ideal elastic body.  It remains valid for a crystal with vacancies.  The reason is that we made sure that $\sigma_{ij}$ is the proper full Cauchy stress matrix as explained in the theory part of the paper.    Note that this does not imply that the volume strain Eq.~\ref{DDJcube}  is  simply equal to the value in the uncoupled system. There is first of all the $\Delta J_0$ eigenstrain offset.  Furthermore,  as discussed in section \ref{sec:closop},  coupling also modifies the bulk modulus.  The vacancy strain interaction  adds an extra $p_0$ term to the elastic bulkmodulus $K_{\mathrm{e}}$ (Eq.~\ref{Kdx}).   Poisson's ratio shows a similar effect.
Taylor expansion of $\nu$ as given in Eq.~\ref{Kdx} produces a first order correction of the ``bare'' elastic value $\nu_{\mathrm{e}}$
 \begin{equation}
 \nu = \nu_{\mathrm{e}}\left(1 +
  \frac{\mu_{\mathrm{e}}}{\lambda_{\mathrm{e}}\left(\lambda_{\mathrm{e}}+ \mu_{\mathrm{e}}\right)} p_0 \right)
  \label{nucubedx}
 \end{equation}
Either increase or decrease of Poisson's ratio  is possible depending on whether particle-strain coupling is repulsive or attractive.  
This perturbation is passed on to the Cauchy pressure Eq.~\ref{pmcube} in the clamped crystal.  

Turning to the chemical pressure,  theory suggests that this quantity should provide a more direct manifestation of  strain-vacancy interaction.    Inserting Eq.~\ref{DDJcube} in Eq.~\ref{selfstress},  keeping only contributions first order in coupling strength $p_0$ gives
 \begin{equation}
 p_{\mathrm{b}} - K \Delta J_0 =  \left(\frac{ p_0}{K} \right)  p_{\mathrm{m}}
 \label{pbwalls}
 \end{equation}
 The corresponding mean elastic pressure is  the difference relative to $p_{\mathrm{m}}$
 \begin{equation}
 p_{\mathrm{em}} + K \Delta J_0 = \left(1 - \frac{p_0}{K} \right) p_{\mathrm{m}}
 \label{pewalls}
 \end{equation}
Eqs.~\ref{pbwalls} and \ref{pewalls}  resemble Eqs~\ref{selfstress0} and \ref{pelast0} for the hydrostatic crystal  with $p$  replaced by $p_{\mathrm{m}}$ underlining the unifying role of mean normal pressure.   What to expect for the deviatoric stress? Chemical pressure is hydrostatic.  Deviatoric stress and chemical stress don't mix.   However expressions Eq.~\ref{devcubex} and \ref{devcubeyz} contain  Poisson's ratio.  Deviatoric stress  is therefore still indirectly affected by particle-strain coupling as a result to the perturbation of Poisson's ratio Eq.~\ref{nucubedx}.  
 
 \begin{equation}
 \Delta \mu(p) - \Delta \mu \left(p=0\right) = \gamma_{\mathrm{b}} \frac{\left(1 + \nu \right)}{3 K} p 
 \end{equation}
 
 How different  is the response of the open system in contact with a reservoir maintained at a  fixed chemical potential $\mu$? The mechanical boundary conditions Eqs.~\ref{cuboundstress} and \ref{cuboundstrain} remain in force.   Adjusting the eigenstrain according to Eq.~\ref{DJ0star} we can solve for the Cauchy stress in the clamped $x$ direction  as we did for the closed system  now using the open system Hook's law Eq.~\ref{Hookmu}.  This leads to the counterpart of Eq.~\ref{cuboundHookdx}. 
\begin{equation}
\epsilon_0^\ast = \frac{1}{E} \left(\sigma_{xx}^\ast +\nu^\ast  p \right)   
+\frac{1}{3K^\ast} \left(p_0 + \alpha \Delta \mu \right)
\label{cuboundHookmu}
\end{equation}
 The eigen strain appearing on both side of Eq.~\ref{cuboundHookmu} cancels.  What is left is the  ``asterisk'' modification of Eq.~\ref{cuboundsigxx}
\begin{equation} 
\sigma_{xx}^\ast = - \nu^\ast p 
\label{cuboundsigxxmu}
\end{equation}
Next evaluating the Cauchy  pressure we find an expression of the same form as Eq.~\ref{pmcube}  with $\nu^\ast$ instead of $\nu$
\begin{equation}
p_{\mathrm{m}}^\ast = \frac{1 + \nu^\ast}{3} p
\label{pmcubemu}
\end{equation}
Eq.~\ref{cuboundsigxxmu},  of course,  is what LC wanted us to start from to arrive right away at \ref{pmcubemu} using Eq.~\ref{pmcube} as template.   However,  agreement is not a foregone conclusion because the composition-strain coupling we use is different and so is the resulting compliance $\nu^\ast$ (see below). 

$\nu^\ast$ is the open system Poisson's ratio of Eq.~\ref{Kmu}.   This expression is almost identical to Eq.~\ref{Kdx} for $\nu$ except that $p_0$ has been exchanged for $\lambda_{\mathrm{b}}$ of Eq.~\ref{lambdab}.  Therefore all we have to do to obtain the first order approximation for $\nu^\ast$  is to make the same substitution in Eq.~\ref{nucubedx}. 
\begin{equation}
 \nu^\ast  = \nu_{\mathrm{e}}\left(1 +
  \frac{\mu_{\mathrm{e}}}{\lambda_{\mathrm{e}}\left(\lambda_{\mathrm{e}}+ \mu_{\mathrm{e}}\right)} \lambda_{\mathrm{b}} \right)
 \end{equation}
 The change in $\nu$ is easily evaluated using the properties of the coefficients Eq.~\ref{taudef}
 \begin{equation}
 \nu^\ast  - \nu = 
-  \frac{\mu_{\mathrm{e}}}{2 \left(\lambda_{\mathrm{e}}+ \mu_{\mathrm{e}}\right)^2} p_0 \tau_{\mathrm{b}}
 \end{equation}
Poisson's ratio in the open system is reduced relative to the closed system value by a correction quadratic in the coupling coefficient $p_0$.  This is consistent with LC theory.  However,  as we saw already for the bulk modulus in section \ref{sec:closop}, both  $\nu$ and $\nu^\ast$ deviate  in first order in $p_0$ from the bare elastic Poisson ratio $\nu_{\mathrm{e}}$.

 Continuing with the evaluation of the volumetric strain we end up again with Eq.~\ref{DDJcube} with all quantities decorated with an asteriks
 \begin{equation}
\Delta J^\ast - \Delta J_0^\ast =  - \frac{p_{\mathrm{m}}^\ast}{K^\ast}
\label{dDJcubemu}
 \end{equation}
There is of course a qualitative difference with the closed system.  This is the absorption which may be small but is not zero.  
This is the cause of the softening of open systems as explained in section \ref{sec:closop}.   An hypothetical  experiment  was carried  out  varying volumetric strain  by an initesimal amount $d \Delta J$ at constant chemical potential.  The result is given in Eq.~\ref{dDxidDJ}.   Strain is a thermodynamic state variable and Eq.~\ref{dDxidDJ} remains valid when strain  is treated as a function of pressure and chemical potential.   Simply substituting Eq.~\ref{dDJcubemu}  in Eq.~\ref{dDxidDJ} should give the absorption induced by an infinitesimal lateral pressure $dp$
\begin{equation}
d \Delta \xi  =  \tau_{\mathrm{b}}  \left(\frac{p_{\mathrm{m}}^\ast}{K^\ast}\right) dp 
\end{equation}
The confined crystal responds  to anisotropic pressure  by releasing particles ($\xi > \xi_0 $) just as it would do under hydrostatic compression.  The driving force is the open system Cauchy pressure $p_{\mathrm{m}}^\ast$ which differs from the closed system value by a small amount proportional to $p_0$.   

\subsection{Bending a prismatic beam} \label{sec:beam}
 Strain fields in elastic bodies under externally applied tractions are in general non-uniform.  This the fundamental characteristic of solid rigidity.  Although a relatively minor effect,  in compressible solids even the volumetric strain fluctuates in space.  In a crystal this is manifested as local contraction or dilatation of cell volume.  The pure bending of a prismatic beam as presented in LL section 4.8  is  a minimal example.  The question examined here concerns the induced vacancy profile.
 
The system is specified by LL as follows. The $z$ axis is chosen along the length of the beam. A load is applied  at its ends at $z=0$ and $z=L$ by equal and opposite bending moments $M_x$ aligned along the $x$ axis.  The lateral surface  is traction free.  Bending shrinks the surface area on one side of the beam and stretches the surface on the opposite side.   In a coordinate frame with the origin at the centroid of $(x,y)$ cross sections the part of the body at $y < 0$ will be under compressive stress and the part at $y > 0$ under tension. The $(y,z$) plane is a vertical plane of symmetry. 
   
This clever choice of geometry enabled  LL to set up a semi-inverse  scheme with only a single non-zero stress tensor component
\begin{equation}
\sigma_{zz} = k y, \, \sigma_{xx} = \sigma_{yy} = \sigma_{xy} = \sigma_{yz} = \sigma_{zx} = 0
\end{equation}
The function $\sigma_{zz} = ky$ is a simple linear model capturing tensile stress for $y>0$ and compressive stress for $y < 0$.  
The parameter $k$ is determined by  imposing an integral equilibrium condition on the $\sigma_{zz}$ normal stress distribution. The result is
\begin{equation}
k = \frac{M_x}{I_x}
\end{equation}
where $I_x$ is the second moment of the cross-sectional area $A$ for the $x$ axis
\begin{equation}
I_x = \int_A y^2 da
\end{equation}
The combined effect of symmetry and force balance leads to two more integral relations
\begin{equation}
 \int_A y da = 0, \quad \int_A xy da = 0
 \label{averagy}
\end{equation}
For detailed proof we refer the LL section 4.8.  As we will see,  the Cauchy pressure $p_{\mathrm{m}}$ as defined Eq.~\ref{pmdef} plays again a central role.  The value inside the deformed beam is
\begin{equation}
 p_{\mathrm{m}} = - \frac{ky}{3},  \quad \overline{p_{\mathrm{m}}} = 0 
 \label{pmbent}
 \end{equation} 
 The spatial average(Eq.~\ref{dJav}) vanishes because of  the system symmetry Eq.~\ref{averagy}.

The system is closed.  The strain is found from Hooke's law Eq.~\ref{Hookdx}. This also introduces the coupling to the vacancy population $\xi$.  
\begin{gather}
\epsilon_{zz} = \frac{k}{E} y + \frac{p_0}{3K} \left(1-  \xi \right)
\nonumber \\
\epsilon_{xx} = \epsilon_{yy} = -  \frac{k \nu} {E} y + \frac{p_0}{3K} \left(1-  \xi \right)
\end{gather}
Adding we find for the volumetric strain
\begin{equation}
 \Delta J  - \Delta J_0 =  - \frac{1}{K} \left( p_{\mathrm{m}} +  p_0 \Delta  \xi  \right)
 \label{DJbeam}
\end{equation}
with $\Delta J_0$ the eigenstrain of the undeformed beam Eq.~\ref{DJ0hook}.   $\Delta \xi$ is the vacancy fluctuation of Eq.~\ref{dxidef}.   

$\Delta J$  and $\Delta \xi$  are non-uniform fields as is clear from the coordinate dependence of the Cauchy pressure Eq.~\ref{pmbent}.  To obtain explict expressions for these fields the mechanical equation Eq.~\ref{DJbeam} is coupled to the chemical equation Eq.~\ref{xidJ}.  This requires evaluation of the avarage volume strain $\overline{\Delta J}$  appearing in Eq.~\ref{xidJ}.  
$\overline{\Delta \xi }= 0$ in a closed system.  $\overline{p_{\mathrm{m}}}$ vanishes on account of the peculiar system symmetry.  Hence averaging  Eq.~\ref{DJbeam} gives 
\begin{equation}
\overline{\Delta J} = \Delta J_0
\label{DJavDJ0}
\end{equation}
converting Eq.~\ref{xidJ} to
\begin{equation} 
\Delta \xi = -\tau_{\mathrm{b}}\left( \Delta J - \Delta J_0 \right)
\label{xidJbent}
\end{equation}
Substituting in  Eq.~\ref{DJbeam} and rearranging gives the solution for the volumetric strain
\begin{equation}
 \Delta J - \Delta J_0 =- \frac{p_{\mathrm{m}}}{\left(K- p_0 \tau_{\mathrm{b}} \right) } =
 - \frac{p_{\mathrm{m}}}{K^\ast } 
 \label{dJbeam}
\end{equation}
As predicted by LC theory,  vacancy strain coupling  is accounted for by the  bulkmodulus $K^\ast$ of Eq.~\ref{dKast} for the open system.   Substituting in Eq.~\ref{xidJbent} gives the vacancy profile induced by the bending
\begin{equation}
\Delta \xi  = \tau_{\mathrm{b}} \left( \frac{p_{\mathrm{m}}}{K^\ast} \right)
\label{dxibeam}
\end{equation}
 Assuming repulsive particle interactions ($\tau_{\mathrm{b}} >0$)  vacancies migrate away from the surface under tension ($y > 0 \rightarrow p_{\mathrm{m}} < 0$) towards the surface under compression ($y<0 \rightarrow p_{\mathrm{m}} > 0 $).  Particles diffuse in opposite direction.  
 
 Equations Eq.~\ref{DJbeam} and \ref{dxibeam} are regular linear response equations with the spatially varying Cauchy pressure $p_{\mathrm{m}}$  as driving  force.   $p_{\mathrm{m}}$ of Eq.~\ref{pmbent} is a purely external force determined by the applied  surface tractions which is why  $p_{\mathrm{m}}$ is denied an asterisk.  The difference  between Eqs.~\ref{DJbeam} and \ref{dxibeam}  is  that  $p_{\mathrm{m}}$ in the equation for $\Delta \xi$ is multiplied by the coupling coefficient $p_0$.  
 According to Eq.~\ref{selfstress} the chemical pressure is the sum of the two.  Adding Eqs.~\ref{DJbeam} and \ref{dxibeam} gives
 \begin{equation}
 \Delta \xi + \Delta J = -  \tau_{\mathrm{s}} \left( \frac{p_{\mathrm{m}}}{K^\ast} \right) + \Delta J_0
 \end{equation}
 Substituting in Eq.~\ref{selfstress} dropping the the second order $p_0 \Delta J_0$ term we obtain
\begin{equation}
p_{\mathrm{b}} - K \Delta J_0 = p_0 \tau_{\mathrm{s}} \left( \frac{p_{\mathrm{m}}}{K^\ast} \right)
\label{pbeamb}
\end{equation}
In the small coupling approximation $\tau_{\mathrm{s}} \rightarrow 1$.   The chemical pressure effectively behaves as a mechanical force at constant composition.  

Pursuing this argument even further,  we can combine Eq.~\ref{pbeamb} with Eq.~\ref{dxibeam} using the small coupling limit Eq.~\ref{taulim} for  $\tau_{\mathrm{b}}$.  The result is
\begin{equation}
\Delta \xi = \frac{\beta c^0}{\rho^0} \left( p_{\mathrm{b}} - K \Delta J_0 \right)
\label{dxipbeam}
\end{equation}
The local vacancy population  is modulated by the fluctuations of the chemical pressure relative to the self stress.  What is remarkable is that this relation is universal in the sense that the prefactor does not involve the composition-strain coupling parameter.   Eq.~\ref{dxipbeam} resembles an ideal gas law for vacancies in the low density regime.  It can be shown that Eq.~\ref{dxipbeam} is generally valid for deformations for which Eq.~\ref{DJavDJ0} holds.  This clearly is a restricted class of systems.   Still Eq.~\ref{dxipbeam} is an illustration of the fundamental relationship between vacancy distribution and chemical pressure.

\subsection{Stretching of a bar by its own weight} \label{sec:bar}
In the third example we study a crystal subject  to  a one-particle body force.  This is the gravitational force.  
The example is a generalization of the system treated in LL section 4.6 allowing for vacacies and diffusion.   The problem also requires  a non-trivial  extension of the non-equilibrium continuum mechanics method of GFA and SM who both omit external one-body forces. The force  density is of the form Eq.~\ref{defb0} 
\begin{equation}
\mathbf{b} =  \rho \,  \,\mathbf{b}_0  = \left(\rho_0 + \Delta \rho \right) \mathbf{b}_0
\label{bdrho}
\end{equation}
The vector $\mathbf{b}_0$ is minus the gradient of the gravitational potential. The number density $\rho$ is written as the sum of the uniform density  $\rho_0$ in the force free state and the redistribution  $\Delta \rho$ due to deformation and diffusion. $\Delta \rho$ is at least first order in $\mathbf{b}_0$.  The self consistent correction $\Delta \rho \mathbf{b}_0$  to the force Eq.~\ref{bdrho} can  therefore be treated as a higher order perturbation and will be neglected in  lowest order approximation.  As a result of the truncation of density the  one body force in  the force balance  Eq.~\ref{bcauchy}
\begin{equation}
\mathrm{div} \boldsymbol{\sigma} + \rho_0 \mathbf{b}_0 = 0
\label{Cauchyrho0}
\end{equation} 
is uniform and constant.  The chemical equilibrium condition Eq.~\ref{chembal} is a balance between  the gradient of the chemical potential and the same  vector $\mathbf{b}_0$. 
\begin{equation}
\mathrm{grad} \, \Delta  \mu =   \mathbf{b}_0
\label{chembalb}
\end{equation}
$\Delta  \mu $ is the linearized chemical potential of Eq.~\ref{mulin}.   Similar to density functional theory,  $\mu$ is an intrinsic chemical potential independent of the external potential.

The model system is a bar of length $L$ and cross-sectional area $A$.  It is assumed that the weight of the bar is carried at the upper end $z=L$.  The lower end at $z=0$ is free.  The gravitational force points downward  opposite to the unit vector $\mathbf{e}_z$ in the z direction.   In the approximation adopted above $\mathbf{b}_0 = - \theta \mathbf{e}_{z}$ 
where $\theta = m g$  is the specific weight.  To determine the stress field we again rely on semi-inverse assumptions used by LL
 \begin{equation}
 \sigma_{xx} = \sigma_{yy} = 0, \quad \sigma_{zz} = \sigma_{zz}(z), 
 \end{equation} 
Shear forces are ignored ($\sigma_{xy} = \sigma_{yz} = \sigma_{zx} = 0$).  Applying the force balance Eq.~\ref{Cauchyrho0} we are left with only a single linear differential equation with is easily solved
\begin{equation}
\frac{d \sigma_{zz}}{d z} - \rho_0 \theta = 0 \quad \rightarrow  \sigma_{zz} = \rho_0 \theta z
\end{equation}
$\sigma_{zz}(0) = 0$ because the dangling surface at the lower end  is traction free.   The Cauchy pressure
\begin{equation}
  p_{\mathrm{m}} = - \frac{\sigma_{zz}}{3}  = - \frac{\rho_0\theta}{3 } z , \quad 
   \overline{p_{\mathrm{m}}} = - \frac{\rho_0 \theta }{6} L
   \label{pmbar}
    \end{equation}
    will again play an important role as driving force for the volumetric strain.  The spatial average $ \overline{p_{\mathrm{m}}}< 0$,  the hanging bar as a whole is under tension. 

The number of particles and lattice sites is fixed.   Applying Hooke's law Eq.~\ref{Hookdx} 
 \begin{gather}
 \epsilon_{zz} =  \frac{\rho_0\theta}{E} z + \frac{p_0}{3K} \left(1 - \xi\right)
 \nonumber \\[4pt]
 \epsilon_{xx} = \epsilon_{yy} = - \nu \frac{ \rho_0 \theta}{E} z+ \frac{p_0}{3K} \left(1 - \xi \right)
 \end{gather}
 we find for the volumetric strain 
 \begin{equation}
 \Delta J =  \Delta J_0 - \frac{1}{K} \left(  p_{\mathrm{m}}  + p_0 \Delta \xi \right)
 \label{DJbar0}
  \end{equation}
 $\Delta J_0$ is the eigenstrain in absence of gravity and surface forces as  determined by the uniform vacancy population $\xi_0$ in this state (see Eq.~\ref{DJ0hook}).  $p_{\mathrm{m}}$ is the Cauchy pressure of Eq.~\ref{pmbar} and
$\Delta \xi$ is  the vacancy profile along the bar defined in Eq.~\ref{dxidef}.  As we saw for beam  bending in section \ref{sec:beam} the effective  force for deformation is the mean normal stress.  However,  the spatial average of the gravitational $p_{\mathrm{m}}$ is not zero but proportional to the length of the bar (see Eq.~\ref{pmbar}).  Averaging Eq.~\ref{DJbar0} over the body of the bar gives the mean volumetric strain.  $\overline{\Delta \xi} = 0$ for closed systems and we obtain
 \begin{equation}
 \overline{\Delta J} = \Delta J_0 - \frac{\overline{p_{\mathrm{m}}}}{K}
  \label{DJbarav}
 \end{equation}
 The average strain exceeds the eigenstrain ($\overline{p_{\mathrm{m}}}< 0$).  All atoms are pulled down. 
 
 Introducing the symmetric dimensionless vertical profile function $g(z)$ and thermal  pressure $p_{\mathrm{s}}$
 \begin{equation}
 g (z) = - \beta \theta \left(z - \frac{L}{2} \right), \quad
 p_{\mathrm{s}}   = \frac{ \rho_0 k_{\mathrm{B}}T}{3} 
 \label{defgps}
\end{equation}
the mean pressure fluctuation can be written as 
\begin{equation}
  p_{\mathrm{m}} - \overline{p_{\mathrm{m}}} =  p_{\mathrm{s}} g
 \label{dpmbar}
\end{equation}
and therefore because $p_{\mathrm{m}}(0)  =0$
\begin{equation}
\overline{p_\mathrm{m}} = - p_{\mathrm{s}} g(0)
\label{pmavbar}
\end{equation}
   Substituting in Eq.~\ref{DJbar0} using Eq.~\ref{DJbarav} the volumetric strain fluctuation can be expressed as
\begin{equation}
\Delta J - \overline{\Delta J}  = - \frac{1}{K} \left(p_{\mathrm{s}} g +  p_0 \Delta  \xi  \right)
\label{DJbar}
  \end{equation} 
   Eq.~\ref{DJbeam} is a special case of Eq.~\ref{DJbar} when average volumetric strain $\overline{\Delta J}$ and  eigenstrain $\Delta J_0$ coincide.

 Continuing along the lines of section \ref{sec:beam},  chemical equilibrium is imposed supplying the second equation for $\Delta \xi$ and $\Delta J$.   For the bent beam this is Eq.~\ref{xidJ} which is a consequence  of the uniformity of the chemical potential in equilibrium.   The equilibrium chemical potential is not uniform in the presence of  external one-body forces (see section \ref{sec:mobil}).  However,  Eq.~\ref{xidJ} is based on Eq.~\ref{mulin} which still applies and therefore also  Eq.~\ref{dcdmu}.  The reason is that  Eq.~\ref{mulin} is a linear approximation for the non-equilibrium intrinsic chemical potential whether uniform or not.    Retracing the  derivation of section \ref{sec:xprof}  we differentiate Eq.~\ref{dcdmu} with respect to $z$,  substitute the gradient  Eq.~\ref{chembalb} of $\mu$,  integrate the result  along $z$ we end up with an equation for $\xi$ containing an unknown 
 integration constant $C$
 \begin{equation}
\xi  = \beta c^0 \tau_{\mathrm{s}} \theta z -   \tau_{\mathrm{b}} \Delta J + C
 \label{dxibarC}
 \end{equation}
As before $C$ is eliminated by substracting the spatial average of $\xi$.  Then using the definitions of Eq.~\ref{defgps} the vacancy profile can be written as 
 \begin{equation}
 \Delta \xi =- \tau_{\mathrm{s}} c^0 g  - \tau_{\mathrm{b}} \left( \Delta J - \overline{\Delta J} \right)
\label{dxibar}
 \end{equation}
complementing Eq.~\ref{DJbar}.  

Before solving Eqs.~\ref{DJbar} and \ref{dxibar} it is intructive to investigate the zero coupling system $p_0 = \tau_{\mathrm{b}} =0, \;  \tau_{\mathrm{s}} = 1$ (see Eq.~\ref{taudef}).  In this limit 
\begin{equation}
\Delta \xi = -c^0 g 
\label{dxigonly}
\end{equation}
 This is the familiar (linearized) barometric  equation.  Of particular interest is the free end at $z=0$.  $g(0) > 0$ and therefore $\Delta \xi (0)   < 0$.  Negative  $\Delta \xi$  means that the vacancy concentration is suppressed ($c < c_0$, see Eq.~\ref{dxidef})).  Substituting  \mbox{$z= L$} gives  the opposite value indicating that a fraction of vacancies has risen to the top pushed by the sinking particles. 
 
 Redistribution of occupation is normal behaviour of a rigid lattice gas subject to an external potential.  However gravity also deforms the lattice itself, even for  $p_0 = 0$.  In that case there is no eigenstrain.  $\Delta J_0 = 0$ in Eq.~\ref{DJbarav}.   Inserting  in Eq.~\ref{DJbar} gives for the gravity-only estimate of the volumetric strain 
 \begin{equation}
 \Delta J = - \frac{p_{\mathrm{s}}}{K} \left(g - g(0) \right) 
 = - \frac{p_{\mathrm{m}}}{K}
 \label{DJgonly}
 \end{equation}
The strain is purely mechanical.  Discounting chemical pressure ($p_0 =0$),  the dangling end is strain free despite the increase in particle density as indicated by the negative value of $\Delta \xi(0)$ in Eq.~\ref{dxigonly}. Tension builds up  towards the depleted suspension point at $z=L$.  

With chemical pressure  included  the modified strain and vacancy profile is found by solving the coupled set of  Eqs.~\ref{DJbar} and \ref{dxibar}.  Substituting Eq.~\ref{dxibar} in Eq.~\ref{DJbar} gives for the strain profile
\begin{equation}
\Delta J - \overline{\Delta J} = - \left( \frac{p_{\mathrm{s}}^\ast }{K^\ast}  \right) g
\label{dJbarask}
\end{equation}
$p_{\mathrm{s}}^\ast $  is the interaction corrected thermal pressure  Eq.~\ref{defgps}. 
\begin{equation}
 p_{\mathrm{s}}^\ast = p_{\mathrm{s}}- p_0 c^0 \tau_{\mathrm{s}}
 \end{equation} 
 As anticipated,  the open system bulkmodulus $K^\ast$ of Eq.~\ref{dKast} has slipped in once again.  
 The resolved  expression for the vacancy profile is
 \begin{equation}
  \Delta \xi = - \tau_{\mathrm{s}} c^0 g + \tau_{\mathrm{b}} \left( \frac{p_{\mathrm{s}}^\ast}{K^\ast} \right)  g
  \label{dxibarask}                                                                                                                            
 \end{equation}
 Vacancy concentrations are low.  $p_{\mathrm{s}}^\ast$ can be assumed to be equal to $p_{\mathrm{s}}$ for most practical purposes.  In this approximation Eq.~\ref{dJbarask} can be recast in a physically more transparant form of a response to the Cauchy stress.  Using Eq.~\ref{dpmbar} we can write
 \begin{equation}
  \Delta J - \Delta J_0    =  - \frac{p_{\mathrm{m}}}{K}  -  \left(\frac{1}{K^\ast} 
   - \frac{1}{K} \right)\left(p_{\mathrm{m}} - \overline{p_{\mathrm{m}}} \right)
\end{equation}
Eq.~\ref{dKinv} for open system compressibility degrades the fluctuation term to a secondary effect.  The leading deformation mechanism  is direct coupling between particle density and strain by the  gravitational  force.   This does not imply that particle-strain coupling can be ignored in first order.   As shown in section \ref{sec:lamehook}  on elastic constants in closed systems the chemical pressure model adds $p_0$  to the  uncoupled bare elastic modulus $K_{\mathrm{e}}$  .

 \section{Summary and outlook} \label{sec:conclook}
 
 Coupling between mechanical and chemical degrees of freedom in the LC theory of alloys is specified in terms composition strain.  In this paper we proposed an alternative mechanism in the form of a chemical pressure determined by the deformed (Eulerian) particle density.   The model in this first exploration is a one-component crystal with vacancies, which in the framework of LC theory can be treated as a binary  alloy.  The spatial particle density  is a product of occupation (composition) and volumetric strain. This gives the chemical pressure a dual dependence leading to more symmetrically structured  governing equations.  The gradient of the chemical pressure acts both as an internal one-body force in the Cauchy equation for the stress and the drift force in Fick's equation for the diffusive current. 
 
 The coupled equations for strain and vacancy profile were obtained using non-equilibrium continuum thermodynamics methods. This lengthy and formal  derivation was deemed necessary for consistent matching between diffusion equation and free energy density.  LC theory as used in material science is based on the Gibbs variational theory of equilibrium thermodynamics.  The effect of deformation is added to the diffusion equation at the end using plausible but non-rigorous arguments.   It was not obvious to us how to adapt this procedure for the chemical pressure based coupling scheme proposed here.   Indeed, the diffusion equation exhibits a novel structure  with the gradient of the chemical pressure directly entering as the drift force and a mobility tensor varying with the principal components of the spatial Cauchy-Green strain tensor.   
 
 Transport was not further investigated in the present paper.  The focus was on  the implications of the novel features of the chemical pressure model on equilibrium properties.   This was analyzed in detail by applying the theory to three examples involving non-isotropic external forces.  Example 1 is a uniform open crystal confined by rigid walls.   The model systems of example 2 and 3 are non-uniform and closed.  Systems 1 and 2 were deformed by surface tractions,  system 3 by an external one body force (gravitation). Systems 1 and 2 complied to the  LC transferability principle. The stress and strain obtained by explicitly coupling mechanical to chemical equilibrium equations  are equal to the results of a purely mechanical treatment using modified elastic moduli.  These open system  effective moduli are derived from the closed system reference value by second perturbation in the effective vacancy-strain coupling strength and are the same for uniform open and heterogenous closed systems. This is again in agreement with LC theory.  
  
 The results for system 3  showed deviations from the transferability rule.  The induced strain is heterogenous, but the linear dependence on the one-body force involved the elastic constants of the closed system.  Furthermore,  the moduli in closed  systems are not simply equal to the purely elastic coefficients in an uncoupled system, but differ from these by a correction first order in coupling strength.  This  applies to all  three examples and is a general feature of the chemical pressure based coupling scheme.  Whether these observations must be interpreted as disagreement with standard LC  was left an open question.    Direct comparison between the two models of composition strain interaction is hindered by ambiguity in the definition of Cauchy stress.   Cauchy stress as understood in this paper is a superposition of elastic and chemical stress.   This  distinction is not explicitly made by LC or may even not be consistent with their approach.  However, as shown in this paper, it matters in the treatment of free boundary conditions of finite bodies which seem to be  less of a concern in the LC  literature.

There is a related fundamental issue not addressed in the current paper.  A detailed comparison to standard LC theory is deferred until this problem is resolved.  The problem is the question of compatibility of strain in systems coupled to composition degrees of freedom.   That there is such a problem is clear from Eq.~\ref{rhosn}.  The ratio of particle density and occupation is volumetric strain and therefore must satisfy Saint Venant's equation to ensure that strain is the gradient of deformation.   Saint Venant's equation  consists of a rather daunting set of second order partial differential equations (see for example LL section  2.9).   The problem was avoided in the examples of section \ref{sec:examples} by considering only strain fields varying linearly with position.  Curved stress and strain fields are ubiquitous in material science.  A much studied example is the stress field generated by a dislocation.  This is also where LC theory has led to controversy (see the discussion about Cottrell atmospheres\cite{nix14,cahn13,mishin16,hirth14a,cahn14,hirth14b}).  In linear elasticity theory Saint Venant's equation can be imposed in terms of compatibility conditions for the stress tensor,  known as the Beltrami-Michell equations (LL section 3.7).  Extension of the Beltrami-Michell equations to include population degrees of freedom is of particular interest and is a priority for future research.  
 
 A further extension under consideration is relaxation of the constraint on the number of lattice sites.   Lattice sites are not rigorously conserved.   They can be created or destroyed at defects such as surfaces,  grain  boundaries and dislocation cores.   Balanced by gain or loss of particles the process  leads to accretion or ablation.  This is the central topic in material science and has become one of the main applications of LC theory\cite{mishin13,mishin15, voorhees24}.   This is in essence a kinetic process for which  the non-equilibrium continuum mechanics method is particularly suitable.  For example Mesarovic has shown how to couple the motion of an interface to supply of new lattice sites filling up the space opened up by the displacement of the surface\cite{mesarovic16}.   It enabled him to model diffusional creep in metals.  This seems to be a good first target for application of the chemical pressure scheme to surface evolution. 
 
 As a second step we can consider coupling chemical pressure to the thermodynamics  of liquid-solid interfaces. Hopefully this  may help the analysis and interpretation of molecular simulation of the effect of  nonhydrostatic stress on planar interfaces\cite{mishin10,frolov10,speck24,speck26} or surface (capillary) stress on curved interfaces\cite{vega20,filion24,frenkel26}.  A further more involved problem could be the migration of grain boundaries \cite{voorhees24,srolovitz18,mishin20,mishin25}.   A necessary condition for these developments is validation of the equilibrium thermodynamics of the chemical pressure model which we hope to have provided in the present publication.

 \appendix

 \section{material free energy imbalance} \label{sec:cnr}
 To verify  Eq.~\ref{freimbR} we transform the spatial integral formulation Eq.~\ref{globalcdh} of the free energy imbalance equation  term by term to integrations over reference volume $\mathcal{P}$. We start with  $\mathcal{W}_0$ of Eq.~\ref{powerbal}.   To make the connection between the spatial and referential representation of mechanical work we use  the fundamental expression of the gradient the velocity in terms of the deformation rate 
 \begin{equation}
\mathrm{grad}  \mathbf{v} =  \overset{\centerdot} {\mathbf{F}}  \mathbf{F}^{-1}
\label{strainrate}
 \end{equation}
 This gives for the work per unit of time exerted by the Cauchy stress  
 \begin{equation}
 \mathbf{T}: \mathrm{grad} \mathbf{v} = \mathbf{T} :  ( \overset{\centerdot} {\mathbf{F}}  \mathbf{F}^{-1} ) 
 = ( \mathbf{T} \mathbf{F}^{-\mathrm{T}}) :   \overset{\centerdot} {\mathbf{F}} 
 \end{equation}
 where we have used the special properties of the double dot contraction
 \begin{multline}
 \mathbf{A}:(\mathbf{B}\mathbf{C}) = \mathrm{tr} ( \mathbf{A}^{\mathrm{T}}\mathbf{B} \mathbf{C})  \\ =
 \mathrm{tr} (\mathbf{C} \mathbf{A}^{\mathrm{T}}\mathbf{B}) = (\mathbf{A}\mathbf{C}^{\mathrm{T}}):\mathbf{B}
 \end{multline}
 This enables us to change over to the Piola stress tensor using Eq.~\ref{Tpiola}
  \begin{equation}
 \mathbf{T}: \mathrm{grad} \mathbf{v} = J^{-1}  \mathbf{T}_{\mathrm{R}} :   \overset{\centerdot} {\mathbf{F}} 
 \label{B1}
 \end{equation}
Subjecting the work rate of the on-body force to a similar treatment we can write using inverse of Eq.~\ref{b0e2l} 
 \begin{equation}
  \mathbf{b}_0 \cdot  \mathbf{h} =  ( \mathbf{F}^{-\mathrm{T}} \mathbf{b}_{0\mathrm{R}})  \cdot \mathbf{h}
  = \mathbf{b}_{0\mathrm{R}} \cdot (\mathbf{F}^{-1} \mathbf{h} )
\end{equation}
Substituting Eq.~\ref{he2l} gives
\begin{equation}
 \mathbf{b}_0 \cdot  \mathbf{h} =  J^{-1} \mathbf{b}_{0\mathrm{R}} \cdot  \mathbf{h}_{\mathrm{R}} 
 \label{B2}
\end{equation}
The prefactor $J^{-1}$ Eqs.~\ref{B1} and \ref{B2} have in common is what is needed to convert volume elements (see Eq.~\ref{dvdae2l}). Putting everything together the result for the reference space representation of the work rate is
 \begin{equation}
\mathcal{W}_0\left( \mathcal{P}_t \right)  = \int_{\mathcal{P}} \left( \mathbf{T}_{\mathrm{R}} : \overset{\centerdot} {\mathbf{F}}  
+ \mathbf{b}_{0\mathrm{R}} \cdot  \mathbf{h}_{\mathrm{R}} \right)   dv_{\mathrm{R}}
\label{w0R}
\end{equation}
The derivation of the stress term was copied from GFA, the diffusion flux term is new. 

The pullback of expression Eq.~\ref{tdifv}  for the chemical energy input is relatively straight forward.  The key point, as stated by GFA, is that a chemical potential is a scalar energy field similar to the external potential $\phi_0$ and invariant under deformation. Its gradient is transformed according to the same rule as given for $\mathbf{b}_0$ in Eq.~\ref{B2} and therefore we have for the flux inproduct  in Eq.~\ref{tdifv}
\begin{equation}
\mathbf{h} \cdot \mathrm{grad} \mu = J^{-1}  \mathbf{h}_{\mathrm{R}} \cdot \nabla \mu
\end{equation}
The material time derivative of occupation in Eq.~\ref{tdifv} has already been dealt with in Eq.~\ref{dNPdR}. 
\begin{equation}
 \mathcal{T}\left( \mathcal{P}_t \right) =  -  \int_{\mathcal{P}}  \left(  \mathbf{h}_{\mathrm{R}} \cdot \nabla \mu 
-  \mu  \overset{\centerdot}{\rho}_{\mathrm{R}}   \right) dv_{\mathrm{R}}
\label{tdifR}
  \end{equation}
  Assembling various work (Eq.~\ref{w0R}) and diffusion terms (Eq.~\ref{tdifR})  we find for  the reference space form of Eq.~\ref{dtglobalcdh} 
\begin{multline}
 \int_{\mathcal{P}}  \overset{\centerdot}{\psi}_{\mathrm{R}} \, dv_{\mathrm{R}} \le 
 \int_{\mathcal{P}} \left( \mathbf{T}_{\mathrm{R}} : \overset{\centerdot} {\mathbf{F}} 
 +  \mu  \overset{\centerdot}{\rho}_{\mathrm{R}}  \right. \\
 \left. -\mathbf{h}_{\mathrm{R}} \cdot \left(\nabla \mu - \mathbf{b}_{0\mathrm{R}} \right) \right)  dv_{\mathrm{R}}
\end{multline}
 where Eq.~\ref{fdote2l} has been applied to convert the left hand side.  Localizing we obtain Eq.~\ref{freimbR}.

  \section{Chemical stress tensor}\label{sec:molstress}
  In this appendix Eq.~\ref{cbpb} for the chemical stress tensor is formally derived from its material definition Eq.~\ref{cbrr2e}.  Applying the product rule to the Cauchy-Green tensor derivative of the material form Eq.~\ref{psibR}  of the site binding density 
  \begin{equation}
 \left( \frac{\partial \hat{\psi}_{\mathrm{bR}}}{\partial \mathbf{C}} \right)
= -  \rho \gamma_b \left( \frac{d J}{d \mathbf{C}} \right)
\label{dpsiRdC}
\end{equation}
$J$ is the square root  of $\det \mathbf{C}$ (Eq.~\ref{detCG}).  The derivative can be evaluated using the cofactor equation for matrix derivative of determinant
\begin{equation}
\frac{ d \det \mathbf{A}}{ d \mathbf{A}} = \left( \det \mathbf{A} \right) \mathbf{A}^{- \mathrm{T}}
\end{equation}
Taking the square root into account 
\begin{equation}
 \frac{d J}{d \mathbf{C}} =  \frac{1}{2} \frac{d \sqrt{\det \mathbf{C}}}{d \mathbf{C}} = \frac{J}{2}  \mathbf{C}^{-1}
\end{equation}
Substituting in Eq.~\ref{cbrr2e} together with Eq.~\ref{dpsiRdC}  we obtain 
\begin{equation}
\mathbf{T}_{\mathrm{b}} = - \rho \gamma_b  \mathbf{FC}^{-1} \mathbf{F}^{\mathrm{T}} 
\end{equation}
Inserting Eq.~\ref{CGdef} for $\mathbf{C}$ the rules of tensor calculus  reduce the product to the unit tensor 
\begin{equation}
 \mathbf{FC}^{-1} \mathbf{F}^{\mathrm{T}} = \mathbf{F} \left(\mathbf{F}^{-1} \mathbf{F}^{-\mathrm{T}}
 \right)\mathbf{F}^{\mathrm{T}} = \mathbf{1}
\end{equation}
The result is the simple isotropic form of Eq.~\ref{cbgb} for the chemical component of the Cauchy stress.


\begin{thebibliography}{66}%
\makeatletter
\providecommand \@ifxundefined [1]{%
 \@ifx{#1\undefined}
}%
\providecommand \@ifnum [1]{%
 \ifnum #1\expandafter \@firstoftwo
 \else \expandafter \@secondoftwo
 \fi
}%
\providecommand \@ifx [1]{%
 \ifx #1\expandafter \@firstoftwo
 \else \expandafter \@secondoftwo
 \fi
}%
\providecommand \natexlab [1]{#1}%
\providecommand \enquote  [1]{``#1''}%
\providecommand \bibnamefont  [1]{#1}%
\providecommand \bibfnamefont [1]{#1}%
\providecommand \citenamefont [1]{#1}%
\providecommand \href@noop [0]{\@secondoftwo}%
\providecommand \href [0]{\begingroup \@sanitize@url \@href}%
\providecommand \@href[1]{\@@startlink{#1}\@@href}%
\providecommand \@@href[1]{\endgroup#1\@@endlink}%
\providecommand \@sanitize@url [0]{\catcode `\\12\catcode `\$12\catcode
  `\&12\catcode `\#12\catcode `\^12\catcode `\_12\catcode `\%12\relax}%
\providecommand \@@startlink[1]{}%
\providecommand \@@endlink[0]{}%
\providecommand \url  [0]{\begingroup\@sanitize@url \@url }%
\providecommand \@url [1]{\endgroup\@href {#1}{\urlprefix }}%
\providecommand \urlprefix  [0]{URL }%
\providecommand \Eprint [0]{\href }%
\providecommand \doibase [0]{https://doi.org/}%
\providecommand \selectlanguage [0]{\@gobble}%
\providecommand \bibinfo  [0]{\@secondoftwo}%
\providecommand \bibfield  [0]{\@secondoftwo}%
\providecommand \translation [1]{[#1]}%
\providecommand \BibitemOpen [0]{}%
\providecommand \bibitemStop [0]{}%
\providecommand \bibitemNoStop [0]{.\EOS\space}%
\providecommand \EOS [0]{\spacefactor3000\relax}%
\providecommand \BibitemShut  [1]{\csname bibitem#1\endcsname}%
\let\auto@bib@innerbib\@empty
\bibitem [{\citenamefont {Balluffi}\ \emph {et~al.}(2005)\citenamefont
  {Balluffi}, \citenamefont {Allen},\ and\ \citenamefont
  {Carter}}]{balluffi05}%
  \BibitemOpen
  \bibfield  {author} {\bibinfo {author} {\bibfnamefont {R.~W.}\ \bibnamefont
  {Balluffi}}, \bibinfo {author} {\bibfnamefont {S.~M.}\ \bibnamefont
  {Allen}},\ and\ \bibinfo {author} {\bibfnamefont {W.~C.}\ \bibnamefont
  {Carter}},\ }\href@noop {} {\emph {\bibinfo {title} {{K}inetics of
  {M}aterials}}}\ (\bibinfo  {publisher} {{W}iley-{I}nterscience},\ \bibinfo
  {address} {Hoboken, New Jersey},\ \bibinfo {year} {2005})\BibitemShut
  {NoStop}%
\bibitem [{\citenamefont {Paul}\ \emph {et~al.}(2014)\citenamefont {Paul},
  \citenamefont {Laurila}, \citenamefont {Vesa~Vuorinen},\ and\ \citenamefont
  {Divinski}}]{divinski14}%
  \BibitemOpen
  \bibfield  {author} {\bibinfo {author} {\bibfnamefont {A.}~\bibnamefont
  {Paul}}, \bibinfo {author} {\bibfnamefont {T.}~\bibnamefont {Laurila}},
  \bibinfo {author} {\bibfnamefont {V.}~\bibnamefont {Vesa~Vuorinen}},\ and\
  \bibinfo {author} {\bibfnamefont {S.~V.}\ \bibnamefont {Divinski}},\
  }\href@noop {} {\emph {\bibinfo {title} {{T}hermodynamics, Diffusion and the
  {K}irkendall Effect in Solids}}}\ (\bibinfo  {publisher} {{S}pringer},\
  \bibinfo {address} {Heidelberg},\ \bibinfo {year} {2014})\BibitemShut
  {NoStop}%
\bibitem [{\citenamefont {Cai}\ and\ \citenamefont {Nix}(2016)}]{nix16}%
  \BibitemOpen
  \bibfield  {author} {\bibinfo {author} {\bibfnamefont {W.}~\bibnamefont
  {Cai}}\ and\ \bibinfo {author} {\bibfnamefont {W.~D.}\ \bibnamefont {Nix}},\
  }\href@noop {} {\emph {\bibinfo {title} {{I}mperfections in Crystalline
  Solids}}}\ (\bibinfo  {publisher} {{C}ambridge {U}niversity {P}ress},\
  \bibinfo {address} {Cambridge UK},\ \bibinfo {year} {2016})\BibitemShut
  {NoStop}%
\bibitem [{\citenamefont {Larch\'{e}}\ and\ \citenamefont
  {Cahn}(1973)}]{cahn73}%
  \BibitemOpen
  \bibfield  {author} {\bibinfo {author} {\bibfnamefont {F.~C.}\ \bibnamefont
  {Larch\'{e}}}\ and\ \bibinfo {author} {\bibfnamefont {J.~W.}\ \bibnamefont
  {Cahn}},\ }\bibfield  {title} {\bibinfo {title} {{A} linear theory of
  thermochemical equilibrium of solids under stress},\ }\href
  {https://doi.org/10.1016/0001-6160(73)90021-7} {\bibfield  {journal}
  {\bibinfo  {journal} {{A}cta {M}etall.}\ }\textbf {\bibinfo {volume} {21}},\
  \bibinfo {pages} {1051} (\bibinfo {year} {1973})}\BibitemShut {NoStop}%
\bibitem [{\citenamefont {Eshelby}(1961)}]{eshelby61}%
  \BibitemOpen
  \bibfield  {author} {\bibinfo {author} {\bibfnamefont {J.~D.}\ \bibnamefont
  {Eshelby}},\ }\bibfield  {title} {\bibinfo {title} {{E}lastic inclusions and
  inhomogeneities.},\ }\href@noop {} {\bibfield  {journal} {\bibinfo  {journal}
  {{P}rog. {S}olid {M}ech.}\ }\textbf {\bibinfo {volume} {2}},\ \bibinfo
  {pages} {89} (\bibinfo {year} {1961})}\BibitemShut {NoStop}%
\bibitem [{\citenamefont {Cai}\ \emph {et~al.}(2014)\citenamefont {Cai},
  \citenamefont {Sills}, \citenamefont {Barnett},\ and\ \citenamefont
  {Nix}}]{nix14}%
  \BibitemOpen
  \bibfield  {author} {\bibinfo {author} {\bibfnamefont {W.}~\bibnamefont
  {Cai}}, \bibinfo {author} {\bibfnamefont {R.}~\bibnamefont {Sills}}, \bibinfo
  {author} {\bibfnamefont {D.}~\bibnamefont {Barnett}},\ and\ \bibinfo {author}
  {\bibfnamefont {W.}~\bibnamefont {Nix}},\ }\bibfield  {title} {\bibinfo
  {title} {{M}odeling a distribution of point defects as misfitting inclusions
  in stressed solids},\ }\href {https://doi.org/10.1016/j.jmps.2014.01.015}
  {\bibfield  {journal} {\bibinfo  {journal} {{J}. {P}hys. {M}ech. {S}olids}\
  }\textbf {\bibinfo {volume} {66}},\ \bibinfo {pages} {154} (\bibinfo {year}
  {2014})}\BibitemShut {NoStop}%
\bibitem [{\citenamefont {Larch\'{e}}\ and\ \citenamefont
  {Cahn}(1978)}]{cahn78}%
  \BibitemOpen
  \bibfield  {author} {\bibinfo {author} {\bibfnamefont {F.}~\bibnamefont
  {Larch\'{e}}}\ and\ \bibinfo {author} {\bibfnamefont {J.~W.}\ \bibnamefont
  {Cahn}},\ }\bibfield  {title} {\bibinfo {title} {{A} nonlinear theory of
  thermochemical equilibrium of solids under stress},\ }\href
  {https://doi.org/10.1016/0001-6160(78)90201-8} {\bibfield  {journal}
  {\bibinfo  {journal} {{A}cta {M}etall.}\ }\textbf {\bibinfo {volume} {26}},\
  \bibinfo {pages} {53} (\bibinfo {year} {1978})}\BibitemShut {NoStop}%
\bibitem [{\citenamefont {Larche}\ and\ \citenamefont {Cahn}(1982)}]{cahn82}%
  \BibitemOpen
  \bibfield  {author} {\bibinfo {author} {\bibfnamefont {F.~C.}\ \bibnamefont
  {Larche}}\ and\ \bibinfo {author} {\bibfnamefont {J.~W.}\ \bibnamefont
  {Cahn}},\ }\bibfield  {title} {\bibinfo {title} {{T}he effect of self-stress
  on diffusion in solids},\ }\href
  {https://doi.org/10.1016/0001-6160(82)90023-2} {\bibfield  {journal}
  {\bibinfo  {journal} {{A}cta {M}etall.}\ }\textbf {\bibinfo {volume} {30}},\
  \bibinfo {pages} {1835} (\bibinfo {year} {1982})}\BibitemShut {NoStop}%
\bibitem [{\citenamefont {Larch\'{e}}\ and\ \citenamefont
  {Cahn}(1985)}]{cahn85}%
  \BibitemOpen
  \bibfield  {author} {\bibinfo {author} {\bibfnamefont {F.~C.}\ \bibnamefont
  {Larch\'{e}}}\ and\ \bibinfo {author} {\bibfnamefont {J.~W.}\ \bibnamefont
  {Cahn}},\ }\bibfield  {title} {\bibinfo {title} {{T}he interaction of
  composition and stress in crystalline solids},\ }\href
  {https://doi.org/10.1016/0001-6160(85)90077-X} {\bibfield  {journal}
  {\bibinfo  {journal} {{A}cta {M}etall.}\ }\textbf {\bibinfo {volume} {33}},\
  \bibinfo {pages} {331} (\bibinfo {year} {1985})}\BibitemShut {NoStop}%
\bibitem [{\citenamefont {Mullins}(1984)}]{mullins84}%
  \BibitemOpen
  \bibfield  {author} {\bibinfo {author} {\bibfnamefont {W.}~\bibnamefont
  {Mullins}},\ }\bibfield  {title} {\bibinfo {title} {{T}hermodynamic
  equilibrium of a crystalline sphere in a fluid},\ }\href
  {https://doi.org/10.1063/1.447779} {\bibfield  {journal} {\bibinfo  {journal}
  {{J}. {C}hem. {P}hys.}\ }\textbf {\bibinfo {volume} {81}},\ \bibinfo {pages}
  {1436} (\bibinfo {year} {1984})}\BibitemShut {NoStop}%
\bibitem [{\citenamefont {Mullins}\ and\ \citenamefont
  {Sekerka}(1985)}]{mullins85}%
  \BibitemOpen
  \bibfield  {author} {\bibinfo {author} {\bibfnamefont {W.~W.}\ \bibnamefont
  {Mullins}}\ and\ \bibinfo {author} {\bibfnamefont {R.~F.}\ \bibnamefont
  {Sekerka}},\ }\bibfield  {title} {\bibinfo {title} {{On} the thermodynamics
  of crystalline solids},\ }\href {https://doi.org/10.1063/1.448644} {\bibfield
   {journal} {\bibinfo  {journal} {{J}. {C}hem. {P}hys.}\ }\textbf {\bibinfo
  {volume} {82}},\ \bibinfo {pages} {5192} (\bibinfo {year}
  {1985})}\BibitemShut {NoStop}%
\bibitem [{\citenamefont {Leo}\ and\ \citenamefont
  {Sekerka}(1989)}]{sekerka89}%
  \BibitemOpen
  \bibfield  {author} {\bibinfo {author} {\bibfnamefont {P.~H.}\ \bibnamefont
  {Leo}}\ and\ \bibinfo {author} {\bibfnamefont {R.~F.}\ \bibnamefont
  {Sekerka}},\ }\bibfield  {title} {\bibinfo {title} {{T}he interaction of
  composition and stress in crystalline solids},\ }\href
  {https://doi.org/10.1016/0001-6160(89)90184-3} {\bibfield  {journal}
  {\bibinfo  {journal} {{A}cta {M}etall.}\ }\textbf {\bibinfo {volume} {37}},\
  \bibinfo {pages} {3119} (\bibinfo {year} {1989})}\BibitemShut {NoStop}%
\bibitem [{\citenamefont {Frolov}\ and\ \citenamefont
  {Mishin}(2010{\natexlab{a}})}]{mishin10}%
  \BibitemOpen
  \bibfield  {author} {\bibinfo {author} {\bibfnamefont {T.}~\bibnamefont
  {Frolov}}\ and\ \bibinfo {author} {\bibfnamefont {Y.}~\bibnamefont
  {Mishin}},\ }\bibfield  {title} {\bibinfo {title} {{E}ffect of nonhydrostatic
  stresses on solid-fluid equilibrium. {I}. {B}ulk thermodynamics},\ }\href
  {https://doi.org/10.1103/PhysRevB.82.174113} {\bibfield  {journal} {\bibinfo
  {journal} {{P}hys. {R}ev. {B}}\ }\textbf {\bibinfo {volume} {82}},\ \bibinfo
  {pages} {174113} (\bibinfo {year} {2010}{\natexlab{a}})}\BibitemShut
  {NoStop}%
\bibitem [{\citenamefont {Frolov}\ and\ \citenamefont
  {Mishin}(2012)}]{mishin12}%
  \BibitemOpen
  \bibfield  {author} {\bibinfo {author} {\bibfnamefont {T.}~\bibnamefont
  {Frolov}}\ and\ \bibinfo {author} {\bibfnamefont {Y.}~\bibnamefont
  {Mishin}},\ }\bibfield  {title} {\bibinfo {title} {{T}hermodynamics of
  coherent interfaces under mechanical stresses. i. theory},\ }\href
  {https://doi.org/10.1103/PhysRevB.85.224106} {\bibfield  {journal} {\bibinfo
  {journal} {{P}hys. {R}ev. {B}}\ }\textbf {\bibinfo {volume} {85}},\ \bibinfo
  {pages} {224106} (\bibinfo {year} {2012})}\BibitemShut {NoStop}%
\bibitem [{\citenamefont {Cahn}(2013)}]{cahn13}%
  \BibitemOpen
  \bibfield  {author} {\bibinfo {author} {\bibfnamefont {J.~W.}\ \bibnamefont
  {Cahn}},\ }\bibfield  {title} {\bibinfo {title} {{T}hermodynamic aspects of
  {C}ottrell atmospheres},\ }\href
  {https://doi.org/10.1080/14786435.2013.793853} {\bibfield  {journal}
  {\bibinfo  {journal} {{P}hilos. {M}ag. {A}}\ }\textbf {\bibinfo {volume}
  {93}},\ \bibinfo {pages} {3741} (\bibinfo {year} {2013})}\BibitemShut
  {NoStop}%
\bibitem [{\citenamefont {Mishin}\ and\ \citenamefont {Cahn}(2016)}]{mishin16}%
  \BibitemOpen
  \bibfield  {author} {\bibinfo {author} {\bibfnamefont {Y.}~\bibnamefont
  {Mishin}}\ and\ \bibinfo {author} {\bibfnamefont {J.~W.}\ \bibnamefont
  {Cahn}},\ }\bibfield  {title} {\bibinfo {title} {{T}hermodynamics of
  {C}ottrell atmospheres tested by atomistic simulations},\ }\href
  {https://doi.org/10.1016/j.actamat.2016.07.013} {\bibfield  {journal}
  {\bibinfo  {journal} {{A}cta {M}ater.}\ }\textbf {\bibinfo {volume} {117}},\
  \bibinfo {pages} {197} (\bibinfo {year} {2016})}\BibitemShut {NoStop}%
\bibitem [{\citenamefont {Mishin}\ \emph {et~al.}(2013)\citenamefont {Mishin},
  \citenamefont {Warren}, \citenamefont {Sekerka},\ and\ \citenamefont
  {Boettinger}}]{mishin13}%
  \BibitemOpen
  \bibfield  {author} {\bibinfo {author} {\bibfnamefont {Y.}~\bibnamefont
  {Mishin}}, \bibinfo {author} {\bibfnamefont {J.~A.}\ \bibnamefont {Warren}},
  \bibinfo {author} {\bibfnamefont {R.~F.}\ \bibnamefont {Sekerka}},\ and\
  \bibinfo {author} {\bibfnamefont {W.~J.}\ \bibnamefont {Boettinger}},\
  }\bibfield  {title} {\bibinfo {title} {{I}rreversible thermodynamics of creep
  in crystalline solids},\ }\href {https://doi.org/10.1103/PhysRevB.88.184303}
  {\bibfield  {journal} {\bibinfo  {journal} {{P}hys. {R}ev. {B}}\ }\textbf
  {\bibinfo {volume} {88}},\ \bibinfo {pages} {184303} (\bibinfo {year}
  {2013})}\BibitemShut {NoStop}%
\bibitem [{\citenamefont {Mishin}\ \emph {et~al.}(2015)\citenamefont {Mishin},
  \citenamefont {McFadden}, \citenamefont {Sekerka},\ and\ \citenamefont
  {Boettinger}}]{mishin15}%
  \BibitemOpen
  \bibfield  {author} {\bibinfo {author} {\bibfnamefont {Y.}~\bibnamefont
  {Mishin}}, \bibinfo {author} {\bibfnamefont {G.~B.}\ \bibnamefont
  {McFadden}}, \bibinfo {author} {\bibfnamefont {R.~F.}\ \bibnamefont
  {Sekerka}},\ and\ \bibinfo {author} {\bibfnamefont {W.~J.}\ \bibnamefont
  {Boettinger}},\ }\bibfield  {title} {\bibinfo {title} {{S}harp interface
  model of creep deformation in crystalline solids},\ }\href
  {https://doi.org/10.1103/PhysRevB.92.064113} {\bibfield  {journal} {\bibinfo
  {journal} {{P}hys. {R}ev. {B}}\ }\textbf {\bibinfo {volume} {92}},\ \bibinfo
  {pages} {064113} (\bibinfo {year} {2015})}\BibitemShut {NoStop}%
\bibitem [{\citenamefont {Chadwick}\ and\ \citenamefont
  {Voorhees}(2024)}]{voorhees24}%
  \BibitemOpen
  \bibfield  {author} {\bibinfo {author} {\bibfnamefont {A.~F.}\ \bibnamefont
  {Chadwick}}\ and\ \bibinfo {author} {\bibfnamefont {P.~W.}\ \bibnamefont
  {Voorhees}},\ }\bibfield  {title} {\bibinfo {title} {{E}ffects of vacancy
  transport and surface adsorption on grain boundary migration in pure
  metals},\ }\href {https://doi.org/10.1103/PhysRevMaterials.8.023602}
  {\bibfield  {journal} {\bibinfo  {journal} {{P}hys. {R}ev. {M}ater.}\
  }\textbf {\bibinfo {volume} {8}},\ \bibinfo {pages} {023602} (\bibinfo {year}
  {2024})}\BibitemShut {NoStop}%
\bibitem [{\citenamefont {Voorhees}\ and\ \citenamefont
  {Johnson}(2004)}]{voorhees04}%
  \BibitemOpen
  \bibfield  {author} {\bibinfo {author} {\bibfnamefont {P.~W.}\ \bibnamefont
  {Voorhees}}\ and\ \bibinfo {author} {\bibfnamefont {W.~C.}\ \bibnamefont
  {Johnson}},\ }\bibfield  {title} {\bibinfo {title} {{T}he thermodynamics of
  elastically stressed crystals},\ }\href
  {https://doi.org/10.1016/S0081-1947(04)80003-1} {\bibfield  {journal}
  {\bibinfo  {journal} {{S}olid {S}tate {P}hys.}\ }\textbf {\bibinfo {volume}
  {59}},\ \bibinfo {pages} {1} (\bibinfo {year} {2004})}\BibitemShut {NoStop}%
\bibitem [{\citenamefont {Sprik}(2025)}]{sprik25}%
  \BibitemOpen
  \bibfield  {author} {\bibinfo {author} {\bibfnamefont {M.}~\bibnamefont
  {Sprik}},\ }\bibfield  {title} {\bibinfo {title} {{T}hermodynamics of a
  compressible lattice gas crystal: Generalized {G}ibbs-{D}uhem equation and
  adsorption},\ }\href {https://doi.org/10.1063/5.0283508} {\bibfield
  {journal} {\bibinfo  {journal} {{J}. {C}hem. {P}hys.}\ }\textbf {\bibinfo
  {volume} {163}},\ \bibinfo {pages} {114702} (\bibinfo {year}
  {2025})}\BibitemShut {NoStop}%
\bibitem [{\citenamefont {Gurtin}\ \emph {et~al.}(2010)\citenamefont {Gurtin},
  \citenamefont {Fried},\ and\ \citenamefont {Anand}}]{gurtin10}%
  \BibitemOpen
  \bibfield  {author} {\bibinfo {author} {\bibfnamefont {M.~E.}\ \bibnamefont
  {Gurtin}}, \bibinfo {author} {\bibfnamefont {E.}~\bibnamefont {Fried}},\ and\
  \bibinfo {author} {\bibfnamefont {L.}~\bibnamefont {Anand}},\ }\href@noop {}
  {\emph {\bibinfo {title} {{T}he Mechanics and Thermodynamics of Continua}}}\
  (\bibinfo  {publisher} {{C}ambridge University Press},\ \bibinfo {address}
  {Cambridge},\ \bibinfo {year} {2010})\BibitemShut {NoStop}%
\bibitem [{\citenamefont {Mesarovic}(2016)}]{mesarovic16}%
  \BibitemOpen
  \bibfield  {author} {\bibinfo {author} {\bibfnamefont {S.~D.}\ \bibnamefont
  {Mesarovic}},\ }\bibfield  {title} {\bibinfo {title} {{L}attice continuum and
  diffusional creep},\ }\href {https://doi.org/10.1098/rspa.2016.0039}
  {\bibfield  {journal} {\bibinfo  {journal} {{P}roc. {R}. {S}oc. {A}}\
  }\textbf {\bibinfo {volume} {472}},\ \bibinfo {pages} {20160039} (\bibinfo
  {year} {2016})}\BibitemShut {NoStop}%
\bibitem [{\citenamefont {Berdichevsky}\ \emph {et~al.}(1997)\citenamefont
  {Berdichevsky}, \citenamefont {Hazzledine},\ and\ \citenamefont
  {Shoykhet}}]{berdichevsky97}%
  \BibitemOpen
  \bibfield  {author} {\bibinfo {author} {\bibfnamefont {V.}~\bibnamefont
  {Berdichevsky}}, \bibinfo {author} {\bibfnamefont {P.}~\bibnamefont
  {Hazzledine}},\ and\ \bibinfo {author} {\bibfnamefont {B.}~\bibnamefont
  {Shoykhet}},\ }\bibfield  {title} {\bibinfo {title} {{M}icromechanics of
  diffusional creep},\ }\href {https://doi.org/10.1016/S0020-7225(97)00005-0}
  {\bibfield  {journal} {\bibinfo  {journal} {{I}nt. {J}. {E}ngng. {S}ci.}\
  }\textbf {\bibinfo {volume} {35}},\ \bibinfo {pages} {2003} (\bibinfo {year}
  {1997})}\BibitemShut {NoStop}%
\bibitem [{\citenamefont {Garikipati}\ \emph {et~al.}(2001)\citenamefont
  {Garikipati}, \citenamefont {Bassman},\ and\ \citenamefont
  {Deal}}]{garikipati01}%
  \BibitemOpen
  \bibfield  {author} {\bibinfo {author} {\bibfnamefont {K.}~\bibnamefont
  {Garikipati}}, \bibinfo {author} {\bibfnamefont {L.}~\bibnamefont
  {Bassman}},\ and\ \bibinfo {author} {\bibfnamefont {M.}~\bibnamefont
  {Deal}},\ }\bibfield  {title} {\bibinfo {title} {{A} lattice-based
  micromechanical continuum formulation for stress-driven mass transport in
  polycrystalline solids},\ }\href
  {https://doi.org/10.1016/S0022-5096(00)00081-8} {\bibfield  {journal}
  {\bibinfo  {journal} {{J}. {M}ech. {P}hys. {S}olids}\ }\textbf {\bibinfo
  {volume} {49}},\ \bibinfo {pages} {1209} (\bibinfo {year}
  {2001})}\BibitemShut {NoStop}%
\bibitem [{\citenamefont {Hong}\ \emph {et~al.}(2008)\citenamefont {Hong},
  \citenamefont {Zhao}, \citenamefont {Zhou},\ and\ \citenamefont
  {Suo}}]{zhao08}%
  \BibitemOpen
  \bibfield  {author} {\bibinfo {author} {\bibfnamefont {W.}~\bibnamefont
  {Hong}}, \bibinfo {author} {\bibfnamefont {X.}~\bibnamefont {Zhao}}, \bibinfo
  {author} {\bibfnamefont {J.}~\bibnamefont {Zhou}},\ and\ \bibinfo {author}
  {\bibfnamefont {Z.}~\bibnamefont {Suo}},\ }\bibfield  {title} {\bibinfo
  {title} {{A} theory of coupled diffusion and large deformation in polymeric
  gels},\ }\href {https://doi.org/10.1016/j.jmps.2007.11.010} {\bibfield
  {journal} {\bibinfo  {journal} {{J}. {M}ech. {P}hys. {S}olids}\ }\textbf
  {\bibinfo {volume} {56}},\ \bibinfo {pages} {1779} (\bibinfo {year}
  {2008})}\BibitemShut {NoStop}%
\bibitem [{\citenamefont {Doi}(2009)}]{doi09}%
  \BibitemOpen
  \bibfield  {author} {\bibinfo {author} {\bibfnamefont {M.}~\bibnamefont
  {Doi}},\ }\bibfield  {title} {\bibinfo {title} {{G}el dynamics},\ }\href
  {https://doi.org/10.1143/JPSJ.78.052001} {\bibfield  {journal} {\bibinfo
  {journal} {{J}. {P}hys. {S}oc. {J}pn}\ }\textbf {\bibinfo {volume} {78}},\
  \bibinfo {pages} {052001} (\bibinfo {year} {2009})}\BibitemShut {NoStop}%
\bibitem [{\citenamefont {Baek}\ and\ \citenamefont
  {Srinivasa}(2004)}]{srinivasa2004}%
  \BibitemOpen
  \bibfield  {author} {\bibinfo {author} {\bibfnamefont {S.}~\bibnamefont
  {Baek}}\ and\ \bibinfo {author} {\bibfnamefont {A.~R.}\ \bibnamefont
  {Srinivasa}},\ }\bibfield  {title} {\bibinfo {title} {{D}iffusion of a fluid
  through an elastic solid undergoing large deformation},\ }\href
  {https://doi.org/10.1016/S0020-7462(02)00153-1} {\bibfield  {journal}
  {\bibinfo  {journal} {{I}nt. {J} {N}on {L}inear {M}ech.}\ }\textbf {\bibinfo
  {volume} {39}},\ \bibinfo {pages} {201} (\bibinfo {year} {2004})}\BibitemShut
  {NoStop}%
\bibitem [{\citenamefont {Baek}\ and\ \citenamefont {Pence}(2011)}]{pence2011}%
  \BibitemOpen
  \bibfield  {author} {\bibinfo {author} {\bibfnamefont {S.}~\bibnamefont
  {Baek}}\ and\ \bibinfo {author} {\bibfnamefont {T.~J.}\ \bibnamefont
  {Pence}},\ }\bibfield  {title} {\bibinfo {title} {{I}nhomogeneous deformation
  of elastomer gels in equilibrium under saturated and unsaturated
  conditions},\ }\href {https://doi.org/10.1016/j.jmps.2010.12.013} {\bibfield
  {journal} {\bibinfo  {journal} {{J}. {M}ech. {P}hys. {S}olids}\ }\textbf
  {\bibinfo {volume} {59}},\ \bibinfo {pages} {561} (\bibinfo {year}
  {2011})}\BibitemShut {NoStop}%
\bibitem [{\citenamefont {Templet}\ and\ \citenamefont
  {Steigmann}(2013)}]{steigmann2013}%
  \BibitemOpen
  \bibfield  {author} {\bibinfo {author} {\bibfnamefont {G.~J.}\ \bibnamefont
  {Templet}}\ and\ \bibinfo {author} {\bibfnamefont {D.~J.}\ \bibnamefont
  {Steigmann}},\ }\bibfield  {title} {\bibinfo {title} {{On} the theory of
  diffusion and swelling in finitely deforming elastomers},\ }\href
  {https://doi.org/10.2140/memocs.2013.1.105} {\bibfield  {journal} {\bibinfo
  {journal} {{M}ath. {M}ech. {C}omplex {S}yst.}\ }\textbf {\bibinfo {volume}
  {1}},\ \bibinfo {pages} {105} (\bibinfo {year} {2013})}\BibitemShut {NoStop}%
\bibitem [{\citenamefont {Chester}\ and\ \citenamefont
  {Anand}(2010)}]{anand10}%
  \BibitemOpen
  \bibfield  {author} {\bibinfo {author} {\bibfnamefont {S.~A.}\ \bibnamefont
  {Chester}}\ and\ \bibinfo {author} {\bibfnamefont {L.}~\bibnamefont
  {Anand}},\ }\bibfield  {title} {\bibinfo {title} {{A} coupled theory of fluid
  permeation and large deformations for elastomeric materials},\ }\href
  {https://doi.org/10.1016/j.jmps.2010.07.020} {\bibfield  {journal} {\bibinfo
  {journal} {{J}. {M}ech. {P}hys. {S}olids}\ }\textbf {\bibinfo {volume}
  {58}},\ \bibinfo {pages} {1879} (\bibinfo {year} {2010})}\BibitemShut
  {NoStop}%
\bibitem [{\citenamefont {Chester}\ and\ \citenamefont
  {Anand}(2011)}]{anand11}%
  \BibitemOpen
  \bibfield  {author} {\bibinfo {author} {\bibfnamefont {S.~A.}\ \bibnamefont
  {Chester}}\ and\ \bibinfo {author} {\bibfnamefont {L.}~\bibnamefont
  {Anand}},\ }\bibfield  {title} {\bibinfo {title} {{A} thermo-mechanically
  coupled theory for fluid permeation in elastomeric materials: Application to
  thermally responsive gels},\ }\href
  {https://doi.org/10.1016/j.jmps.2011.07.005} {\bibfield  {journal} {\bibinfo
  {journal} {{J}. {M}ech. {P}hys. {S}olids}\ }\textbf {\bibinfo {volume}
  {59}},\ \bibinfo {pages} {1978} (\bibinfo {year} {2011})}\BibitemShut
  {NoStop}%
\bibitem [{\citenamefont {Anand}(2015)}]{anand15}%
  \BibitemOpen
  \bibfield  {author} {\bibinfo {author} {\bibfnamefont {L.}~\bibnamefont
  {Anand}},\ }\bibfield  {title} {\bibinfo {title} {2014 {D}rucker medal paper:
  {A} derivation of the theory of linear poroelasticity from chemoelasticity},\
  }\href {https://doi.org/10.1115/1.4031049} {\bibfield  {journal} {\bibinfo
  {journal} {{J}. {A}ppl. {M}ech.}\ }\textbf {\bibinfo {volume} {82}},\
  \bibinfo {pages} {111005} (\bibinfo {year} {2015})}\BibitemShut {NoStop}%
\bibitem [{\citenamefont {Morro}(2016)}]{morro16}%
  \BibitemOpen
  \bibfield  {author} {\bibinfo {author} {\bibfnamefont {A.}~\bibnamefont
  {Morro}},\ }\bibfield  {title} {\bibinfo {title} {{D}iffusion in mixtures of
  reacting thermoelastic solids},\ }\href
  {https://doi.org/10.1007/s10659-015-9547-0} {\bibfield  {journal} {\bibinfo
  {journal} {{J}. {E}last.}\ }\textbf {\bibinfo {volume} {123}},\ \bibinfo
  {pages} {59} (\bibinfo {year} {2016})}\BibitemShut {NoStop}%
\bibitem [{\citenamefont {Sprik}(2021)}]{sprik21c}%
  \BibitemOpen
  \bibfield  {author} {\bibinfo {author} {\bibfnamefont {M.}~\bibnamefont
  {Sprik}},\ }\bibfield  {title} {\bibinfo {title} {{C}hemomechanical
  equilibrium at the interface between a simple elastic solid and its liquid
  phase},\ }\href {https://doi.org/10.1063/5.0073316} {\bibfield  {journal}
  {\bibinfo  {journal} {{J}. {C}hem. {P}hys.}\ }\textbf {\bibinfo {volume}
  {155}},\ \bibinfo {pages} {244701} (\bibinfo {year} {2021})}\BibitemShut
  {NoStop}%
\bibitem [{\citenamefont {Sprik}(2024)}]{sprik24}%
  \BibitemOpen
  \bibfield  {author} {\bibinfo {author} {\bibfnamefont {M.}~\bibnamefont
  {Sprik}},\ }\bibfield  {title} {\bibinfo {title} {{O}n the chemical potential
  and grand potential density of solids under non-hydrostatic stress},\ }\href
  {https://doi.org/10.1080/00268976.2024.2441390} {\bibfield  {journal}
  {\bibinfo  {journal} {{Mol}. {P}hys.}\ }\textbf {\bibinfo {volume} {asap}},\
  \bibinfo {pages} {e2441390} (\bibinfo {year} {2024})}\BibitemShut {NoStop}%
\bibitem [{\citenamefont {Sekerka}(2015)}]{sekerka15book}%
  \BibitemOpen
  \bibfield  {author} {\bibinfo {author} {\bibfnamefont {R.~F.}\ \bibnamefont
  {Sekerka}},\ }\href {https://doi.org/10.1016/C2014-0-03233-9} {\emph
  {\bibinfo {title} {{T}hermal {P}hysics}}}\ (\bibinfo  {publisher}
  {{E}lsevier},\ \bibinfo {address} {Amsterdam},\ \bibinfo {year}
  {2015})\BibitemShut {NoStop}%
\bibitem [{\citenamefont {Lubarda}\ and\ \citenamefont
  {Lubarda}(2020)}]{lubarda20}%
  \BibitemOpen
  \bibfield  {author} {\bibinfo {author} {\bibfnamefont {M.~V.}\ \bibnamefont
  {Lubarda}}\ and\ \bibinfo {author} {\bibfnamefont {V.~A.}\ \bibnamefont
  {Lubarda}},\ }\href@noop {} {\emph {\bibinfo {title} {{I}ntermediate Solid
  Mechanics}}}\ (\bibinfo  {publisher} {{C}ambridge University Press},\
  \bibinfo {address} {Cambridge},\ \bibinfo {year} {2020})\BibitemShut
  {NoStop}%
\bibitem [{\citenamefont {Domb}(1956)}]{domb56}%
  \BibitemOpen
  \bibfield  {author} {\bibinfo {author} {\bibfnamefont {C.}~\bibnamefont
  {Domb}},\ }\bibfield  {title} {\bibinfo {title} {{S}pecific heats of
  compressible lattices and the theory of melting},\ }\href
  {https://doi.org/10.1063/1.1743060} {\bibfield  {journal} {\bibinfo
  {journal} {{J}. {C}hem. {P}hys.}\ }\textbf {\bibinfo {volume} {25}},\
  \bibinfo {pages} {783} (\bibinfo {year} {1956})}\BibitemShut {NoStop}%
\bibitem [{\citenamefont {Baker}\ and\ \citenamefont {Essam}(1970)}]{essam70}%
  \BibitemOpen
  \bibfield  {author} {\bibinfo {author} {\bibfnamefont {G.~A.}\ \bibnamefont
  {Baker}}\ and\ \bibinfo {author} {\bibfnamefont {J.~W.}\ \bibnamefont
  {Essam}},\ }\bibfield  {title} {\bibinfo {title} {{E}ffects of lattice
  compressibility on critical behavior},\ }\href
  {https://doi.org/10.1103/PhysRevLett.24.447} {\bibfield  {journal} {\bibinfo
  {journal} {{P}hys. {R}ev. {L}ett.}\ }\textbf {\bibinfo {volume} {24}},\
  \bibinfo {pages} {447} (\bibinfo {year} {1970})}\BibitemShut {NoStop}%
\bibitem [{\citenamefont {Oitmaa}\ and\ \citenamefont
  {Barber}(1975)}]{oitmaa75}%
  \BibitemOpen
  \bibfield  {author} {\bibinfo {author} {\bibfnamefont {J.}~\bibnamefont
  {Oitmaa}}\ and\ \bibinfo {author} {\bibfnamefont {M.~N.}\ \bibnamefont
  {Barber}},\ }\bibfield  {title} {\bibinfo {title} {{O}n the critical
  behaviour of an ising system with lattice coupling},\ }\href
  {https://doi.org/10.1088/0022-3719/8/21/036} {\bibfield  {journal} {\bibinfo
  {journal} {{J}. {P}hys. C: {S}olid {S}tate {P}hys.}\ }\textbf {\bibinfo
  {volume} {8}},\ \bibinfo {pages} {3653} (\bibinfo {year} {1975})}\BibitemShut
  {NoStop}%
\bibitem [{\citenamefont {Henriques}\ and\ \citenamefont
  {Salinas}(1987)}]{salinas87}%
  \BibitemOpen
  \bibfield  {author} {\bibinfo {author} {\bibfnamefont {V.~B.}\ \bibnamefont
  {Henriques}}\ and\ \bibinfo {author} {\bibfnamefont {S.~R.}\ \bibnamefont
  {Salinas}},\ }\bibfield  {title} {\bibinfo {title} {{E}ffective spin
  hamiltonians for compressible ising models},\ }\href
  {https://doi.org/10.1088/0022-3719/20/16/014} {\bibfield  {journal} {\bibinfo
   {journal} {{J}. {P}hys. C: {S}olid {S}tate {P}hys.}\ }\textbf {\bibinfo
  {volume} {20}},\ \bibinfo {pages} {2415} (\bibinfo {year}
  {1987})}\BibitemShut {NoStop}%
\bibitem [{\citenamefont {Cerdeiri\~{n}a}\ \emph {et~al.}(2016)\citenamefont
  {Cerdeiri\~{n}a}, \citenamefont {Orkoulas},\ and\ \citenamefont
  {Fisher}}]{fisherm16}%
  \BibitemOpen
  \bibfield  {author} {\bibinfo {author} {\bibfnamefont {C.~A.}\ \bibnamefont
  {Cerdeiri\~{n}a}}, \bibinfo {author} {\bibfnamefont {G.}~\bibnamefont
  {Orkoulas}},\ and\ \bibinfo {author} {\bibfnamefont {M.~E.}\ \bibnamefont
  {Fisher}},\ }\bibfield  {title} {\bibinfo {title} {{C}ompressible cell gas
  models for asymmetric fluid criticality},\ }\href
  {https://doi.org/10.1103/PhysRevLett.116.040601} {\bibfield  {journal}
  {\bibinfo  {journal} {{P}hys. {R}ev. {L}ett.}\ }\textbf {\bibinfo {volume}
  {116}},\ \bibinfo {pages} {040601} (\bibinfo {year} {2016})}\BibitemShut
  {NoStop}%
\bibitem [{\citenamefont {Cerdeiri\~{n}a}\ and\ \citenamefont
  {Orkoulas}(2017)}]{cerdeirina17}%
  \BibitemOpen
  \bibfield  {author} {\bibinfo {author} {\bibfnamefont {C.~A.}\ \bibnamefont
  {Cerdeiri\~{n}a}}\ and\ \bibinfo {author} {\bibfnamefont {G.}~\bibnamefont
  {Orkoulas}},\ }\bibfield  {title} {\bibinfo {title} {{C}ompressible cell gas
  models for asymmetric fluid criticality},\ }\href
  {https://doi.org/10.1103/PhysRevE.95.032105} {\bibfield  {journal} {\bibinfo
  {journal} {{P}hys. {R}ev. {E}}\ }\textbf {\bibinfo {volume} {95}},\ \bibinfo
  {pages} {032105} (\bibinfo {year} {2017})}\BibitemShut {NoStop}%
\bibitem [{\citenamefont {Campa}\ \emph {et~al.}(2009)\citenamefont {Campa},
  \citenamefont {Dauxois},\ and\ \citenamefont {Ruffo}}]{ruffo09}%
  \BibitemOpen
  \bibfield  {author} {\bibinfo {author} {\bibfnamefont {A.}~\bibnamefont
  {Campa}}, \bibinfo {author} {\bibfnamefont {T.}~\bibnamefont {Dauxois}},\
  and\ \bibinfo {author} {\bibfnamefont {S.}~\bibnamefont {Ruffo}},\ }\bibfield
   {title} {\bibinfo {title} {{S}tatistical mechanics and dynamics of solvable
  models with long-range interactions},\ }\href
  {https://doi.org/10.1016/j.physrep.2009.07.001} {\bibfield  {journal}
  {\bibinfo  {journal} {{P}hys. {R}ep.}\ }\textbf {\bibinfo {volume} {480}},\
  \bibinfo {pages} {57} (\bibinfo {year} {2009})}\BibitemShut {NoStop}%
\bibitem [{\citenamefont {Campa}\ \emph {et~al.}(2014)\citenamefont {Campa},
  \citenamefont {Dauxois}, \citenamefont {Fanelli},\ and\ \citenamefont
  {Ruffo}}]{ruffo14}%
  \BibitemOpen
  \bibfield  {author} {\bibinfo {author} {\bibfnamefont {A.}~\bibnamefont
  {Campa}}, \bibinfo {author} {\bibfnamefont {T.}~\bibnamefont {Dauxois}},
  \bibinfo {author} {\bibfnamefont {D.}~\bibnamefont {Fanelli}},\ and\ \bibinfo
  {author} {\bibfnamefont {S.}~\bibnamefont {Ruffo}},\ }\href@noop {} {\emph
  {\bibinfo {title} {{P}hysics of long-range interacting systems.}}}\ (\bibinfo
   {publisher} {{O}xford {U}niversity {P}ress},\ \bibinfo {address} {Oxford},\
  \bibinfo {year} {2014})\BibitemShut {NoStop}%
\bibitem [{\citenamefont {Frechette}\ \emph {et~al.}(2020)\citenamefont
  {Frechette}, \citenamefont {Dellago},\ and\ \citenamefont
  {Geissler}}]{dellago20}%
  \BibitemOpen
  \bibfield  {author} {\bibinfo {author} {\bibfnamefont {L.~B.}\ \bibnamefont
  {Frechette}}, \bibinfo {author} {\bibfnamefont {C.}~\bibnamefont {Dellago}},\
  and\ \bibinfo {author} {\bibfnamefont {P.~L.}\ \bibnamefont {Geissler}},\
  }\bibfield  {title} {\bibinfo {title} {{O}rigin of mean-field behavior in an
  elastic ising model},\ }\href {https://doi.org/10.1103/PhysRevB.102.024102}
  {\bibfield  {journal} {\bibinfo  {journal} {{P}hys. {R}ev. B}\ }\textbf
  {\bibinfo {volume} {102}},\ \bibinfo {pages} {024102} (\bibinfo {year}
  {2020})}\BibitemShut {NoStop}%
\bibitem [{\citenamefont {Frechette}\ \emph {et~al.}(2021)\citenamefont
  {Frechette}, \citenamefont {Dellago},\ and\ \citenamefont
  {Geissler}}]{dellago21}%
  \BibitemOpen
  \bibfield  {author} {\bibinfo {author} {\bibfnamefont {L.~B.}\ \bibnamefont
  {Frechette}}, \bibinfo {author} {\bibfnamefont {C.}~\bibnamefont {Dellago}},\
  and\ \bibinfo {author} {\bibfnamefont {P.~L.}\ \bibnamefont {Geissler}},\
  }\bibfield  {title} {\bibinfo {title} {{E}lastic forces drive nonequilibrium
  pattern formation in a model of nanocrystal ion exchange},\ }\href
  {https://doi.org/10.1073/pnas.2114551118} {\bibfield  {journal} {\bibinfo
  {journal} {{P}roc. {N}atl. {A}cad. {S}ci. U.S.A}\ }\textbf {\bibinfo {volume}
  {118}},\ \bibinfo {pages} {e2114551118} (\bibinfo {year} {2021})}\BibitemShut
  {NoStop}%
\bibitem [{\citenamefont {Evans}(1979)}]{evans79}%
  \BibitemOpen
  \bibfield  {author} {\bibinfo {author} {\bibfnamefont {R.}~\bibnamefont
  {Evans}},\ }\bibfield  {title} {\bibinfo {title} {{T}he nature of the
  liquid-vapour interface and other topics in the statistical mechanics of
  nonuniform, classical fluids},\ }\href
  {https://doi.org/10.1080/00018737900101365} {\bibfield  {journal} {\bibinfo
  {journal} {{A}dv. {P}hys.}\ }\textbf {\bibinfo {volume} {28}},\ \bibinfo
  {pages} {143} (\bibinfo {year} {1979})}\BibitemShut {NoStop}%
\bibitem [{\citenamefont {Hansen}\ and\ \citenamefont
  {McDonald}(2013)}]{hansen13}%
  \BibitemOpen
  \bibfield  {author} {\bibinfo {author} {\bibfnamefont {J.-P.}\ \bibnamefont
  {Hansen}}\ and\ \bibinfo {author} {\bibfnamefont {I.~R.}\ \bibnamefont
  {McDonald}},\ }\href@noop {} {\emph {\bibinfo {title} {{T}heory of Simple
  Liquids}}},\ \bibinfo {edition} {4th}\ ed.\ (\bibinfo  {publisher}
  {{A}cademic {P}ress},\ \bibinfo {address} {Oxford},\ \bibinfo {year}
  {2013})\BibitemShut {NoStop}%
\bibitem [{\citenamefont {Epstein}\ and\ \citenamefont
  {Goriely}(2012)}]{epstein12}%
  \BibitemOpen
  \bibfield  {author} {\bibinfo {author} {\bibfnamefont {M.}~\bibnamefont
  {Epstein}}\ and\ \bibinfo {author} {\bibfnamefont {A.}~\bibnamefont
  {Goriely}},\ }\bibfield  {title} {\bibinfo {title} {{S}elf-diffusion in
  remodeling and growth},\ }\href {https://doi.org/10.1007/s00033-011-0150-3}
  {\bibfield  {journal} {\bibinfo  {journal} {{Z}. {A}ngew. {M}ath. {P}hys.}\
  }\textbf {\bibinfo {volume} {63}},\ \bibinfo {pages} {339} (\bibinfo {year}
  {2012})}\BibitemShut {NoStop}%
\bibitem [{\citenamefont {de~Groot}\ and\ \citenamefont
  {Mazur}(2011)}]{degrootmazur11}%
  \BibitemOpen
  \bibfield  {author} {\bibinfo {author} {\bibfnamefont {S.~R.}\ \bibnamefont
  {de~Groot}}\ and\ \bibinfo {author} {\bibfnamefont {P.}~\bibnamefont
  {Mazur}},\ }\href@noop {} {\emph {\bibinfo {title} {{N}on-{E}quilibrium
  {T}hermodynamics}}}\ (\bibinfo  {publisher} {{D}over},\ \bibinfo {address}
  {New York},\ \bibinfo {year} {2011})\BibitemShut {NoStop}%
\bibitem [{\citenamefont {Coleman}\ and\ \citenamefont
  {Noll}(1963)}]{Coleman1963}%
  \BibitemOpen
  \bibfield  {author} {\bibinfo {author} {\bibfnamefont {B.~D.}\ \bibnamefont
  {Coleman}}\ and\ \bibinfo {author} {\bibfnamefont {W.}~\bibnamefont {Noll}},\
  }\bibfield  {title} {\bibinfo {title} {{T}he thermodynamics of elastic
  materials with heat conduction and viscosity},\ }\href
  {https://doi.org/10.1007/BF01262690} {\bibfield  {journal} {\bibinfo
  {journal} {{A}rch. Rational Mech. Anal.}\ }\textbf {\bibinfo {volume} {13}},\
  \bibinfo {pages} {167} (\bibinfo {year} {1963})}\BibitemShut {NoStop}%
\bibitem [{\citenamefont {Di~Nucci}\ \emph {et~al.}(2024)\citenamefont
  {Di~Nucci}, \citenamefont {Michele},\ and\ \citenamefont
  {Di~Risio}}]{dirisio24}%
  \BibitemOpen
  \bibfield  {author} {\bibinfo {author} {\bibfnamefont {C.}~\bibnamefont
  {Di~Nucci}}, \bibinfo {author} {\bibfnamefont {S.}~\bibnamefont {Michele}},\
  and\ \bibinfo {author} {\bibfnamefont {M.}~\bibnamefont {Di~Risio}},\
  }\bibfield  {title} {\bibinfo {title} {{D}ecomposition of the mechanical
  stress tensor: from the compressible navier–stokes equation to a turbulent
  potential flow model},\ }\href {https://doi.org/10.1007/s00707-024-03961-8}
  {\bibfield  {journal} {\bibinfo  {journal} {{A}cta {M}ech.}\ }\textbf
  {\bibinfo {volume} {235}},\ \bibinfo {pages} {4639} (\bibinfo {year}
  {2024})}\BibitemShut {NoStop}%
\bibitem [{\citenamefont {Hirth}(2014{\natexlab{a}})}]{hirth14a}%
  \BibitemOpen
  \bibfield  {author} {\bibinfo {author} {\bibfnamefont {J.~P.}\ \bibnamefont
  {Hirth}},\ }\bibfield  {title} {\bibinfo {title} {{O}n definitions and
  assumptions in the dislocation theory for solid solutions},\ }\href
  {https://doi.org/10.1080/14786435.2014.951707} {\bibfield  {journal}
  {\bibinfo  {journal} {{P}hilos. {M}ag. {A}}\ }\textbf {\bibinfo {volume}
  {94}},\ \bibinfo {pages} {3162} (\bibinfo {year}
  {2014}{\natexlab{a}})}\BibitemShut {NoStop}%
\bibitem [{\citenamefont {Cahn}(2014)}]{cahn14}%
  \BibitemOpen
  \bibfield  {author} {\bibinfo {author} {\bibfnamefont {J.~W.}\ \bibnamefont
  {Cahn}},\ }\bibfield  {title} {\bibinfo {title} {{R}eprise: partial chemical
  strain dislocations and their role in pinning dislocations to their
  atmospheres},\ }\href {https://doi.org/10.1080/14786435.2014.951711}
  {\bibfield  {journal} {\bibinfo  {journal} {{P}hilos. {M}ag. {A}}\ }\textbf
  {\bibinfo {volume} {94}},\ \bibinfo {pages} {3170} (\bibinfo {year}
  {2014})}\BibitemShut {NoStop}%
\bibitem [{\citenamefont {Hirth}(2014{\natexlab{b}})}]{hirth14b}%
  \BibitemOpen
  \bibfield  {author} {\bibinfo {author} {\bibfnamefont {J.~P.}\ \bibnamefont
  {Hirth}},\ }\bibfield  {title} {\bibinfo {title} {Response to comments},\
  }\href {https://doi.org/10.1080/14786435.2014.952255} {\bibfield  {journal}
  {\bibinfo  {journal} {{P}hilos. {M}ag. {A}}\ }\textbf {\bibinfo {volume}
  {94}},\ \bibinfo {pages} {3177} (\bibinfo {year}
  {2014}{\natexlab{b}})}\BibitemShut {NoStop}%
\bibitem [{\citenamefont {Frolov}\ and\ \citenamefont
  {Mishin}(2010{\natexlab{b}})}]{frolov10}%
  \BibitemOpen
  \bibfield  {author} {\bibinfo {author} {\bibfnamefont {T.}~\bibnamefont
  {Frolov}}\ and\ \bibinfo {author} {\bibfnamefont {Y.}~\bibnamefont
  {Mishin}},\ }\bibfield  {title} {\bibinfo {title} {{E}ffect of nonhydrostatic
  stresses on solid-fluid equilibrium. {II}. {I}nterface thermodynamics},\
  }\href {https://doi.org/10.1103/PhysRevB.82.174114} {\bibfield  {journal}
  {\bibinfo  {journal} {{P}hys. {R}ev. {B}}\ }\textbf {\bibinfo {volume}
  {82}},\ \bibinfo {pages} {174114} (\bibinfo {year}
  {2010}{\natexlab{b}})}\BibitemShut {NoStop}%
\bibitem [{\citenamefont {Mazzucchelli}\ \emph {et~al.}(2024)\citenamefont
  {Mazzucchelli}, \citenamefont {Moulas}, \citenamefont {Kaus},\ and\
  \citenamefont {Speck}}]{speck24}%
  \BibitemOpen
  \bibfield  {author} {\bibinfo {author} {\bibfnamefont {M.~L.}\ \bibnamefont
  {Mazzucchelli}}, \bibinfo {author} {\bibfnamefont {E.}~\bibnamefont
  {Moulas}}, \bibinfo {author} {\bibfnamefont {B.~J.~P.}\ \bibnamefont
  {Kaus}},\ and\ \bibinfo {author} {\bibfnamefont {T.}~\bibnamefont {Speck}},\
  }\bibfield  {title} {\bibinfo {title} {{F}luid-mineral equilibrium under
  nonhydrostatic stress: {I}nsight from molecular dynamics},\ }\href
  {https://doi.org/10.2475/001c.92881} {\bibfield  {journal} {\bibinfo
  {journal} {{A}m. {J}. {S}ci.}\ }\textbf {\bibinfo {volume} {324}},\ \bibinfo
  {pages} {1} (\bibinfo {year} {2024})}\BibitemShut {NoStop}%
\bibitem [{\citenamefont {Mazzucchelli}\ \emph {et~al.}(2026)\citenamefont
  {Mazzucchelli}, \citenamefont {Moulas}, \citenamefont {Schmalholz},
  \citenamefont {Kaus},\ and\ \citenamefont {Speck}}]{speck26}%
  \BibitemOpen
  \bibfield  {author} {\bibinfo {author} {\bibfnamefont {M.~L.}\ \bibnamefont
  {Mazzucchelli}}, \bibinfo {author} {\bibfnamefont {E.}~\bibnamefont
  {Moulas}}, \bibinfo {author} {\bibfnamefont {S.~M.}\ \bibnamefont
  {Schmalholz}}, \bibinfo {author} {\bibfnamefont {B.~J.~P.}\ \bibnamefont
  {Kaus}},\ and\ \bibinfo {author} {\bibfnamefont {T.}~\bibnamefont {Speck}},\
  }\bibfield  {title} {\bibinfo {title} {{I}nstability of fluid‐mineral
  equilibrium under non‐hydrostatic stress investigated with molecular
  dynamics},\ }\href {https://doi.org/10.1029/2025JB033520} {\bibfield
  {journal} {\bibinfo  {journal} {{J}. {G}eophys. {R}es. {S}olid {E}arth}\
  }\textbf {\bibinfo {volume} {131}},\ \bibinfo {pages} {e2025JB033520}
  (\bibinfo {year} {2026})}\BibitemShut {NoStop}%
\bibitem [{\citenamefont {Montero~de Hijes}\ \emph {et~al.}(2020)\citenamefont
  {Montero~de Hijes}, \citenamefont {Shi}, \citenamefont {Noya}, \citenamefont
  {Santiso}, \citenamefont {Gubbins}, \citenamefont {Sanz},\ and\ \citenamefont
  {Vega}}]{vega20}%
  \BibitemOpen
  \bibfield  {author} {\bibinfo {author} {\bibfnamefont {P.}~\bibnamefont
  {Montero~de Hijes}}, \bibinfo {author} {\bibfnamefont {K.}~\bibnamefont
  {Shi}}, \bibinfo {author} {\bibfnamefont {E.~G.}\ \bibnamefont {Noya}},
  \bibinfo {author} {\bibfnamefont {E.~E.}\ \bibnamefont {Santiso}}, \bibinfo
  {author} {\bibfnamefont {K.~E.}\ \bibnamefont {Gubbins}}, \bibinfo {author}
  {\bibfnamefont {E.}~\bibnamefont {Sanz}},\ and\ \bibinfo {author}
  {\bibfnamefont {C.}~\bibnamefont {Vega}},\ }\bibfield  {title} {\bibinfo
  {title} {{T}he {Y}oung-{L}aplace equation for a solid-liquid interfaces},\
  }\href {https://doi.org/10.1063/5.0032602} {\bibfield  {journal} {\bibinfo
  {journal} {{J}. {C}hem. {P}hys.}\ }\textbf {\bibinfo {volume} {153}},\
  \bibinfo {pages} {191102} (\bibinfo {year} {2020})}\BibitemShut {NoStop}%
\bibitem [{\citenamefont {de~Jager}\ \emph {et~al.}(2024)\citenamefont
  {de~Jager}, \citenamefont {Vega}, \citenamefont {Montero~de Hijes},
  \citenamefont {Smallenburg},\ and\ \citenamefont {Filion}}]{filion24}%
  \BibitemOpen
  \bibfield  {author} {\bibinfo {author} {\bibfnamefont {M.}~\bibnamefont
  {de~Jager}}, \bibinfo {author} {\bibfnamefont {C.}~\bibnamefont {Vega}},
  \bibinfo {author} {\bibfnamefont {P.}~\bibnamefont {Montero~de Hijes}},
  \bibinfo {author} {\bibfnamefont {F.}~\bibnamefont {Smallenburg}},\ and\
  \bibinfo {author} {\bibfnamefont {L.}~\bibnamefont {Filion}},\ }\bibfield
  {title} {\bibinfo {title} {{S}tatistical mechanics of crystal nuclei of hard
  spheres},\ }\href {https://doi.org/10.1063/5.0226862} {\bibfield  {journal}
  {\bibinfo  {journal} {{J}. {C}hem. {P}hys.}\ }\textbf {\bibinfo {volume}
  {161}},\ \bibinfo {pages} {184501} (\bibinfo {year} {2024})}\BibitemShut
  {NoStop}%
\bibitem [{\citenamefont {Frenkel}(2026)}]{frenkel26}%
  \BibitemOpen
  \bibfield  {author} {\bibinfo {author} {\bibfnamefont {D.}~\bibnamefont
  {Frenkel}},\ }\bibfield  {title} {\bibinfo {title} {{T}he second gibbs
  paradox},\ }\href {https://doi.org/10.1063/5.0314217} {\bibfield  {journal}
  {\bibinfo  {journal} {{J}. {C}hem. {P}hys.}\ }\textbf {\bibinfo {volume}
  {164}},\ \bibinfo {pages} {094503} (\bibinfo {year} {2026})}\BibitemShut
  {NoStop}%
\bibitem [{\citenamefont {Jian~Hana}\ \emph {et~al.}(2018)\citenamefont
  {Jian~Hana}, \citenamefont {Thomas},\ and\ \citenamefont
  {Srolovitz}}]{srolovitz18}%
  \BibitemOpen
  \bibfield  {author} {\bibinfo {author} {\bibfnamefont {J.}~\bibnamefont
  {Jian~Hana}}, \bibinfo {author} {\bibfnamefont {S.~L.}\ \bibnamefont
  {Thomas}},\ and\ \bibinfo {author} {\bibfnamefont {D.~J.}\ \bibnamefont
  {Srolovitz}},\ }\bibfield  {title} {\bibinfo {title} {{G}rain-boundary
  kinetics: A unifed approach},\ }\href
  {https://doi.org/10.1016/j.pmatsci.2018.05.004} {\bibfield  {journal}
  {\bibinfo  {journal} {{P}rog. {M}ater. {S}ci.}\ }\textbf {\bibinfo {volume}
  {98}},\ \bibinfo {pages} {386} (\bibinfo {year} {2018})}\BibitemShut
  {NoStop}%
\bibitem [{\citenamefont {McFadden}\ \emph {et~al.}(2020)\citenamefont
  {McFadden}, \citenamefont {Boettinger},\ and\ \citenamefont
  {Mishin}}]{mishin20}%
  \BibitemOpen
  \bibfield  {author} {\bibinfo {author} {\bibfnamefont {G.~B.}\ \bibnamefont
  {McFadden}}, \bibinfo {author} {\bibfnamefont {W.~J.}\ \bibnamefont
  {Boettinger}},\ and\ \bibinfo {author} {\bibfnamefont {Y.}~\bibnamefont
  {Mishin}},\ }\bibfield  {title} {\bibinfo {title} {{E}ffect of vacancy
  creation and annihilation on grain boundary motion},\ }\href
  {https://doi.org/10.1016/j.actamat.2019.11.044} {\bibfield  {journal}
  {\bibinfo  {journal} {{A}cta {M}ater.}\ }\textbf {\bibinfo {volume} {185}},\
  \bibinfo {pages} {66} (\bibinfo {year} {2020})}\BibitemShut {NoStop}%
\bibitem [{\citenamefont {Hussein}\ and\ \citenamefont
  {Mishin}(2025)}]{mishin25}%
  \BibitemOpen
  \bibfield  {author} {\bibinfo {author} {\bibfnamefont {O.}~\bibnamefont
  {Hussein}}\ and\ \bibinfo {author} {\bibfnamefont {Y.}~\bibnamefont
  {Mishin}},\ }\bibfield  {title} {\bibinfo {title} {{A} model of full
  thermodynamic stabilization of nanocrystalline alloys},\ }\href
  {https://doi.org/10.1016/j.actamat.2025.121545} {\bibfield  {journal}
  {\bibinfo  {journal} {{A}cta {M}ater.}\ }\textbf {\bibinfo {volume} {301}},\
  \bibinfo {pages} {121545} (\bibinfo {year} {2025})}\BibitemShut {NoStop}%
\end{thebibliography}

%

\end{document}